\documentclass[prd,aps,
nofootinbib,floatfix,
superscriptaddress]{revtex4}
\usepackage{graphicx}
\usepackage{epsfig}
\usepackage{rotating}
\usepackage{amssymb}
\usepackage{subfigure}
\usepackage{dsfont}
\usepackage{psfrag}
\usepackage{amsmath,euscript,array,mathrsfs}
\usepackage{slashed}
\usepackage{array}
\usepackage{youngtab}
\usepackage{color}
\usepackage{bbold}

\usepackage{hyperref}
\hypersetup{
colorlinks = true,
linkcolor = blue,
linktocpage = true,
citecolor = blue
}

\newcommand{\beq}{\begin{equation}}
\newcommand{\eeq}{\end{equation}}
\newcommand{\beqs}{\begin{eqnarray}}
\newcommand{\eeqs}{\end{eqnarray}}
\newcommand{\Tr}{{\rm Tr}}

\newcommand{\dd}{\mbox{d}}

\newcommand{\SigmaSlash}{{\Sigma}\hspace{-5.5pt}{\slash}}
\newcommand{\XiSlash}{{\Xi}\hspace{-5.5pt}{\slash}}

\newcommand{\be}{\begin{equation}}
\newcommand{\ee}{\end{equation}}
\newcommand{\ba}{\begin{array}}
\newcommand{\ea}{\end{array}}

\newcommand{\BigCurlyLeft}{\left\{ \rule{-0.1cm}{0.75cm} \right.}
\newcommand{\BigCurlyRight}{\left. \rule{-0.1cm}{0.75cm} \right\}}

\begin{document}

\title{On Holographic Vacuum Misalignment}
\author{Daniel Elander}
\affiliation{Porto, Portugal}
\author{Ali Fatemiabhari}
\email{a.fatemiabhari.2127756@swansea.ac.uk}
\affiliation{Department of Physics, Faculty  of Science and Engineering,
Swansea University,
Singleton Park, SA2 8PP, Swansea, Wales, UK}
\author{Maurizio Piai}
\affiliation{Department of Physics, Faculty  of Science and Engineering,
Swansea University,
Singleton Park, SA2 8PP, Swansea, Wales, UK}

\date{\today}

\begin{abstract}

We develop a bottom-up holographic model that provides the dual description of a strongly coupled field theory, in which the spontaneous breaking of an approximate global symmetry yields the $SO(5)/SO(4)$ coset relevant to minimal composite-Higgs models. The gravity background is completely regular and smooth, and has an end of space that mimics confinement on the field theory side. We add to the gravity description a set of localised boundary terms, that introduce additional symmetry-breaking effects, capturing those that would result from coupling the dual strongly coupled field theory to an external, weakly coupled sector. Such terms encapsulate the gauging of a subgroup of the global $SO(5)$ symmetry of the dual field theory, as well as additional explicit symmetry-breaking effects. 
We show how to combine spurions and  gauge fixing and how to take the appropriate limits, so as to respect gauge principles and avoid violations of unitarity.

The interplay of bulk and boundary-localised couplings leads to the breaking of the $SO(5)$ symmetry to either its $SO(4)$ or $SO(3)$ subgroup, via vacuum misalignment.  
In field theory terms, the model describes the spontaneous breaking of a $SO(4)$ gauge symmetry to its $SO(3)$ subgroup. We expose the implications of the higgsing phenomenon by computing the spectrum of fluctuations of the model, which we interpret in four-dimensional field-theory terms, for a few interesting choices of parameters. We conclude by commenting on the additional steps needed to build a realistic composite Higgs model. 

\end{abstract}

\maketitle

\tableofcontents

\section{Introduction}
\label{Sec:introduction}

The discovery of the Higgs boson~\cite{Aad:2012tfa,Chatrchyan:2012xdj} provides a compelling argument 
for the study of composite Higgs models (CHMs)~\cite{Kaplan:1983fs,Georgi:1984af,Dugan:1984hq}:
they extend the standard model (SM) of particle physics and can be
tested by Large Hadron Collider (LHC) experiments.
In this context, Higgs fields emerge as composite pseudo-Nambu-Goldstone bosons (PNGBs), 
 in the
 weakly-coupled effective field theory (EFT) description of a more fundamental theory.
Reviews can be found in Refs.~\cite{Panico:2015jxa,Witzel:2019jbe,Cacciapaglia:2020kgq},
and the summary tables in Refs.~\cite{Ferretti:2013kya,Ferretti:2016upr,Cacciapaglia:2019bqz}
provide an interesting classification of possible fundamental field theory origins for CHMs, 
amenable (in principle) to non-perturbative numerical studies
with lattice gauge theory.

The  literature on
 model-building  and phenomenological 
studies (see, e.g., Refs.~\cite{
Katz:2005au,
Barbieri:2007bh,Lodone:2008yy,Gripaios:2009pe,Mrazek:2011iu,Marzocca:2012zn,
Grojean:2013qca,Barnard:2013zea,Cacciapaglia:2014uja,Ferretti:2014qta,Arbey:2015exa,Cacciapaglia:2015eqa,
vonGersdorff:2015fta,Feruglio:2016zvt,DeGrand:2016pgq,
Fichet:2016xvs,Galloway:2016fuo,Agugliaro:2016clv,Belyaev:2016ftv,Bizot:2016zyu,Csaki:2017cep,Chala:2017sjk,
Golterman:2017vdj,Serra:2017poj,Csaki:2017jby,Alanne:2017rrs,Alanne:2017ymh,Sannino:2017utc,Alanne:2018wtp,Bizot:2018tds,
Cai:2018tet,Agugliaro:2018vsu,Cacciapaglia:2018avr,
Gertov:2019yqo,
Ayyar:2019exp,
Cacciapaglia:2019ixa,
BuarqueFranzosi:2019eee,
Cacciapaglia:2019dsq,
Cacciapaglia:2020vyf,
Dong:2020eqy,
Cacciapaglia:2021uqh,Ferretti:2022mpy,
Banerjee:2022izw,
Cai:2022zqu,Cacciapaglia:2024wdn}),  is complemented by an expanding 
body of   numerical lattice calculations,
in theories with gauge group $SU(2)$~\cite{
Hietanen:2014xca,Detmold:2014kba,Arthur:2016dir,Arthur:2016ozw,
Pica:2016zst,Lee:2017uvl,Drach:2017btk,Drach:2020wux,Drach:2021uhl,Bowes:2023ihh},
 $Sp(4)$~\cite{Bennett:2017kga,Lee:2018ztv,Bennett:2019jzz,Bennett:2019cxd, Lucini:2021xke,Bennett:2021mbw,Bennett:2022yfa,Lee:2022xbp,Bennett:2023wjw,Bennett:2023rsl,Bennett:2023gbe,Mason:2023ixv,Bennett:2023mhh,Bennett:2023qwx,
 Hsiao:2023nyn,Bennett:2024cqv, Bennett:2024wda, Dengler:2024maq}
and $SU(4)$~\cite{Ayyar:2017qdf,Ayyar:2018zuk,Ayyar:2018ppa,
 Ayyar:2018glg,Cossu:2019hse,Shamir:2021frg,Lupo:2021nzv,
 DelDebbio:2022qgu,Hasenfratz:2023sqa}.
Results for the  $SU(3)$ theory with $N_f=8$ Dirac fermions~\cite{
Aoki:2014oha,
Appelquist:2016viq,
Aoki:2016wnc,
Gasbarro:2017fmi,
Appelquist:2018yqe,LSD:2023uzj,LatticeStrongDynamics:2023bqp,Ingoldby:2023mtf}
have  been reinterpreted in terms of new CHMs, embedded in the dilaton EFT framework~\cite{Appelquist:2020bqj,
Appelquist:2022qgl}---see also Refs.~\cite{Vecchi:2015fma,Ma:2015gra,BuarqueFranzosi:2018eaj}.
But  the microscopic origin 
of CHMs with  minimal $SO(5)/SO(4)$ coset
is more obscure---see for instance Ref.~\cite{Caracciolo:2012je}.

Central to the CHM model-building programme
 is  the absence of new physics signals in direct and indirect searches,
  indicating that the scale of new phenomena, $f$, is higher
  than the electroweak scale, $v$.
This little hierarchy
 originates in the strong-coupling dynamics, as a consequence of  
 small destabilising perturbations of the vacuum, due to 
 perturbative, weak interactions with an external sector.
 Rephrasing the title of the classical work in Ref.~\cite{Peskin:1980gc},
 the electroweak scale is  suppressed by what is called
the vacuum misalignment angle, $\theta\sim v/f \ll 1$---see, e.g., Sect.~2.2.1 of Ref.~\cite{Cacciapaglia:2020kgq},
and references therein.
In practical terms, phenomenological constraints can be satisfied with a
 rather modest suppression of $\theta$, and only a moderate tuning of
 parameters, which adds to the appeal of CHMs.
Yet, computing $\theta$ from first principles requires 
 non-perturbative methods.

A new alternative avenue for the study of non-perturbative phenomena
opened within the context of string theory and supergravity, with the discovery of 
 gauge-gravity dualities (holography)~\cite{
Maldacena:1997re,Gubser:1998bc,Witten:1998qj,
Aharony:1999ti}.  The strongly coupled regime of   special field theories admits 
an equivalent description as a weakly coupled gravity theory in higher dimensions.
For example, 
type-IIA
supergravity backgrounds, in which the geometry includes  a shrinking internal circle, 
provide a holographic description of
 linear confinement in Yang-Mills theories~\cite{Witten:1998zw},
allowing to study glueball spectra~\cite{Brower:2000rp},
 chiral symmetry breaking~\cite{Karch:2002sh,Babington:2003vm}, and
 masses of  mesons~\cite{Sakai:2004cn,
Sakai:2005yt,Kruczenski:2003be,Nunez:2003cf,Erdmenger:2007cm}.\footnote{
See also the critical discussions in Refs.~\cite{Faedo:2017fbv,Elander:2018gte}}
Also  within type IIB supergravity, backgrounds  exist in which
 a whole portion of internal space, the base of the conifold~\cite{Candelas:1989js}, 
 shrinks
to zero size, giving rise to a rich phenomenology~\cite{
Klebanov:1998hh,Klebanov:2000nc,Klebanov:2000hb,Chamseddine:1997nm,
Maldacena:2000yy,Butti:2004pk},
including the possibility of a light dilaton~\cite{
Elander:2009pk,
Elander:2012yh,
Elander:2014ola,
Elander:2017cle,Elander:2017hyr}.
Simpler bottom-up holographic models dispense with the microscopic origin of the
gravity theory,  complementing lattice explorations
of CHMs~\cite{Erdmenger:2020lvq,Erdmenger:2020flu,Elander:2020nyd,Elander:2021bmt,Erdmenger:2023hkl,Erdmenger:2024dxf}.
It is indeed in this context 
that CHMs based on the minimal $SO(5)/SO(4)$ coset 
have been first developed~\cite{Contino:2003ve,
Agashe:2004rs,
Agashe:2005dk,
Agashe:2006at,
Contino:2006qr,
Falkowski:2008fz,
Contino:2010rs,
Contino:2011np}.

A holographic CHM
derived from  a fundamental gravity theory 
 more accurately predicts the properties of   composite states.
A step towards such a construction
was pursued in Ref.~\cite{Elander:2021kxk},  by exploiting the classical work on 
$S^4$ compactifications of maximal supergravity in $D=11$ dimensions~\cite{
Pilch:1984xy,
Nastase:1999cb,
Pernici:1984xx,
Pernici:1984zw,
Cvetic:1999xp,
Lu:1999bc,
Cvetic:2000ah,
Samtleben:2005bp},
its reductions to $D=7$ dimensions,
and its symmetry-breaking backgrounds~\cite{Campos:2000yu,Elander:2013jqa},
which lead to the appearance of the minimal $SO(5)/SO(4)$ coset.
Reference~\cite{Elander:2021kxk} is  the first  exploratory work, in the larger space of supergravity theories, that features a variety of different coset structures and gauge symmetries---a useful
review on gauged supergravities is  Ref.~\cite{Samtleben:2008pe}.

In this paper, we present the next stage of development of the minimal realisation of
 holographic CHMs. Our motivation for embarking on this task is that 
the ambitious programme  of embedding a holographic CHM within the fully rigorous context of top-down holography
 requires additional preliminary 
work on the formalism, before making contact with model-building and phenomenological aspects.
In the next subsection, we  elaborate on why this is so, and clarify what are the
 objectives of the
present investigation, in which we focus on  developing and testing the formalism  with
a semi-realistic bottom-up model.
The simpler construction we propose 
  captures much of the physically interesting aspects of the CHM programme and is interesting and useful in itself, as a stand-alone model.

\subsection{A roadmap towards top-down holographic composite Higgs}
\label{Sec:roadmap}

The complete construction of a top-down holographic CHM  would involve the following steps.

\begin{itemize}

\item[1)] Identify a fundamental gravity theory (the low energy description of which may be given by supergravity),
providing the dual description of a
 field theory in which spontaneous symmetry breaking involves the
$G/H$ coset relevant to a CHM of interest.

\item[2)]   Find  regular gravity solutions that 
holographically describe  confinement in the dual four-dimensional field theory.

\item[3)] Compute the (holographically renormalised) free energy, and the spectrum of fluctuations, dual to the bound
states of the field theory. Verify the absence of tachyons or other signals of instability.
 
\item[4)]   Extend the gravity theory so that
in the field-theory interpretation a subgroup of the global symmetry is
gauged, with coupling strength weak enough to allow for the perturbative treatment to be viable.

\item[5)]  Extend the gravity theory to implement explicit breaking of the field-theory global symmetry,
compatibly with gauge principles and unitarity. Additional auxiliary fields (spurions)
may be needed.

\item[6)]  Perform a vacuum alignment analysis, within the gravity theory, 
to determine the field-theory vacuum structure.

\item[7)]  Verify that no pathologies emerge in the mass spectrum in the presence of 
 symmetry-breaking terms.

\item[8)]  Couple the theory to standard-model fields,
and identify viable regions of parameter space. 
This step might also involve the introduction of top partial compositeness~\cite{Kaplan:1991dc} (see also 
the discussions in Refs.~\cite{Grossman:1999ra,Gherghetta:2000qt,Lodone:2008yy,
Chacko:2012sy,Grojean:2013qca}).

\end{itemize}

Points 1), 2), and 3)  are  addressed in Ref.~\cite{Elander:2021kxk}; 
the minimal $SO(5)/SO(4)$ coset, relevant to CHMs, emerges
 within maximal supergravity in $D=7$ dimensions, which  lifts to type-IIA supergravity in $D=10$ dimensions.
The spectrum of fluctuations of   bosons, upon compactification on a 2-torus,
is computed~\cite{Elander:2021kxk} for backgrounds  dual to a
confining, four-dimensional field theory,  by exploiting the formalism
 in Refs.~\cite{Bianchi:2003ug,Berg:2005pd,Berg:2006xy,Elander:2009bm,Elander:2010wd}---see
also Refs.~\cite{Elander:2014ola,
Elander:2017cle,Elander:2017hyr,Elander:2020csd,Elander:2020fmv,Roughley:2021suu,Elander:2018aub}.

A  simpler bottom-up model, written in $D=6$ dimensions~\cite{Elander:2022ebt,Elander:2023aow}, readdresses points
 1), 2), and 3), with several
technical advantages: the spectrum is simpler,  the action contains only 
essential fields, with canonical normalisations and minimal interactions.
The model is free from the complexities descending from supersymmetry in higher dimensions,
while retaining the essential features of interest (confinement and symmetry breaking).\footnote{The model presented in this paper
 is a development of the one in Ref.~\cite{Elander:2022ebt}, which, in turn,
is the bottom-up holographic realisation of a mechanism qualitatively similar to the one discovered in the top-down constructions in Refs.~\cite{Elander:2020ial,Elander:2020fmv,Elander:2021wkc}.
Yet, the CHM context of interest here is profoundly different, and so is the range of parameters in the model
that is relevant for current physics considerations. 
To be more explicit, conversely to what is done in Ref.~\cite{Elander:2022ebt}, we are not going to further explore the relation between
the emergence of a classical instability in some range of the parameter space of the model with the 
ideas of Ref.~\cite{Kaplan:2009kr} (nor the more general arguments in Refs.~\cite{Breitenlohner:1982jf,Cohen:1988sq}), 
related  to unitarity bounds, operator dimensions, walking dynamics, complex fixed points, 
and spontaneous breaking of scale invariance---see the extensive discussions in Refs.~\cite{Pomoni:2009joh,Jensen:2010ga,Kutasov:2011fr,Gorbenko:2018ncu,Gorbenko:2018dtm,Pomarol:2019aae,
Faedo:2019nxw,Faedo:2021ksi,Zwicky:2023krx,Pomarol:2023xcc}. In the numerical examples to appear later in the paper, we focus attention on regions of parameter space where the background solutions of interest are stable, 
and pass
both local as well as global stability tests. In particular, in the regions of parameter space of interest, there is no light dilaton (the 
PNGB associated with dilatations) in the spectrum
 of bound states of the dual field theory.}

In this paper, we address points 4), 5), 6), and 7) for the  bottom-up model in Refs.~\cite{Elander:2022ebt,Elander:2023aow},
 in a way that can be generalised to more complicated cosets and geometries.
The $SO(4)$ subgroup of the (field-theory) global symmetry is gauged weakly, and a spurion field
introduces explicit breaking of $SO(5)$ to $SO(4)$. We discuss how to implement 
 gauge-fixing within this setting. Both new features are controlled by a boundary-localised action in the gravity theory.
The interplay with the background dynamics---vacuum (mis)alignment---determines
 whether  the mass spectra
display
$SO(5)\rightarrow SO(4)$ breaking (and unbroken gauged $SO(4)$) or
$SO(5)\rightarrow SO(3)$ breaking (with the gauged $SO(4)$ subgroup higgsed to $SO(3)$).

In contrast to the earliest bottom-up models~\cite{Contino:2010rs}, the
smooth and regular geometry deviates substantially from AdS, and the mass gap is due to the existence of an
endpoint of the radial (holographic) direction~\cite{Witten:1998zw}, while symmetry breaking is triggered by bulk fields, similar to so-called soft-wall models previously considered in the literature~\cite{Karch:2006pv,Falkowski:2008fz,Batell:2008me}. For these reasons, some formal developments warrant special attention, and are the main topic of this paper.
The consequences of holographic vacuum (mis)alignment can be illustrated
within this simple model to highlight
 general results that apply to a large variety of holographic realisations of CHMs.
We  detail necessary, unconventional yet rigorous elements of the formalism, and
subtleties in the (weak) gauging of the  symmetry that generalise
 holographic renormalisation~\cite{Bianchi:2001kw,
Skenderis:2002wp,Papadimitriou:2004ap}. 
We defer
the construction  of a realistic bottom-up holographic CHM~\cite{EWSB}, obtained by
 gauging  the 
standard-model $SU(2)_L\times U(1)_Y\subset SO(4)\times U(1)_{B-L}$---the Abelian factor is related to 
baryon, $B$, and lepton, $L$, quantum numbers---and by adding fermions, either in the bulk or localised at the boundary.

The paper is organised as follows. In Sect.~\ref{Sec:model}, we introduce the six-dimensional bottom-up gravity
model of interest and its $SO(5)$ gauge symmetry.
The profile of a bulk scalar field breaks such symmetry.
A shrinking circle in the geometry mimics confinement on the field theory side.
 We introduce the elements needed to make the exposition self-contained, and  fix  notation and conventions, but 
dispense with details, that may be found in Refs.~\cite{Elander:2022ebt,Elander:2023aow}.
We digress in Sect.~\ref{Sec:SO(5)oSO(4)}, to discuss a rather general description, in effective field theory terms,
of non-linear sigma models based on the  $SO(5)/SO(4)$ coset, in particular in order to clarify the subtle differences between gauge and global symmetries, and explicit and spontaneous symmetry breaking.
In Sect.~\ref{Sec:addboundaryterms}, we return  to the gravity theory. In order to weakly gauge, in the dual field theory,
  an $SO(4)$ subgroup of the $SO(5)$ global symmetry,
we add appropriate boundary terms to the gravity description.
We discuss the implications for the vacuum of the theory, and the resulting symmetry breaking pattern, in Sect.~\ref{Sec:vacuum}.
 In Sect.~\ref{Sec:spectrum}, we display the mass spectrum of fluctuations of the gravity theory, 
which we interpret as bound states of the dual field theory.
We outline further work that we leave for future investigations in Sect.~\ref{Sec:outlook}.
We relegate to the Appendix extensive amounts of technical details.

\section{The gravity model}
\label{Sec:model}

The bottom-up holographic model of interest~\cite{Elander:2023aow} is built by coupling gravity in $D=6$ dimensions 
 to a bulk scalar field, $\mathcal X$, transforming in the vector, real  representation, $5$,
  of a gauged $SO(5)$ symmetry. 
One of the non-compact space-time dimensions, denoted as $\rho$, is interpreted as the holographic direction.
We restrict attention to background solutions with asymptotically AdS$_6$ geometry,
for large values of the holographic direction---the ultra-violet (UV) regime
of the (putative) dual field theory.
Another one of the space-like dimensions is compactified on a circle that shrinks smoothly to zero size at a finite value, $\rho=\rho_o$, of the holographic direction---the infra-red (IR) regime.
The presence of an end to the space introduces a mass gap in the dual field theory interpretation, 
mimicking the effect of confinement in the four-dimensional field theory.
A subclass of backgrounds exists in which $\mathcal X$ acquires a non-trivial profile, breaking spontaneously the $SO(5)$ gauge symmetry of the gravity theory to its $SO(4)$ subgroup.

\subsection{Six-dimensional action}
\label{Sec:6}

\begin{table}
\caption{Field content of the model, organised in terms of  irreducible representations of the symmetries in 
$D=6$ dimensions ($SO(5)$ multiplets), as well as $D=5$ dimensions ($SO(4)$ multiplets, 
assuming $\langle {\mathcal X} \rangle \neq 0$), and $D=4$ dimensions
($SO(3)$ multiplets, assuming $\langle \vec\pi\rangle \neq 0$). In the case of the language in  $D=4$ dimensions, we indicate the field content in terms of 
the massive representations of the Poincar\'e group, and keep into account the degrees of freedom of
gauge-invariant combinations only. The irreducible representations for which we indicate $N_{\rm dof}= -$ 
 refer to cases where
 the degrees of freedom
 have  been included in propagating degrees of freedom of other fields,
to build gauge invariant combinations---for explanations, see the body of the paper.
For space-time indexes, $\hat{M}=0,\,1,\,2,\,3,\,5,\,6$, while $M=0,\,1,\,2,\,3,\,5$, and $\mu=0,\,1,\,2,\,3$.
For indexes of the internal symmetry, $\alpha, \beta=1,\,\cdots,\,5$,  while $A=1,\,\cdots,\,10$,  $\hat{A}=1,\,\cdots,\,4$, and $\bar{A}=6,\,\cdots,\,10$.
When the $SO(4)$ symmetry is broken to $SO(3)$, we use indexes $\hat{\cal A}=1,\,2,\,3$,
$\tilde{\cal A}=5,\,6,\,7$, and $\bar{\cal A}=8,\,9,\,10$.
}
\label{Fig:Fields}
\begin{center}
\begin{tabular}{|c|c|c||c|c|c||c|c|c|}
\hline\hline
\multicolumn{3}{|c||}{{
$D=6$, $SO(5)$,
}} &
\multicolumn{3}{|c||}{{
$D=5$, $SO(4)$,
}} &
\multicolumn{3}{|c|}{{
$D=4$, $SO(3)$,
}}
\cr
\multicolumn{3}{|c||}{{
{\rm massless irreps.}
}} &
\multicolumn{3}{|c||}{{
{\rm massless irreps.}
}} &
\multicolumn{3}{|c|}{{
{\rm massive irreps.}
}}
\cr
\hline\hline
{\rm Field} & $SO(5)$ & $N_{\rm dof}$  
&{\rm Field} & $SO(4)$ & $N_{\rm dof}$  
&{\rm Field} & $SO(3)$ & $N_{\rm dof}$  \cr
\hline\hline
$\hat{g}_{\hat{M}\hat{N}}$ & $1$ & $9$ &
$g_{MN}$ & $1$ & $5$ &
$g_{\mu\nu}$ & $1$ & $5$ \cr
 & &  &
 & &  &
$g_{\mu5}$ & $1$ & $-$ \cr
 & &  &
 & &  &
$g_{55}$ & $1$ & $-$ \cr
 & &  &
$\chi_M$ & $1$ & $3$ &
$\chi_{\mu}$ & $1$ & $3$ \cr
 & &  &
 & &  &
$\chi_{5}$ & $1$ & $-$ \cr
 & & &
$\chi$ & $1$ & $1$ &
$\chi$ & $1$ & $1$ \cr
\hline
${\mathcal X}_{\alpha}$ & $5$ & $5$ &
$\phi$ & $1$ & $1$ &
$\phi$ & $1$ & $1$ \cr
&&&
$\pi^{\hat{A}}$ & $4$ & $4$ &
$\pi^{\hat{\cal A}}$ & $3$ & $3$ \cr
&&&
&&&
$\pi^{4}$ & $1$ & $1$ \cr
\hline
${\cal A}_{\hat{M}\,\alpha}{}^{\beta}$ & $10$ & $40$ &
${\cal A}_{M}^{\,\,\,\hat{A}}$  & $4$ & $12$ &
${\cal A}_{\mu}^{\,\,\,\hat{\cal A}}$  & $3$ & $9$ \cr
&&&
&&&
${\cal A}_{\mu}^{\,\,\,4}$  & $1$ & $3$ \cr
&&&
&&&
${\cal A}_{5}^{\,\,\,\hat{\cal A}}$  & $3$ & $-$ \cr
&&&
&&&
${\cal A}_{5}^{\,\,\,4}$  & $1$ & $-$ \cr
&&&
${\cal A}_{6}^{\,\,\,\hat{A}}$  & $4$ & $4$ &
${\cal A}_{6}^{\,\,\,\hat{\cal A}}$  & $3$ & $3$ \cr
&&&
&&&
${\cal A}_{6}^{\,\,\,4}$  & $1$ & $1$ \cr
&&&
${\cal A}_{M}^{\,\,\,\bar{A}}$  & $6$ & $18$ &
${\cal A}_{\mu}^{\,\,\,\tilde{\cal A}}$  & $3$ & $9$ \cr
&&&
&&&
${\cal A}_{\mu}^{\,\,\,\bar{\cal A}}$  & $3$ & $9$ \cr
&&&
&&&
${\cal A}_{5}^{\,\,\,\tilde{\cal A}}$  & $3$ & $-$ \cr
&&&
&&&
${\cal A}_{5}^{\,\,\,\bar{\cal A}}$  & $3$ & $-$ \cr
&&&
${\cal A}_{6}^{\,\,\,\bar{A}}$  & $6$ & $6$ &
${\cal A}_{6}^{\,\,\,\tilde{\cal A}}$  & $3$ & $3$ \cr
&&&
&&&
${\cal A}_{6}^{\,\,\,\bar{\cal A}}$  & $3$ & $3$ \cr
\hline
$P_5{}_{\alpha}$&5&5&
$P_5{}_{\hat{A}}$&4&4&
$P_5{}_{\hat{\cal A}}$&3&3\cr
&&&
&&&
$P_5{}_{4}$&1&1\cr
&&&
$P_5{}_{5}$&1&1&
$P_5{}_{5}$&1&1\cr
\hline\hline
\end{tabular}
\end{center}
\end{table}

We review the essential features of the model 
 to keep the presentation self-contained and explain the notation---see Table~\ref{Fig:Fields}---without 
 discussing details that can be found in Ref.~\cite{Elander:2022ebt}, in respect to which
we rescale the action in $D=6$ dimensions by an overall factor of $\frac{1}{2\pi}$, to lighten the notation in matching it
to lower dimensions:
\begin{align}
	\mathcal S_6^{(bulk)} &= \frac{1}{2\pi}\int \dd^6 x \sqrt{-\hat g_6} \, \bigg\{ \frac{\mathcal R_6}{4} - \frac{1}{2} \hat g^{\hat M \hat N}\left( D_{\hat M} \mathcal X \right)^T D_{\hat N} \mathcal X - \mathcal V_6(\mathcal X) - \frac{1}{2} \Tr \left[ \hat g^{\hat M \hat P} \hat g^{\hat N \hat Q} \mathcal F_{\hat M \hat N} \mathcal F_{\hat P \hat Q} \right] \bigg\} \,.
\end{align}
The six-dimensional space-time indexes are denoted by $\hat M = 0,1,2,3,5,6$. The six-dimensional metric, $\hat g_{\hat M \hat N}$, has determinant $\hat g_6$,  and signature mostly $+$. The six-dimensional Ricci scalar is denoted as $\mathcal R_6$. 

The components,
${\mathcal X}_{\alpha}$, of the scalar transforming in the $5$ of $SO(5)$, are labelled by Greek indexes, $\alpha = 1,\, \cdots, \,5$.  The $SO(5)$ gauge field is denoted by $\mathcal A_{\hat M}$.
The covariant derivatives are defined as follows:
\beqs
	\left( D_{\hat M} \mathcal X \right)_\alpha &\equiv \partial_{\hat M} \mathcal X_\alpha + i { g } \mathcal A_{\hat M \, \alpha}{}^\beta \mathcal X_\beta \,, 
	\eeqs
and the field-strength tensors are
	\beqs
	\mathcal F_{\hat M \hat N \, \alpha}{}^\beta &\equiv 2 \left( \partial_{[\hat M} \mathcal A_{\hat N] \, \alpha}{}^\beta + i { g }  \mathcal A_{[\hat M \, \alpha}{}^\gamma \mathcal A_{\hat N] \, \gamma}{}^\beta \right) \,,
\eeqs
where antisymmetrisation is defined as $[n_1 n_2] \equiv \frac{1}{2} \left( n_1 n_2 - n_2 n_1 \right)$.
The bulk gauge coupling, ${g} $, is a free parameter.
One can, equivalently, write covariant derivatives and field-strength tensors in terms of the generators, $t^A$ ($A=1,\cdots,10$) of $SO(5)$, normalised so that $\Tr (t^A t^B) = \frac{1}{2} \delta^{AB}$. We exhibit an explicit basis for these generators in Appendix~\ref{Sec:so5}.

The bulk scalar potential, $\mathcal V_6(\mathcal X)$, is manifestly $SO(5)$ invariant, as it depends only on the combination $\phi\equiv \sqrt{\mathcal X^T\mathcal X}$. 
With the convenient choice of superpotential, ${\mathcal W}_6$, adopted in Ref.~\cite{Elander:2023aow}:
\beq
	\mathcal W_6 \equiv -2 - \frac{\Delta}{2} {\mathcal X}^T{\mathcal X} \,=\,-2 - \frac{\Delta}{2} \phi^2\,,
\eeq
one finds the potential to be the following:
\beq
\label{eq:VfromW}
	\mathcal V_6 \equiv \frac{1}{2} \sum_{\alpha}\left(\frac{\partial {\mathcal W}_6}
	{\partial{\mathcal X}_{\alpha}}\right)^2	
	- \frac{5}{4} \mathcal W_6^2 = -5 - \frac{\Delta (5 - \Delta)}{2} \phi^2 - \frac{5 \Delta^2}{16} \phi^4 \,,
\eeq
We adopt this form of the potential, $\mathcal V_6$, for its simplicity, and note that its origin in terms of a superpotential, $\mathcal W_6$, does not make the model supersymmetric: there are no fermionic fields, nor do the backgrounds discussed in this paper originate from solving first-order equations derived from $\mathcal W_6$.

The general configuration of the scalar field, $\mathcal X$, can be parametrised as
\beqs
\label{eq:Xdecomp}
	\mathcal X &\equiv \exp \left[ 2 i \sum_{\hat{A}} \pi^{\hat A} t^{\hat A} \right] \, \mathcal X_0\,\phi  \,, \qquad
	{\rm where} \qquad \mathcal X_0 \equiv (0, 0, 0, 0, 1)^T \,,
\eeqs
with ${\hat A}=1,\,\cdots,\,4$, labelling  the generators of the $SO(5)/SO(4)$ coset.
 If $\langle \phi \rangle \neq 0$,  one can also write $\mathcal X$ explicitly, as
\beq
	\mathcal X =  \phi  \left( \sin(|\vec\pi|) \frac{\vec\pi}{|\vec\pi|}, \cos(|\vec\pi|) \right)^T\,,
\eeq
in terms of the four PNGBs, $\vec\pi = (\pi^1,\,\pi^2,\,\pi^3,\,\pi^4)$,
spanning the $SO(5)/SO(4)$ coset.

\subsection{Dimensional reduction and background solutions}
\label{Sec:backgroundsolutions}

We find it convenient to write the
 background solutions  of interest  by first dimensionally reducing the theory 
to $D=5$ dimensions. The sixth dimension is a circle, parameterised by an angle,
 $0 \leq \eta < 2\pi$.
The metric ansatz is
\begin{align}
\label{eq:metricansatz}
	\dd s_6^2 &= e^{- 2\chi} \dd s_5^2 + e^{6\chi} \left(\dd \eta + \chi_M \dd x^M \right)^2 \,,
\end{align}
where the five-dimensional space-time index is $M = 0,1,2,3,5$, and the five-dimensional metric is 
\beq
\label{eq:5dmetricansatz}
	\dd s_5^2 = \dd r^2 + e^{2A(r)} \dd x_{1,3}^2 = e^{2\chi(\rho)} \dd \rho^2 + e^{2A(\rho)} \dd x_{1,3}^2 \,.
\eeq
In the background, the warp factor, $A$, as well as the scalars, $\mathcal X_\alpha$ and $\chi$, 
 depend only on the holographic coordinate, $\rho$, whereas $\mathcal A_6 = 0$, and $\mathcal A_M = 0 = \chi_M$. The background equations of motion for $\mathcal X_\alpha(\rho)$, $\chi(\rho)$, and $A(\rho)$ are given by
\begin{align}
\label{Eq:bg1}
	\partial_\rho^2 \mathcal X_\alpha + (4 \partial_\rho A - \partial_\rho \chi) \partial_\rho \mathcal X_\alpha &=
	\frac{\partial{ \mathcal V_6}}{\partial {\mathcal X}_{\alpha}} \,, \\
	\label{Eq:bg2}
	\partial_\rho^2 \chi + (4 \partial_\rho A - \partial_\rho \chi) \partial_\rho \chi &= - \frac{\mathcal V_6}{3} \,, \\
	\label{Eq:bg3}
	3 (\partial_\rho A)^2 - \frac{1}{2} \partial_\rho \mathcal X_\alpha \partial_\rho \mathcal X_\alpha - 3 (\partial_\rho \chi)^2 &= - \mathcal V_6 \,.
\end{align}

In Refs.~\cite{Elander:2022ebt,Elander:2023aow}, radial profiles of the background fields, $\phi(\rho)$, $\chi(\rho)$, and $A(\rho)$, satisfying Eqs.~(\ref{Eq:bg1}--\ref{Eq:bg3}), and resulting in regular geometries, are identified and referred to (with some abuse of language) as confining solutions.\footnote{In contrast with Ref.~\cite{Witten:1998zw}, we work within the context of bottom-up holography, and our model does not contain strings.}
We repeat here their IR expansions, that can be used to construct the full solutions numerically, by setting up the boundary conditions in the vicinity of the coordinate, $\rho = \rho_o$, at which the space ends, and by expanding in powers of the small difference, $\rho-\rho_o$:
\beqs
\label{Eq:IR1}
	\phi(\rho) &=&
\phi_I - \frac{1}{16} \Delta \phi_I \left( 20 + \Delta 
	 \left( 5 \phi_I^2 - 4 \right) \right) (\rho - \rho_o)^2 
	 + \mathcal O\left((\rho - \rho_o)^4\right) \,, \\
	\label{Eq:IR2}
	\chi(\rho) &=&
 \chi_I + \frac{1}{3} \log(\rho - \rho_o) + \frac{1}{288} 
 \left( -80 + 8 \left( \Delta - 5 \right) \Delta \phi_I^2 - 5 \Delta^{\chi} \phi_I^4 \right) 
 (\rho - \rho_o)^2 + O\left((\rho - \rho_o)^4\right) \,, \\
	\label{Eq:IR3}
	A(\rho) &= &
A_I + \frac{1}{3} \log(\rho - \rho_o) + \frac{7}{576} \left( 80 +\Delta \phi_I^2 \left( 40 + \Delta
 \left( 5 \phi_I^2 - 8 \right) \right) \right) (\rho - \rho_o)^2 + \mathcal O\left((\rho - \rho_o)^4\right) \,.
\eeqs
Here, $\phi_I$, $\chi_I$, and $A_I$ are integration constants, and we set $\chi_I=0$ to avoid a conical singularity in the plane described in polar coordinates by $(\rho,\eta)$. In those solutions, $\langle \vec\pi\rangle = 0$. In this paper, we adopt the same solutions for 
 $\phi(\rho)$, $\chi(\rho)$, and $A(\rho)$, but we allow for $\langle \vec\pi\rangle \neq 0$, as we shall see in Sect.~\ref{Sec:addboundaryterms}.

 We also reproduce here the UV expansions for the background solutions, written in terms of $z\equiv e^{-\rho}$. These depend non-trivially on the value of $\Delta$. Defining $\Delta_J={\rm min} \left(\Delta,\,5-\Delta\right)$, and $\Delta_V=5-\Delta_J$, they take the generic form
\beqs
\label{eq:backgroundUVexp1}
	\phi(z) &=& \phi_{J} z^{\Delta_J}\,+\,\cdots\, + \phi_{V} z^{\Delta_V} \,+\,\cdots \,, \\
\label{eq:backgroundUVexp2}
	\chi(z) &=& \chi_U - \frac{1}{3} \log(z)  \,+\,\cdots + ({\chi_5}+\cdots) z^5  \,+\,\cdots \,, \\
\label{eq:backgroundUVexp3}
	A(z) &=& A_U - \frac{4}{3} \log(z)  \,+\,\cdots \,.
\eeqs
The integration constants, $\phi_J$ and $\phi_V$, appear at leading and subleading order, respectively. Another integration constant, $\chi_5$, is related to the radial dependence of the size of the compactified circle. Finally, $A_U$ and $\chi_U$ only appear trivially, as overall factors, and in the following, we set them to zero without 
loss of generality---see Ref.~\cite{Elander:2020ial}.

The circle reduction of the six-dimensional model leads to the following five-dimensional bulk action~\cite{Elander:2023aow}:
\beqs
\label{eq:action5d}
	\mathcal S_5^{(bulk)} 	&=&
	\int \dd^5 x \sqrt{-g_5} \, \BigCurlyLeft
	\frac{R}{4} - 3g^{MN}  \partial_M \chi \partial_N \chi 
	- \frac{g^{MN}}{2}   \left( D_M \mathcal X \right)^T (D_N \mathcal X)
	- g^{MN} e^{-6\chi}\Tr \left[D_M \mathcal A_6 \frac{}{}D_N \mathcal A_6\right]
	\\
	&&   - e^{-2\chi} \mathcal V_6
	 - \frac{1}{2} g^2 e^{-8\chi} \mathcal X^T \mathcal A_6^2 \mathcal X - \frac{1}{16} e^{8\chi} g^{MP} g^{NQ} F^{(\chi)}_{MN} F^{(\chi)}_{PQ} - \frac{1}{2} e^{2\chi} \Tr \left[ g^{MP} g^{NQ} \mathcal F_{MN} \mathcal F_{PQ} \right] 
\nonumber \\
	&&
	- i g \,g^{MN} \chi_M \mathcal X^T \mathcal A_6 D_N \mathcal X
	- 2 e^{2\chi} g^{MN} g^{OP} \chi_M \Tr \left( \mathcal F_{NO} D_P \mathcal A_6 \right)
		- \frac{1}{2} g^2 g^{MN} \chi_M \chi_N \mathcal X^T \mathcal A_6^2 \mathcal X
	\nonumber \\
&&
	+ e^{2\chi} g^{MP} g^{NQ} \chi_M \chi_N \Tr \left( D_P \mathcal A_6 D_Q \mathcal A_6 \right)
	- e^{2\chi} g^{MN} g^{PQ} \chi_M \chi_N \Tr \left( D_P \mathcal A_6 D_Q \mathcal A_6 \right)
	\BigCurlyRight  \nonumber\,.
\eeqs
We exhibit this action purely in the interest of completeness, and anticipate  that, in Sect.~\ref{Sec:addboundaryterms}, we will rewrite it, in a form more suitable for the computation of spectra, by expanding it in powers of small fluctuations, and approximating it by retaining only quadratic order in said fluctuations.

\section{Effective Field Theory}
\label{Sec:SO(5)oSO(4)}

\begin{center}
   \begin{figure}
     \includegraphics[width=0.50\linewidth]{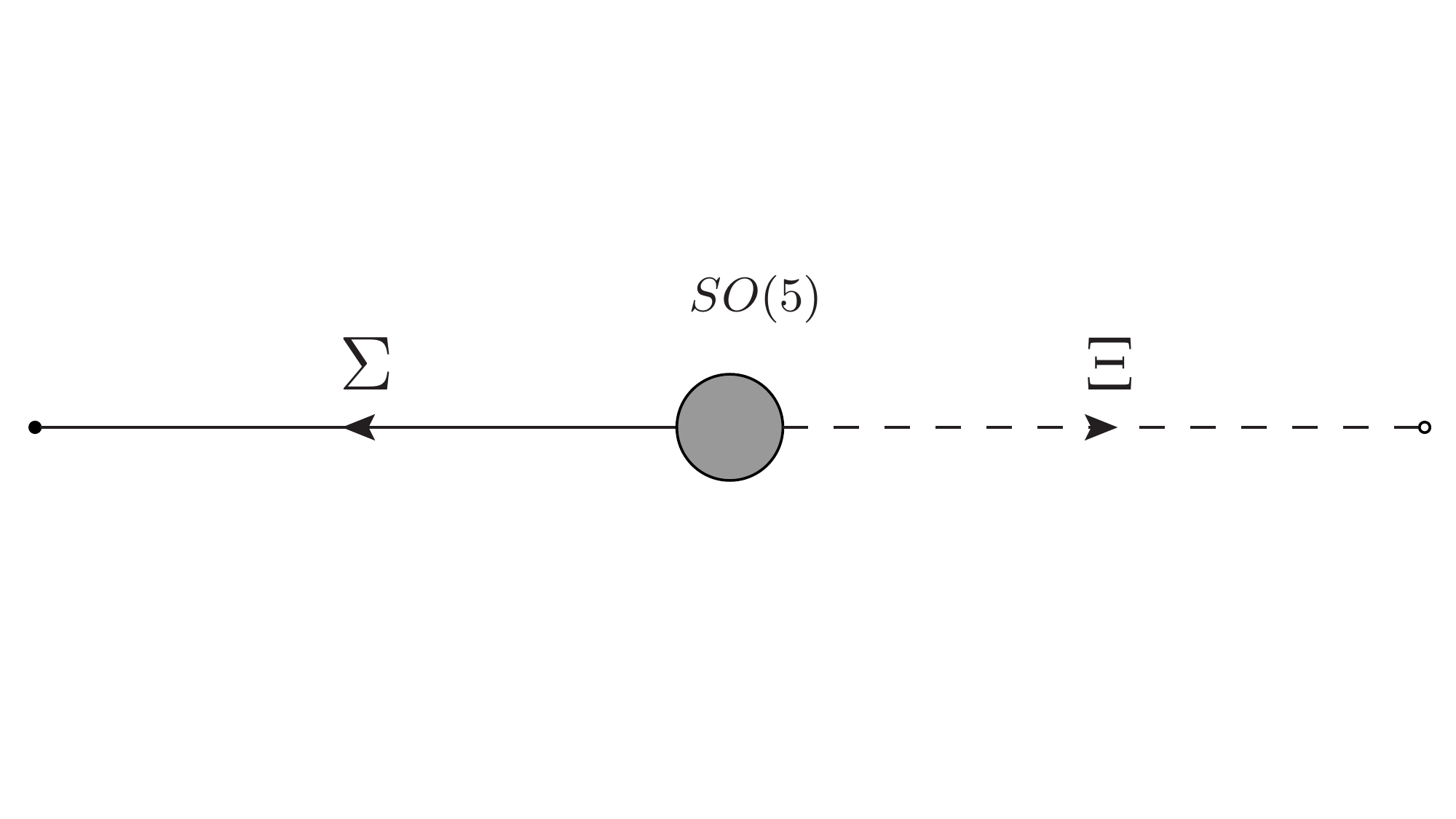}
   \caption{Moose diagram representing  
   the $SO(5)/SO(4)$  EFT---figure generated with axodraw2~\cite{Collins:2016aya}.}
\label{Fig:EFTSO(5)}
     \end{figure}
\end{center}

The purpose of this section is to clarify, within the language of four-dimensional effective field theories, the role of explicit and spontaneous breaking of 
internal continuous symmetries. In particular, we want to expose differences and commonalities between the treatment 
of global and local symmetries.
We occasionally recall definitions and notational conventions introduced earlier on in such a way that this short section is self-contained. It should be stressed that this section does not provide the precise EFT,
 low-energy description of the theory of interest in the rest of the paper, and we do not attempt to match the two, as doing so would go beyond current purposes.

We start by writing  the 
non-linear sigma-model Lagrangian density 
capturing the
long distance dynamics associated with 
the spontaneous breaking of a gauged $SO(5)$ symmetry 
 to its $SO(3)$ subgroup.
The field content consists of two (real) fields, $\Sigma$  and ${\Xi}$, both transforming in the $5$ of $SO(5)$, so that, under the action of a symmetry transformation:
\beqs
(\Sigma,\,{\Xi})&\rightarrow & (U\,\Sigma,\,U\,{\Xi})\,,
\eeqs
where $U\in SO(5)$ is a group element (a special, orthogonal, real matrix).

As stated in Sect.~\ref{Sec:model}, we denote as $t^A$ the generators of $SO(5)$, normalised so that  $\Tr\, (t^{{A}}t^B)=\frac{1}{2}\delta^{AB}$,
and use, where appropriate, the choice of basis in  Appendix~\ref{Sec:so5}.
We gauge the $SO(5)$ symmetry, by defining $A_{\mu}\equiv \sum_AA_{\mu}^{{A}}t^{{A}}$, so that the covariant derivatives are
\beqs
D_{\mu} \Sigma &\equiv & \partial_{\mu} \Sigma \,+\,i \,g\,A_{\mu}\Sigma\,,\\
D_{\mu} {\Xi} &\equiv & \partial_{\mu }{\Xi} \,+\,i \,g\,A_{\mu}{\Xi}\,,\\
F_{\mu\nu}&\equiv& \partial_{\mu}A_{\nu} - \partial_{\nu} A_{\mu} + i \, g\left[A_{\mu},\,A_{\nu}\right]\,,
\eeqs
with $g$ the (weak) gauge coupling strength. 
The $SO(5)$-invariant  Lagrangian density is the following:\footnote{An interaction term of the form 
$\Sigma^T F_{\mu\nu}F^{\mu\nu} \Sigma$ has been omitted for simplicity. As it changes cubic and quartic interactions in a non-trivial way, it must be added in a complete analysis, more general than the one discussed here.}
\beqs
{\cal L}_{SO(5)}&\equiv& \frac{f^2}{2}\left[\left(D_{\mu}\Sigma\right)^T\left(D^{\mu}\Sigma\right)\right]
\,+\,\frac{\kappa^2 f^2}{2}\left[\left(D_{\mu}{\Xi}\right)^T\left(D^{\mu}{\Xi}\right)\right]\nonumber\\
&&
\label{Eq:L5}
\,-\,\frac{1}{2}\Tr\left[F_{\mu\nu}F^{\mu\nu}\right]
\,+\,{(1-\tilde{\kappa}^2)}{}\,{\Xi}^T\,F_{\mu\nu}\,F^{\mu\nu}\,{\Xi}\\
&&\,-\,\lambda_{\Sigma}f^4\left(\Sigma^T\Sigma -1\right)^2
\,-\,\lambda_{\Xi}f^4\left({\Xi}^T{\Xi} -1\right)^2\,
-\,{\cal V}_{SO(5)} (\Sigma,\,\Xi)\,,\nonumber
\eeqs
where $f$ is the scale of the theory, and the couplings are denoted by $\kappa$, $\tilde{\kappa}$, $\lambda_{\Sigma}$, and $\lambda_{\Xi}$.
We will return to the $SO(5)$-invariant potential, ${\cal V}_{SO(5) }(\Sigma,\,\Xi)$, and the important physical parameters it encodes, in due course.

We require the  divergence of the couplings $\lambda_{\Sigma}\rightarrow +\infty$ and 
$\lambda_{{\cal V}}\rightarrow +\infty$, 
hence enforcing  the non-linear constraints $\Sigma^T\Sigma=1={\Xi}^T{\Xi}$.
Both these vacuum expectation values (VEVs) break $SO(5)$ to a  $SO(4)$ subgroup,
which may differ in the two cases.
We conventionally adapt our choice of basis for $SO(5)$ so that 
we denote as $t^{\bar{A}}$, with $\bar{A}=5,\,\cdots,\,10$, the generators that
 obey the relation $t^{\bar{A}}\,\langle {\Xi} \rangle=0$,
and generate an $SO(4)$ subgroup of $SO(5)$, while 
$t^{\hat{A}}$, with $\hat{A}=1,\,\cdots,\,4$,
are the four generators describing the corresponding $SO(5)/SO(4)$ coset.
 We then define  $\varsigma=\sum_{\hat{A}}\varsigma^{\hat {A}}t^{\hat{A}}$ and $\varrho=\sum_{\hat{A}}\varrho^{\hat{A}}t^{\hat{A}}$, to
parameterise the two scalar fields as
\beqs
\label{Eq:Sig}
\Sigma &=&e^{\frac{2 i}{f} {\varsigma}}\,\left(\begin{array}{c} 
0 \cr 0 \cr 0 \cr 0 \cr 1\end{array} \right)\,,
\qquad
{\rm and}
\qquad
{\Xi} \,=\, e^{\frac{2 i}{ f} \varrho}\,\left(\begin{array}{c} 
0 \cr 0 \cr 0 \cr 0 \cr 1 \end{array} \right)\,.
\eeqs
If the VEVs  are aligned, the $SO(4)$ symmetry is unbroken.
If otherwise, an $SO(3)$ symmetry is left intact, and it can be used to show that 
the most general  vacuum can be written as 
\beqs
\langle \Sigma \rangle =
\left(\begin{array}{c} 
0 \cr 0 \cr 0 \cr \sin \left(\frac{v}{f}\right) \cr \cos  \left(\frac{v}{f}\right) \end{array} \right)\,,
 \qquad  {\rm and} \qquad
\langle {\Xi} \rangle =\left(\begin{array}{c} 
0 \cr 0 \cr 0 \cr 0 \cr 1 \end{array} \right)\,.
\label{Eq:VEV5}
\eeqs
General values of the misalignment angle, $\frac{v}{f}$, lead to the breaking of $SO(4)$ to  $SO(3)$.
The vacuum in Eq.~(\ref{Eq:VEV5}) is given by the choice of parameters
$\langle \varrho \rangle=0$ and $\langle \varsigma \rangle= v\, t^4$.

With all of the above in place, by imposing the non-linear constraints, the first two lines of 
Eq.~(\ref{Eq:L5})  provide the leading-order, two-derivative terms of the 
EFT that determines all the two-point functions involving gauge fields.
 We ignore 
higher-derivative terms, that give small corrections to observable quantities computed at small energy.
To make any further progress, we discuss the properties expected of the last potential term
in Eq.~(\ref{Eq:L5}),
${\cal V}_{SO(5)}(\Sigma,\,\Xi)$, that controls vacuum (mis)alignment and spontaneous breaking of $SO(4)$ to $SO(3)$.
For current  purposes, the only two quantities of interest are the position of the minimum of the potential, and its second derivative  at said minimum. The former controls the vacuum misalignment angle, the latter the 
mass of scalar excitations. As we ignore all interaction terms and higher-order interactions,
rather than worrying about power-counting and other subtleties, 
we adopt a simplified,  illustrative choice, in the remainder of this section.
(We will discuss  more realistic, physically motivated choices in Sect.~\ref{Sec:top}.)
We write the potential as
\beqs
{\cal V}_{SO(5)}&=&\lambda \frac{f^4}{2}\left({\Xi}^T \Sigma-\cos\theta\right)^2\,,
\label{Eq:VSO5}
\eeqs
where $\theta$ and $\lambda$ are treated as free parameters, that can be traded for  the VEV and mass.
To this purpose, we replace the parametrisation of the vacuum in Eqs.~(\ref{Eq:VEV5}), to study, as a function of $v$, the resulting static potential:
\beqs
\mathcal V_{\rm static}&=&-\left.{\cal L}_{SO(5)}\frac{}{}\right|_{\varrho=0=A_{\mu},\,\varsigma =v t^4 }\,=\,
\frac{\lambda f^4}{2}\left(\cos\left(\frac{v}{f}\right)-\cos \theta\right)^2\,.
\eeqs
The minimum of the potential is at $v=\theta f$, and the second derivative evaluated at the minimum 
yields 
\beqs
\left.\frac{\partial^2 \mathcal V_{\rm static}}{\partial v^2}\right|_{v=\theta f}&=&\lambda f^2 \sin^2\theta\,>\,0
\,~~~~({\rm for}\,\lambda>0)\,.
\eeqs

The mass matrix,  ${\cal M}^2_0$, of the spin-0 states, evaluated at the minimum, ${v=\theta f}$, is
\beqs
{\cal M}^2_0&=&
\lambda f^2\sin^2 \theta
\left(
\begin{array}{cccccccc}
0 & 0 & 0 & 0 & 0 & 0 & 0 & 0 \cr
0 & 0 & 0 & 0 & 0 & 0 & 0 & 0 \cr
0 & 0 & 0 & 0 & 0 & 0 & 0 & 0 \cr
0 & 0 & 0 & 1 & 0 & 0 & 0 & -\frac{1}{{\kappa}} \cr
0 & 0 & 0 & 0 & 0 & 0 & 0 & 0 \cr
0 & 0 & 0 & 0 & 0 & 0 & 0 & 0 \cr
0 & 0 & 0 & 0 & 0 & 0 & 0 & 0 \cr
0 & 0 & 0 &  -\frac{1}{{\kappa}}  & 0 & 0 & 0 &  \frac{1}{{\kappa^2}}  \cr
\end{array}
 \right)\,,
\eeqs
written in the basis $(\varsigma^1,\,\varsigma^2,\,\varsigma^3,\,\varsigma^4,\,\varrho^1,\,\varrho^2,\,\varrho^3,\,\varrho^4)$.
The dependence on
 $\kappa$ comes from  the non-canonical normalisation of the kinetic terms in ${\cal L}_{SO(5)}$.
 For $\theta\neq 0$, the seven resulting massless states are exact Nambu-Goldstone
bosons associated with the breaking $SO(5)\rightarrow
 SO(4)\rightarrow SO(3)$, while one scalar has mass squared given by $m_{\pi}^2=\frac{1+\kappa^2}{\kappa^2}\lambda f^2 \sin^2(\theta)$.

The mass matrix for the gauge fields, ${\cal M}^2_1$,  evaluated at the minimum, ${v=\theta f}$,  obeys
\beqs
\frac{4{\cal M}^2_1}{g^2f^2}&=&
\left(
\begin{array}{cccccccccc}
 \frac{\cos (2 \theta )+2 \kappa ^2+1}{2 \tilde{\kappa}^2} & 0 & 0 & 0 &
   \frac{\sin (2\theta )}{2\tilde{\kappa}} & 0 & 0 & 0 & 0 & 0 \\
 0 & \frac{\cos (2 \theta )+2 \kappa ^2+1}{2 \tilde{\kappa}^2} & 0 & 0 & 0 &
   \frac{\sin (2\theta )}{2\tilde{\kappa}} & 0 & 0 & 0 & 0 \\
 0 & 0 & \frac{\cos (2 \theta )+2 \kappa ^2+1}{2 \tilde{\kappa}^2} & 0 & 0 & 0 &
   \frac{\sin (2\theta )}{2\tilde{\kappa}} & 0 & 0 & 0 \\
 0 & 0 & 0 & \frac{\kappa ^2+1}{\tilde{\kappa}^2} & 0 & 0 & 0 & 0 & 0 & 0 \\
 \frac{\sin (2\theta )}{2\tilde{\kappa}} & 0 & 0 & 0 & \sin ^2(\theta
   ) & 0 & 0 & 0 & 0 & 0 \\
 0 & \frac{\sin (2\theta )}{2\tilde{\kappa}} & 0 & 0 & 0 & \sin
   ^2(\theta ) & 0 & 0 & 0 & 0 \\
 0 & 0 & \frac{\sin (2\theta )}{2\tilde{\kappa}} & 0 & 0 & 0 & \sin
   ^2(\theta ) & 0 & 0 & 0 \\
 0 & 0 & 0 & 0 & 0 & 0 & 0 & 0 & 0 & 0 \\
 0 & 0 & 0 & 0 & 0 & 0 & 0 & 0 & 0 & 0 \\
 0 & 0 & 0 & 0 & 0 & 0 & 0 & 0 & 0 & 0 \\
\end{array}
\right)\,,
\eeqs
written in the basis $(A_{\mu}^1,\,A_{\mu}^2,\,A_{\mu}^3,\,A_{\mu}^4,A_{\mu}^5,\,A_{\mu}^6,\,A_{\mu}^7,\,A_{\mu}^8,\,A_{\mu}^9,\,A_{\mu}^{10})$.
The factors of $\tilde{\kappa}^2$ descend from the fact that the kinetic terms for the gauge bosons, in the vacuum with $\langle{ \Xi}\rangle \neq 0$,
are normalised by the kinetic matrix ${\rm diag}\,\left(\tilde{\kappa}^2,\,\tilde{\kappa}^2,\,\tilde{\kappa}^2,\,\tilde{\kappa}^2,\,1,\,1,\,1,\,1,\,1,\,1\right)$.
The three massless states, $A_{\mu}^8,\,A_{\mu}^9,$ and $A_{\mu}^{10}$,  are associated with the unbroken $SO(3)$.
For $\theta=0$, only $(A_{\mu}^1,\,A_{\mu}^2,\,A_{\mu}^3,\,A_{\mu}^4)$ are massive. But for $v=f \theta\neq 0$, seven gauge fields acquire a mass, 
and the seven massless pions provide the longitudinal polarisations necessary for the Higgs mechanism.  Only one massive real scalar remains in the physical spectrum.
The gauge fixing condition, defining the unitary gauge, is $\varsigma^1=\varsigma^2=\varsigma^3=0=\varrho^1=\varrho^2=\varrho^3$, and $\varsigma^4+\kappa \varrho^4=0$.

So far, all the symmetries are local, and only spontaneous symmetry breaking is present.
Yet, this scenario is of general validity, and encompasses also the case of global symmetries, and their explicit breaking.
We show in the remainder of this section how to take appropriate limits and 
 recover more general symmetry-breaking patterns. 
 In particular, we want to describe the case in which only the $SO(4)$ subgroup of $SO(5)$ is gauged, 
and there is an additional, independent source of
explicit breaking of $SO(5)$ to $SO(4)$. Their 
combined effect is to trigger, via vacuum misalignment,
the further spontaneous breaking to $SO(3)$.
We proceed as follows, in reference to  Eq.~(\ref{Eq:L5}).

\begin{itemize}
\item The limits $\lambda_{\Sigma},\, \lambda_{\Xi}\rightarrow +\infty$ 
impose  the non-linear constraints, $\Sigma^T\Sigma=1={\Xi}^T{\Xi}$. We also find it convenient to redefine the constant $\kappa$, by the rescaling $\kappa \equiv a \,\tilde{\kappa} $.

\item When $\tilde{\kappa}\gg 1$, the coefficients of kinetic terms of $A_{\mu}^1$, $A_{\mu}^2$, $A_{\mu}^3$, and $A_{\mu}^4$ are large, hence the couplings of these four fields are small, 
and the mass matrices are approximately diagonal.  
We take a second limit,
$\tilde{\kappa}\rightarrow +\infty$, to find 
\beqs \label{eq:li}
{\cal M}^2_0&\rightarrow&
\lambda f^2\sin^2\theta\,\,{\rm diag}\,\left(\frac{}{}0,\,0,\,0,\,1,\,0,\,0,\,0,\,0\frac{}{}\right)\,,\\
{\cal M}^2_1&\rightarrow&\frac{g^2 f^2}{4}\,\,
{\rm diag}\,\left(\frac{}{}a^2,\,a^2,\,a^2,\,a^2,\,\sin^2(\theta),\,\sin^2(\theta),\,\sin^2(\theta),
\,0,\,0,\,0\frac{}{}\right)\,.
\eeqs
The unitary gauge  is defined by setting
$\varsigma^1=\varsigma^2=\varsigma^3=0=\varrho^1=\varrho^2=\varrho^3=\varrho^4=0$, while $\varsigma^4$ remains in the theory as a
physical spin-0 field.

\item Holding fixed $g$ and $f$ (besides $\lambda$ and $\theta$), for very large choices of $a \gg 1$, the four gauge bosons, $A_{\mu}^{\hat{A}}$, that live in the $SO(5)/SO(4)$ coset, decouple from the physical
scalar, $\varsigma^4$, and become parametrically heavy.
They can  be trivially integrated  out of the EFT.
We hence take the third limit $a\rightarrow +\infty$. The EFT field content now consists of
 the scalar $\varsigma^4$, with mass
$m_{\pi}^2=\lambda f^2 \sin^2(\theta)$, three massive gauge fields, with mass $\frac{1}{4}g^2f^2 \sin^2(\theta)$, corresponding to the spontaneous breaking 
$SO(4)\rightarrow SO(3)$, and denoted as $A_{\mu}^5$, $A_{\mu}^6$, and $A_{\mu}^7$ (their longitudinal components are provided by $\varsigma^1$, $\varsigma^2$ and $\varsigma^3$), 
and, finally, the massless gauge bosons $A_{\mu}^8$, $A_{\mu}^9$, and $A_{\mu}^{10}$.
\end{itemize}

The result of this process is equivalent to adopting the following Lagrangian density:
\beqs
{\cal L}_{{SO(4)}}&\equiv& \frac{f^2}{2}\left[\left(\tilde{D}_{\mu}\Sigma\right)^T\left(\tilde{D}^{\mu}\Sigma\right)\right]
\,-\,\frac{1}{4}\sum_{A=5}^{10}\left.F_{\mu\nu}F^{\mu\nu}\right.
\,-\,\lambda_{\Sigma}f^4\left(\Sigma^T\Sigma -1\right)^2
-{\cal V}_{SO(4)}(\Sigma)\,,
\label{Eq:spur}
\eeqs
which has been obtained by replacing the second field, $\Xi$, with a spurion field, $P_5=\left(0,\,0,\,0,\,0,\,1\right)$.
The covariant derivative is now restricted to $SO(4)$:
\beqs
\tilde{D}_{\mu} \Sigma &\equiv & \partial_{\mu} \Sigma \,+\,i \,g\,\sum_{A=5}^{10}A_{\mu}^At^A\Sigma\,,
\eeqs
as are the kinetic kinetic terms for the gauge field.
By taking $\lambda_{\Sigma}\rightarrow +\infty$ one imposes the non-linear constraint $\Sigma^T\Sigma = 1$.
From the choice in Eq.~(\ref{Eq:VSO5}), one finds that 
  ${\cal V}_{SO(4)}=\lambda \frac{f^4}{2}\left(P_5^T \Sigma-\cos\theta\right)^2$.
This potential term leads to vacuum misalignment, and also provides a mass for $\varsigma^4$.
In this Lagrangian, the global $SO(5)$ symmetry is broken explicitly, both by the gauging of an $SO(4)$ subgroup, 
and by the coupling to the spurion, $P_5$.

In summary, as is well known, in the presence of an admixture of explicit and spontaneous breaking of a set of continuous global symmetries,
one must pay attention to gauge only unbroken subgroups, as prescribed by the Higgs mechanism. Yet, one may be able to elegantly describe the whole system in terms of only gauge symmetries, undergoing spontaneous breaking, without ever referring to explicit symmetry breaking. If this can be arranged, the case of interest can then be recovered by taking the appropriate limits in the space of parameters.
Caution must be applied to the order of limits one takes for the parameters,  in such a way that no violation of unitarity ensues,
no ghosts or negative norm states remain,  
and the theory is weakly coupled, at all stages of the analysis.

\subsection{External fields,  Coleman-Weinberg potential, and vacuum misalignment}
\label{Sec:top}

The choice of potential in Eq.~(\ref{Eq:VSO5}) has a certain appeal, both for its simplicity, and for the fact that it induces vacuum misalignment,
while also suppressing the mass of the associated scalar field, $\varsigma^4$. But it is not realistic.
As anticipated, since we are interested only in the vacuum misalignment angle and the mass of the scalar,
and only in the two-point functions, not in interaction terms, the detailed functional form
of the potential is not important for our purposes. Yet, it may be instructive to demonstrate how a more realistic potential may emerge dynamically.
We devote this short subsection to demonstrating one simple example of such a potential.

To build such an example, we couple the EFT to external fermions, by borrowing some ideas from the discussion of Eq.~(116) in Ref.~\cite{Contino:2010rs},
but with major simplifications.
First,  we write the couplings in such a way as to preserve an $SO(4)\sim SU(2)_L\times SU(2)_R$ subgroup of $SO(5)$.
Second, as we are not attempting here to implement a version of top partial compositeness, we do not couple the external fermions to bulk fermions, that would represent baryons in the strongly coupled dual field theory.
In its stead, we realise a simpler mechanism for fermion mass generation, reminiscent at the algebraic level
 of the one adopted in the literature on technicolor~\cite{Weinberg:1975gm,Susskind:1978ms},
 extended technicolor theories~\cite{Dimopoulos:1979es,Eichten:1979ah}, and
 walking technicolor~\cite{Holdom:1984sk,Yamawaki:1985zg,Appelquist:1986an}
 (see also the reviews in Refs.~\cite{Chivukula:2000mb,Lane:2002wv,Hill:2002ap,Sannino:2009za,Piai:2010ma}),
by coupling directly a fermion bilinear to a scalar composite operator
of the strongly coupled theory (a meson).
As we are interested only in the symmetry and symmetry-breaking 
patterns emerging, rather than in controlling the  dynamics and predicting 
the natural magnitude of the couplings involved,
this simpler approach is adequate for our current purposes.\footnote{
In a more refined and realistic model, one would want to gauge the standard-model $SU(2)_L\times U(1)_Y \subset SO(4)\times U(1)_{B-L}$ symmetry, and possibly introduce bulk fermion fields transforming in the appropriate representations of the symmetry groups. We defer such steps to future work, with particular reference to the interplay of fermion partial compositeness and vacuum misalignment~\cite{Contino:2010rs,Elander:2021bmt,Banerjee:2023ipb}.}

We start from the local identification $SO(5)\sim Sp(4)$.
We introduce a convenient basis of $4\times 4$  matrices defining the adjoint, $10$, and the
antisymmetric, $5$,  irreducible representations of $Sp(4)$.
We borrow the latter, with some adjustments, from Refs.~\cite{Bennett:2017kga, Bennett:2019cxd}---see also Appendix~\ref{Sec:so5}. 
We write the symplectic matrix, $\Omega$, as follows:
\beqs
\Omega_{\alpha\beta} &\equiv&\left(\begin{array}{cccc}
0 & 0 & 1 & 0 \cr
0 & 0 & 0 & 1 \cr
-1 & 0 & 0 & 0 \cr
0 & -1 & 0 & 0 \cr
\end{array}\right)\,,
\eeqs
and define
 the matrices, $T^A$, that satisfy
\beqs
\Omega \, T^A + T^{A\,T} \Omega\,=\,0\,,\qquad {\rm for} \qquad A=1,\,\cdots,\,10\,,
\eeqs
as  the $10$ generators of $Sp(4)$.
We also introduce
 the Hermitian and traceless matrices $\Gamma^B$ that satisfy the relation
\beqs
\Omega \, \Gamma^B - \Gamma^{B\,T} \Omega\,=\,0\,,\qquad {\rm for} \qquad B=1,\,\cdots,\,5\,,
\eeqs
hence identifying them with the $5$ generators of the coset $SU(4)/Sp(4)$, so that we have a complete basis
of the natural embedding of $Sp(4)$ in $SU(4)$.
We adopt the normalisation $\Tr \, (T^A T^B) = \frac{1}{4}\delta^{AB} = \Tr\, (\Gamma^A\Gamma^B)$.

The components of $\Sigma$, that can be read off Eq.~(\ref{Eq:Sig}), 
are recombined to define the Hermitian matrix
\beqs
{\SigmaSlash}^{\,\alpha}{}_\beta&\equiv&4
\sum_{A=1}^5\Sigma^A\left(\Gamma^A\right)^{\alpha}_{\,\,\,\beta}
\,=\,
\left(\begin{array}{cccc}
\Sigma^5 & \Sigma^1-i\Sigma^2  & 0 & -i \Sigma^3 + \Sigma^4 \cr
\Sigma^1+i\Sigma^2  & -\Sigma^5 & i \Sigma^3 - \Sigma^4 & 0 \cr
0 &  -i \Sigma^3 - \Sigma^4 & \Sigma^5 &\Sigma^1+i\Sigma^2  \cr
i \Sigma^3 + \Sigma^4 & 0 &\Sigma^1- i\Sigma^2  & -\Sigma^5 \cr
\end{array}\right)\,.
\eeqs
Similarly, the components of $\Xi$ are used to define  ${\XiSlash}^{\,\alpha}{}_\beta \equiv
4 \sum_{A=1}^5{\Xi}^A\left(\Gamma^A\right)^{\alpha}_{\,\,\,\beta}$.
Both these matrix-valued fields transform in the adjoint representation: 
\beqs
({\SigmaSlash},\,{\XiSlash})&\rightarrow & (\tilde{U}\,{\SigmaSlash}\tilde{U}^{\dagger},\,\tilde{U}\,{\XiSlash}\tilde{U}^{\dagger})\,,
\eeqs
where $\tilde{U}=\exp(i\sum_{A=1}^{10}\alpha^AT^A)$ are $4\times 4$ unitary matrices
describing $Sp(4)$ transformations.
The combination $\tilde{\Sigma}\equiv \SigmaSlash\,\Omega$  
is a  $4\times 4$ antisymmetric matrix transforming as the $5$ of $Sp(4)$.
The same applies to  $\tilde{\Xi}\equiv\XiSlash\,\Omega$:
\beqs
(\tilde{\Sigma},\,\tilde{\Xi})\,&\rightarrow & (\tilde{U}\,\tilde{\Sigma}\tilde{U}^{T},\,\tilde{U}\,\tilde{\Xi}\tilde{U}^{T})\,.
\eeqs

We introduce 
 chiral fermions, $\psi_{L\,\alpha}$ and $\psi_{R\,\alpha}$,
  formally transforming in the $4$ of $Sp(4)$,
which is also the spinorial representation of $SO(5)$.
We break explicitly the symmetry by writing the fermions as incomplete multiplets:
\beqs
\psi_L&=&\left(\begin{array}{c}
t_L \cr
0 \cr
b_L \cr
0\cr
\end{array}
\right)\,, \qquad {\rm and}\qquad 
\tilde{\psi}_R\,=\,\Omega^{-1} \psi_R\,=\,
\Omega^T
\, \left(\begin{array}{c}
t_R \cr
0 \cr
b_R\cr
0 \cr
\end{array}
\right)
\,=\,\left(\begin{array}{c}
-b_R \cr
0 \cr
t_R\cr
0 \cr
\end{array}
\right)
\,.
\eeqs
Each of the entries  is itself  a 2-component chiral spinor. The notation is suggestive of the fact that in an
extension of the standard model they would represent the top and bottom quarks, respectively, but notice the 
absence of QCD color quantum numbers.

Finally, we add to the Lagrangian density a set of   couplings between scalars and fermions, written as
\beqs
{\cal L}_Y&=&-y {f}\overline{\psi_L}\left(\tilde{\Sigma}\,\frac{}{}
-\frac{}{}\,\tilde{\Xi}\right)\tilde{\psi}_R
\,+\,{\rm h.c.}\,,
\eeqs
where $y$ is a Yukawa coupling.
In the vacuum,  $\langle \tilde{\Sigma} \rangle\neq 0 \neq{\langle \tilde{\Xi} \rangle}$, one finds:
\beqs
{\cal L}_Y&=&-\,\overline{t_L}\left[{y\,f}\left(\cos\left(\frac{v}{f}\right)-1\right)\right] t_R
-\,\overline{b_L}\left[y\,f\left(\cos\left(\frac{v}{f}\right)-1\right)\right] b_R\,+\, {\rm h.c.}\,+\,\cdots\,,
\eeqs
where we omit interactions with the PNGBs.
The resulting   Dirac mass  matrices  break $SU(2)_L\times SU(2)_R\sim SO(4)$, 
but preserves the diagonal subgroup, $SU(2)\sim SO(3)$.
The fermion mass is ${\cal M}_{1/2}={y\,f}{}\left(\cos\left(\frac{v}{f}\right)-1\right)$.
It vanishes  when the vacuum is  aligned, $\langle \tilde{\Sigma}\rangle =\langle \tilde{\Xi} \rangle$.

Perturbatively, at the one-loop level, the presence of the symmetry-breaking terms, encoded in the Yukawa coupling 
and in the gauging of $SO(4)$,
induces a divergent contribution (\`a la Coleman-Weinberg~\cite{Coleman:1973jx}) to the static
 effective potential, ${\cal V}_{\rm CW}$. Defining the matrix $T\equiv{\rm diag}(1,\,0,\,1,\,0)$, the naive result for
  ${\cal V}_{\rm CW}$
   takes the  form
\beqs
{\cal V}_{\rm CW} &=&
\frac{\Lambda^2}{32 \pi^2} {\cal S}\Tr {\cal M}^2
\\
&=&
\left.\frac{\Lambda^2}{32 \pi^2} 
\left(3g^2f^2\sum_{A=5}^{10}\Sigma^T\,t^A\,t^A\,\Sigma
\,-\,
4 y^2 f^2\,\Tr\left[\frac{}{}
\left(\tilde{\Sigma}\,\frac{}{}
-\frac{}{}\,\tilde{\Xi}\right)
T
\left(\tilde{\Sigma}\,\frac{}{}
-\frac{}{}\,\tilde{\Xi}\right)^{\dagger}
T\frac{}{}\right]
\right)\right|_{
\tilde{\Sigma},\,\tilde{\Xi} =\langle \tilde{\Sigma}\rangle,\,\langle \tilde{\Xi} \rangle
}
\,.
\eeqs

This potential depends explicitly on the divergent cutoff, $\Lambda$, of the theory,
which requires the introduction of counter-terms and a choice of subtraction prescription. Doing so requires two free parameters, controlling
the overall size of the terms proportional to $g^2$ and $y^2$, respectively.
By denoting these two free parameters as $C_g, C_t \sim {\cal O}(1)$ (and  estimating
 $\Lambda\sim {\cal O}(4\pi f)$), we conclude that in the Lagrangian density in Eq.~(\ref{Eq:L5}) one
should include the potential
\beqs
\label{Eq:VM}
{\cal V}_{SO(4)}&=&
\left.\frac{3}{2}g^2f^4 C_g\,\sum_{A=5}^{10}\Sigma^T\,t^A\,t^A\,\Sigma
\,-\,
2 y^2 f^4 C_t\,\Tr\left[\frac{}{}
\left(\tilde{\Sigma}\,\frac{}{}
-\frac{}{}\,\tilde{\Xi}\right)
T
\left(\tilde{\Sigma}\,\frac{}{}
-\frac{}{}\,\tilde{\Xi}\right)^{\dagger}
T\frac{}{}\right]
\right|_{
\tilde{\Sigma},\,\tilde{\Xi} =\langle \tilde{\Sigma}\rangle,\,\langle \tilde{\Xi} \rangle
}
\\
&=&
\frac{9}{8}g^2f^4 C_g\,\sin^2\left(\frac{v}{f}\right)
-{4}y^2f^4 C_t\,\left(\cos\left(\frac{v}{f}\right)-1\right)^2
\,.
\eeqs

While  the resulting potential in Eq.~(\ref{Eq:VM}) is different from Eq.~(\ref{Eq:VSO5}), it is not difficult to convince oneself that a potential with the structure  in Eq.~(\ref{Eq:VM}) will ultimately lead to vacuum misalignment, and yield the 
same leading-order,  long distance features as they emerge from Eq.~(\ref{Eq:VSO5})---as we are interested only in the vacuum and in the mass of the scalar excitation around the vacuum, not in the interactions or the details of the potential away from its minimum.
Furthermore, there are other possible choices for fermion
sector that one can make, that  lead to different 1-loop potentials.
As anticipated, it is also possible to marry this model with fermion partial compositeness, in which case the divergencies can be milder, or even absent, depending on details that vary between different models.
We close here this digression, and return to the gravity theory.

\section{Boundary terms and action to quadratic order}
\label{Sec:addboundaryterms}

We return in this section to the higher-dimensional gravity model of interest and to its dual field theory interpretation. We want to build a holographic model such that its long-distance behaviour reproduces the qualitative features of the EFT described in Sect.~\ref{Sec:SO(5)oSO(4)}, at least at the level of low-momentum two-point functions and light particle spectrum. 
We implement the (weak) gauging of an $SO(4)$ subgroup of the $SO(5)$ global symmetry of the four-dimensional field theory. The further explicit breaking of $SO(5)$ to $SO(4)$ is realised through the addition of a set of interaction terms that are localised at the boundary of the five-dimensional gravity geometry obtained in Sect.~\ref{Sec:backgroundsolutions}.
We do not specify the short-distance origin of such terms, neither in field-theory, nor in higher-dimensional gravity  terms,  and leave such tasks for future work.   In particular, we keep our treatment of the explicit symmetry breaking terms general, rather than specifying the external sector and performing a perturbative effective potential analysis a la Coleman-Weinberg~\cite{Coleman:1973jx}. As discussed in Sect.~\ref{Sec:top},  we are interested in the misalignment angle and mass for the lightest scalar excitation, while  the shape of the effective potential and the higher-order couplings are beyond the scope of this paper.

In the gravity description, the $SO(5)$ symmetry is gauged, and hence
 one is not allowed to write terms that explicitly break it. As illustrated in Sect.~\ref{Sec:SO(5)oSO(4)},  this difficulty can be overcome if one writes the full action in a manifestly $SO(5)$-invariant way, by introducing a new field, $P_5$---transforming in the $5$ of $SO(5)$ and localised at the UV boundary. Explicit breaking of $SO(5)$ to $SO(4)$ is recovered (without violating unitarity or introducing other pathologies) by taking the appropriate limits, that decouple the additional degrees of freedom.

The boundary terms added to the gravity theory are necessary in order to gauge a subgroup of the corresponding global symmetry in the dual field theory interpretation. They also introduce explicit breaking of global symmetries in the way that triggers vacuum misalignment and spontaneous breaking of the $SO(4)$ gauge symmetry in the dual field theory down to its $SO(3)$ subgroup. The gravity background solutions of interest break (spontaneously) the $SO(5)$ symmetry to $SO(4)$, due to the radial profile of the bulk scalar field, $\phi$. The boundary conditions for the background fields select a (constant) value of $\pi^{\hat A}$, which further breaks (spontaneously) $SO(4)$ to $SO(3)$. The radial profiles of all other background fields have been discussed in Sect.~\ref{Sec:model}, and remain the same as in Ref.~\cite{Elander:2023aow}. We end this section by expanding the action of the model to quadratic order around the new background solutions, and writing it in a form suitable to the computation of the spectrum of fluctuations in Sect.~\ref{Sec:spectrum}.

\subsection{Boundary-localised interactions}
\label{Sec:bcloc}

In the five-dimensional theory, we introduce boundaries at finite values of the radial 
directions, $\rho=\rho_i$, with $i=1,\,2$, to 
serve as regulators; 
our calculations are performed within the restricted range
 range $\rho_1<\rho<\rho_2$, yet physical results are obtained by taking the limits $\rho_1 \rightarrow \rho_o$ and 
$\rho_2 \rightarrow \infty$. As in Table~\ref{Fig:Fields}, we denote boundary space-time indexes by $\mu = 0, 1, 2, 3$.
We add to the bulk action, $\mathcal S_5^{(bulk)}$, several boundary terms---denoted  as 
$\mathcal S_{{\rm GHY},i}$, $\mathcal S_{\lambda,i}$, 
 $\mathcal S_{P_5,2}$, $\mathcal S_{\mathcal V_4,2}$, $\mathcal S_{\mathcal A,2}$, 
  $\mathcal S_{\chi,2}$, and  $\mathcal S_{\mathcal X,2}$---in order to obtain the complete five-dimensional action, $\mathcal S_5$:
\beq
    \mathcal S_5 = \mathcal S_5^{(bulk)} + \sum_{i=1,2} \Big( \mathcal S_{{\rm GHY},i} + \mathcal S_{\lambda,i} \Big) 
        + \mathcal S_{P_5,2} + \mathcal S_{\mathcal V_4,2} 
    + \mathcal S_{\mathcal A,2} + \mathcal S_{\chi,2 } + \mathcal S_{\mathcal X,2} 
\,.
\eeq
The boundary actions with subscript $i= 1,\,2$ are localised at $\rho = \rho_i$. 
We now proceed to discuss each of these terms, in both gravity and field-theory language.

The Gibbons-Hawking-York boundary actions, $\mathcal S_{{\rm GHY},i}$, take the following form:
\beq
\label{Eq:GHY}
    \mathcal S_{{\rm GHY},i} = (-)^i \int \dd^4 x \sqrt{-\tilde g} \, \frac{K}{2} \bigg|_{\rho = \rho_i} \,,
\eeq
where $\tilde g_{MN}$ is the induced metric on the boundaries, $\tilde{g}$ is its determinant,
 and $K$ is the extrinsic curvature---see Appendix~\ref{Sec:scalars}.
Boundary-localised scalar potentials, 
$\lambda_i(\mathcal X,\chi,\mathcal A_6)$, that  are $SO(5)$ 
invariant,
 enter the action as
\beq
\label{Eq:Slambdai}
    \mathcal S_{\lambda,i} = (-)^i \int \dd^4 x \sqrt{-\tilde g} \, \lambda_i(\mathcal X,\chi,\mathcal A_6) \bigg|_{\rho = \rho_i} \,.
\eeq
These two types of boundary terms are needed
 to make the variational problem well-defined, and ultimately ensure that 
the background solutions can be consistently truncated at the boundaries, $\rho=\rho_i$, in the holographic direction. Details about the structure of the potentials can be found in Ref.~\cite{Elander:2010wd}, but play a marginal role in the following.

The UV-localised actions, $\mathcal S_{P_5,2}$ and $\mathcal S_{\mathcal V_4,2}$, 
 involve the bulk scalar, $\mathcal X$, and a new, boundary-localised field, $P_5$,  transforming as the $5$ of $SO(5)$. 
We refer to $P_5$ as a spurion, because its dynamics is frozen (in the appropriate limit).
These terms have the same qualitative structure and implications as the second, sixth and seventh terms of the 
Lagrangian density in Eq.~(\ref{Eq:L5}). Explicitly, we write them as follows:
\begin{align}
\label{eq:SP5}
	\mathcal S_{P_5,2} =& \int \dd^4 x \sqrt{-\tilde g} \, \bigg\{ -\frac{1}{2} K_5 \, \tilde g^{\mu\nu} \left( D_\mu P_5 \right) D_\nu P_5 - \lambda_5 \left( P_5^T P_5 - v_5^2 \right)^2 \bigg\} \bigg|_{\rho = \rho_2} \,, \\
\label{eq:SV}
	\mathcal S_{\mathcal V_4,2} =& - \int \dd^4 x \sqrt{-\tilde g} \, \mathcal V_4(\mathcal X, \chi, P_5) \bigg|_{\rho = \rho_2} \,,
\end{align}
where $K_5$, $\lambda_5$, and $v_5$ are free parameters. The potential, $\mathcal V_4(\mathcal X, \chi, P_5)$, is $SO(5)$ invariant. It  depends (besides $\chi$) on two invariants, $\phi \equiv \sqrt{\mathcal X^T \mathcal X}$ and $\psi \equiv \mathcal X^T P_5$. 

The background equation for a boundary-localised, constant $P_5$ is 
\beq
    4 \lambda_5 (P_5^TP_5 - v_5^2) P_{5 \, \alpha} + \frac{\partial \mathcal V_4}{\partial \psi} \mathcal X_\alpha = 0 \,.
\eeq
We take the limit $\lambda_5 \rightarrow \infty$, so that $|P_5| = v_5$, freezing one component of the spurion, $P_5$. In the next section, we will further discuss a limit that involves $K_5$, and which decouples the four remaining spurion degrees of freedom. Because of $SO(5)$ invariance, without loss of generality, we fix the background value of $P_5$ to be along its fifth component:
\beq
	\overline{P_5} \equiv (0,0,0,0,v_5)^T \,.
\eeq
We assume there exists a value of $\psi$ with $\partial_\psi \mathcal V_4 = 0$. Since $\mathcal V_4$ is a function of  $SO(5)$ invariants, $\phi$ and $\psi$,  then $\mathcal V_4(\chi, \mathcal X, \overline{P_5}) = \mathcal V_4(\chi, \phi, |\vec\pi|)$ is $SO(4)$ invariant. Hence, $\mathcal S_{\mathcal V_4}$ effectively serves as a UV boundary-localised potential for $\mathcal X$ (and $\chi$), that captures the explicit breaking of $SO(5)$ to $SO(4)$, due to an external sector.\footnote{In the following, we will assume that $\mathcal V_4$ has been chosen such that, when evaluated on $P_5 = \overline{P_5}$, it does not depend on $v_5$.}
All of these steps realise, in the context of the five-dimensional gravity theory with boundaries, the mechanism discussed in Sect.~\ref{Sec:SO(5)oSO(4)}.

Next, the boundary-localised action, $S_{A,2}$, realises in the gravity theory the (weak) 
gauging of an $SO(4)\subset SO(5)$ subgroup of the global symmetry of the dual  field theory.
We write
\begin{align}
\label{Eq:SA2}
	\mathcal S_{\mathcal A,2} =
	\int \dd^4 x \sqrt{-\tilde g} \, \bigg\{ - \frac{\hat D_2}{v_5^2} \,
\tilde{g}^{\mu\rho}\tilde{g}^{\nu\sigma} P_5^T \mathcal F_{\mu\nu} \mathcal F_{\rho\sigma} P_5 
- \frac{1}{4} \bar D_2 \,
\tilde{g}^{\mu\rho}\tilde{g}^{\nu\sigma} \left( \mathcal F^A_{\mu\nu} \mathcal F^A_{\rho\sigma} 
- \frac{4}{v_5} P_5^T \mathcal F_{\mu\nu} \mathcal F_{\rho\sigma} P_5 \right) \bigg\} \bigg|_{\rho = \rho_2} \,,
\end{align}
which, after fixing the spurion to its background value, $P_5 = \overline{P_5}$, becomes
\begin{align}
	\mathcal S_{\mathcal A,2} \big|_{P_5 = \overline{P_5}} =
	 \int \dd^4 x \sqrt{-\tilde g} \, \bigg\{ -\frac{1}{4} \hat D_2 \,
\tilde{g}^{\mu\rho}\tilde{g}^{\nu\sigma}{\cal F}^{\hat A}_{\mu\nu}{\cal F}^{\hat A}_{\rho\sigma} 
-\frac{1}{4} \bar D_2 \,
\tilde{g}^{\mu\rho}\tilde{g}^{\nu\sigma}{\cal F}^{\bar A}_{\mu\nu}{\cal F}^{\bar A}_{\rho\sigma} \bigg\} \bigg|_{\rho = \rho_2} \,.
\end{align}
We shall show how the choice of  coefficients, $\hat D_2$ and $\bar D_2$, relates to the $SO(4)$ gauge coupling
in field theory.
We anticipate here that the two terms appearing in this localised action are closely related to the third and fourth terms in the Lagrangian density
of Eq.~(\ref{Eq:L5}).

The circle reduction also left an Abelian symmetry in the five-dimensional gravity theory, which requires the introduction of the  corresponding boundary action for the $U(1)$ gauge field:
\begin{align}
\label{Eq:Schi2}
	\mathcal S_{\chi,2} =
	\int \dd^4 x \sqrt{-\tilde g} \, \bigg\{ - \frac{1}{4} D_{\chi,2} \,
\tilde{g}^{\mu\rho}\tilde{g}^{\nu\sigma}F^{(\chi)}_{\mu\nu}F^{(\chi)}_{\rho\sigma} \bigg\} \bigg|_{\rho = \rho_2} \,,
\end{align}
with $D_{\chi,2}$ a (possibly divergent) constant required by holographic renormalisation. This term will not play a crucial role in the following, but is needed for completeness.

By contrast, the next term is needed for holographic renormalisation and will play an important role in the following,
consisting of a boundary-localised action for the bulk scalar, $\mathcal X$:
\begin{align}
\label{Eq:SX2}
	\mathcal S_{\mathcal X,2} =
	 \int \dd^4 x \sqrt{-\tilde g} \, \bigg\{ - \frac{1}{2} K_{\mathcal X,2} \, \tilde g^{\mu\nu} (D_\mu \mathcal X)^T D_\nu \mathcal X \bigg\} 
	 \bigg|_{\rho = \rho_2} \,.
\end{align}

Finally, the complete action of the model also contains gauge-fixing terms that realise the $R_\xi$ gauge. We relegate their explicit form to the Appendix, as there are no substantive elements of novelty to this technical part, and we follow the formalism in Ref.~\cite{Elander:2018aub}, adapted to include the treatment of the additional, boundary-localised spurion.

\subsection{Boundary conditions for the background solutions}
\label{Sec:bcs}

The bulk equations of motion
 for the background fields, Eqs.~(\ref{Eq:bg1})--(\ref{Eq:bg3}), are not affected by the
addition of localised  terms to action. The boundary conditions differ from Ref.~\cite{Elander:2023aow}, and read as follows:
\begin{align}
\label{eq:BCX}
	\left( \partial_r \mathcal X_\alpha - \frac{\partial \lambda_1}{\partial \mathcal X_\alpha} \right) \bigg|_{\rho_1} &= 0 \,, \qquad \left( \partial_r \mathcal X_\alpha - \frac{\partial \lambda_2}{\partial \mathcal X_\alpha} + \frac{\partial \mathcal V_4}{\partial \mathcal X_\alpha} \right) \bigg|_{\rho_2} = 0 \,, \nonumber \\
	\left( 6 \partial_r \chi - \frac{\partial \lambda_1}{\partial \chi} \right) \bigg|_{\rho_1} &= 0 \,, \qquad
	\left( 6 \partial_r \chi - \frac{\partial \lambda_2}{\partial \chi} + \frac{\partial \mathcal V_4}{\partial \chi} \right) \bigg|_{\rho_2} = 0 \,, \nonumber \\
	\left( \frac{3}{2} \partial_r A + \lambda_1 \right) \bigg|_{\rho_1} &= 0 \,, \qquad \left( \frac{3}{2} \partial_r A + \lambda_2 - \mathcal V_4 \right) \bigg|_{\rho_2} = 0 \,.
\end{align}
Recalling that $\psi = \mathcal X^T P_5$, and replacing $P_5 = \overline{P_5}=(0,0,0,0,v_5)^T$, we can write the second of these equations,
evaluated at the UV boundary, $\rho=\rho_2$,  as follows:
\beq
	0  = \left( \left[ \partial_r \phi - \frac{\partial \lambda_2}{\partial \phi} + \frac{\partial \mathcal V_4}{\partial \phi} \right] \frac{\mathcal X_\alpha}{\phi} + 2 i \partial_r \pi^{\hat A} (t^{\hat A})_\alpha{}^\beta \mathcal X_\beta + \frac{\partial \mathcal V_4}{\partial \psi} \overline{P_5}{}_{\, \alpha} \right) \bigg|_{\rho_2} \,.
\eeq
This is solved by imposing the following algebraic constraints:
\beq
	\partial_r \phi |_{\rho_2} = \left( \frac{\partial \lambda_2}{\partial \phi} - \frac{\partial \mathcal V_4}{\partial \phi} \right) \bigg|_{\rho_2} \,, \qquad \partial_r \pi^{\hat A} |_{\rho_2} = 0 \,, \qquad \frac{\partial \mathcal V_4}{\partial \psi} = 0 \,.
 	\label{Eq:v}
\eeq

The  boundary conditions for $\phi$, $\chi$, and $A$ are only trivially modified by the presence of $\mathcal V_4$, in a way that
can be absorbed into a redefinition of the boundary potentials, $\lambda_i$, hence there is no element of novelty in this respect,
and the solutions are those
displayed in Sect.~\ref{Sec:backgroundsolutions}.\footnote{Suppose that one has obtained background solutions $\phi^{(0)}$, $\chi^{(0)}$, $A^{(0)}$ to the system without $\mathcal V_4$, as in Sect.~\ref{Sec:model}, and that these satisfy the boundary conditions following from a boundary potential $\lambda_2^{(0)}$. Then, after including $\mathcal V_4$, one may choose
\begin{align}
	\lambda_2(\phi,\chi,\mathcal A_6) = & \ \lambda_2^{(0)}(\phi,\chi,\mathcal A_6) + \mathcal V_4(\phi^{(0)},\chi^{(0)},|\vec\pi|=v) \\ & + \left( \phi - \phi^{(0)} \right) \frac{\partial \mathcal V_4}{\partial \phi}(\phi^{(0)},\chi^{(0)},|\vec\pi|=v) + \left( \chi - \chi^{(0)} \right) \frac{\partial \mathcal V_4}{\partial \chi}(\phi^{(0)},\chi^{(0)},|\vec\pi|=v) \,,
\end{align}
such that the same background profiles of $\phi$, $\chi$, and $A$ again satisfy the updated boundary conditions. Hence, the background solutions for $\phi$, $\chi$, and $A$ are exactly the same as in Sect.~\ref{Sec:backgroundsolutions}, irrespectively of the addition of ${\cal V}_4$ at the boundary. }
The only significant difference is that the third of the  conditions~(\ref{Eq:v}) is satisfied by choosing
$\vec\pi$ so that $|\vec\pi| = v$, where the parameter $v$ is related to the vacuum misalignment angle, and governs the spontaneous breaking to $SO(3)$. Without loss of generality, we  assume that only the
 fourth component of $\vec\pi$ is non-zero on 
the background solutions, i.e. $\pi^4 = v$---in analogy with Sect.~\ref{Sec:SO(5)oSO(4)}.

\subsection{Truncation of the action to quadratic order}
\label{Sec:trunc}

Besides identifying interesting gravity backgrounds, associated with the field-theory vacuum, 
 the main objective of this paper is to compute the mass spectrum of
 their fluctuations, which correspond to the field-theory bound states. 
We hence simplify the action by expanding it in powers of the fields that vanish in the background and truncating the resulting action to quadratic order. This approximation retains 
all the information needed to compute two-point functions.\footnote{One has to take extra care that the gauging of $SO(4)$ is sufficiently weak. We will return to this point later, and in the Appendix.} 
We retain full functional dependence of the action on fields having non-trivial profiles: $\phi$, $\chi$, and $g_{MN}$.

As anticipated, we set  the background value of the spurion,  $P_5 = \overline{P_5}$, 
by taking  the limit $\lambda_5 \rightarrow \infty$, so that the fluctuation of its fifth component has infinite mass. We hence only retain its first four components, treating them as perturbations. 
We write the resulting five-dimensional action, truncated at the quadratic order, as follows:
\beq \label{eq:5d}
	\mathcal S_5^{(2)} = \mathcal S_5^{(bulk,2)} + \mathcal S_{P_5,2}^{(2)} + S_{\mathcal V_4,2}^{(2)} 
	+\sum_{i=1,2} \mathcal S_{4,i}^{(2)} \,,
\eeq
where $\mathcal S_5^{(bulk,2)}$ is the bulk part of the action, the boundary actions $\mathcal S_{P_5,2}^{(2)}$
 and $S_{\mathcal V_4,2}^{(2)}$ are localised at $\rho = \rho_2$, while $\mathcal S_{4,i}^{(2)}$ are localised at $\rho = \rho_i$ ($i = 1,2$). In the remainder of this subsection, we display the explicit form of these terms.

In the backgrounds, the non-zero value of  $\pi^4 = v$ breaks spontaneously $SO(4)$ to $SO(3)$.
It is  convenient to use indices adapted to the $SO(3)$ language, namely $\hat{\mathcal A}= 1,2,3$, 
$\tilde{\mathcal A}= 5,6,7$, and 
$\bar{\mathcal A} = 8,9,10$, chosen so that $t^{\bar{\mathcal A}}$ are the unbroken generators of $SO(3)$.
We parametrise the fluctuations of the fourth component of $\pi^{\hat A}$ by writing $\pi^4 = v + \Pi^4$. 
We note that the physical (mass eigen-)states in the spin-1 sector of the theory
result from mixing of the two triplets labelled by the $\hat{\cal A}$ and $\tilde{\cal A}$ indexes.
To simplify the resulting equations, we define the  linear combinations:
\beqs
	\mathcal B_6^{\hat{\mathcal A}} &=& \cos(v) \mathcal A_6^{\hat{\mathcal A}} + \sin(v) \mathcal A_6^{\hat{\mathcal A}+4} \,, \\
	\mathcal B_6^{\tilde{\mathcal A}} &=& - \sin(v) \mathcal A_6^{\tilde{\mathcal A} - 4} + \cos(v) \mathcal A_6^{\tilde{\mathcal A}} \,, \\
	\mathcal B_M{}^{\hat{\mathcal A}} &=& \cos(v) \mathcal A_M{}^{\hat{\mathcal A}} + \sin(v) \mathcal A_M{}^{\hat{\mathcal A}+4} \,, \\
	\mathcal B_M{}^{\tilde{\mathcal A}} &=& - \sin(v) \mathcal A_M{}^{\tilde{\mathcal A}-4} + \cos(v) \mathcal A_M{}^{\tilde{\mathcal A}} \,.
\eeqs
We hence adopt the following choice of basis for the fields (other than the metric) that we allow to fluctuate over the backgrounds:\footnote{For $v = 0$, this basis coincides with the one used in Ref.~\cite{Elander:2023aow} {(up to a trivial reordering of the fields)}.}
\beqs
\label{eq:flucbasis1}
	\Phi^a &=& \{ \phi, \chi \} \,, \\
\label{eq:flucbasis2}
	\Phi^{(0)a} &=& { \{ \mathcal B_6^{\hat{\mathcal A}}, \mathcal A_6^4, \mathcal B_6^{\tilde{\mathcal A}}, \mathcal A_6^{\bar{\mathcal A}} \} } \,, \\
\label{eq:flucbasis3}
	V_M{}^A &=& { \{ \chi_M, \mathcal B_M{}^{\hat{\mathcal A}}, \mathcal A_M{}^4, \mathcal B_M{}^{\tilde{\mathcal A}}, \mathcal A_M{}^{\bar{\mathcal A}} \} } \,, \\
\label{eq:flucbasis4}
	{\cal H}^{(1)}_M{}^A &=& { \left\{ 0, \frac{\sin(v)}{v} \partial_M \pi^{\hat{\mathcal A}} + \frac{g}{2} \mathcal B_M{}^{\hat{\mathcal A}}, \partial_M \Pi^4 + \frac{g}{2} \mathcal A_M{}^4, 0, 0 \right\} } \,.
\eeqs

We 
write explicitly the action, starting  from the bulk part, which  takes the form:
\begin{align}
\label{eq:sigmamodelaction}
	\mathcal S_5^{(bulk,2)} =& \int \dd^5 x \sqrt{-g_5} \, \bigg\{ \frac{R}{4} - \frac{1}{2} g^{MN} G_{ab} \partial_M \Phi^a \partial_N \Phi^b - \mathcal V_5(\Phi^a) \nonumber \\ & - \frac{1}{2} g^{MN} G_{ab}^{(0)} \partial_M \Phi^{(0)a} \partial_N \Phi^{(0)b} - \frac{1}{2} m_{ab}^{(0)2} \Phi^{(0)a} \Phi^{(0)b} \nonumber \\
	& -\frac{1}{2} g^{MN} G^{(1)}_{AB} {\cal H}^{(1)}_M{}^A {\cal H}^{(1)}_N{}^B -\frac{1}{4} g^{MO} g^{NP} H^{(1)}_{AB}F_{MN}{}^{A}F_{OP}{}^B \bigg\} \,.
\end{align}
In this expression, the field strengths are  $F_{MN}{}^A \equiv 2\partial_{[M}V_{N]}{}^{A}$, and the scalar potential is $\mathcal V_5(\phi,\chi) = e^{-2\chi} \mathcal V_6(\phi)$. 
For completeness, we reproduce here all the entries of the sigma-model matrices, that are independent of $v$~\cite{Elander:2023aow}:
\beqs
	& G_{ab} = {\rm diag}\left(1,6\right) \,, \qquad
G^{(0)} = { \left(\begin{array}{c|c}
e^{-6\chi} \mathbb 1_{4 \times 4} &\cr
\hline
&e^{-6\chi} \mathbb 1_{6 \times 6} \cr
\end{array}\right) } \,, \qquad
\frac{m^{(0)2}}{g^2} = { \left(\begin{array}{c|c}
\frac{1}{4} \phi^2 e^{-8\chi} \mathbb 1_{4 \times 4} &\cr
\hline
& \mathbb 0_{6 \times 6} \cr
\end{array}\right) } \,, \nonumber \\
& G^{(1)} = { \left(\begin{array}{c|c|c}
0&&\cr
\hline
& \phi^2 \,\mathbb{1}_{4 \times 4} &\cr
\hline
&& \mathbb 0_{6 \times 6} \cr
\end{array}\right) } \,, \qquad
H^{(1)} = { \left(\begin{array}{c|c|c}
\frac{1}{4}e^{8\chi}&&\cr
\hline
& e^{2\chi}\,\mathbb{1}_{4 \times 4}&\cr
\hline
&& e^{2\chi}\,\mathbb{1}_{6 \times 6} \cr
\end{array}\right) } \,.
\eeqs

We now turn attention to the boundary-localised actions, $S_{P_5,2}^{(2)}$, $S_{\mathcal V_4,2}^{(2)}$, and $\mathcal S_{4,i}^{(2)}$. After defining the variables
\beqs
	\mathcal P_{5 \, \mu}^{(1)}{}^{\hat{A}} &=& \left\{ \partial_\mu P_5^{\hat{\mathcal A}} + \frac{g v_5}{2} \mathcal A_\mu{}^{\hat{\mathcal A}}, \partial_\mu P_5^4 + \frac{g v_5}{2} \mathcal A_\mu{}^4 \right\} \nonumber \\
	&=& \left\{ \partial_\mu P_5^{\hat{\mathcal A}} + \frac{g v_5}{2} \left( \cos(v) \mathcal B_\mu{}^{\hat{\mathcal A}} - \sin(v) \mathcal B_\mu{}^{\hat{\mathcal A}+4}\right), \partial_\mu P_5^4 + \frac{g v_5}{2} \mathcal A_\mu{}^4 \right\} \,,
\eeqs
we find that we can write
\beq
	S_{P_5}^{(2)} = \int \dd^4 x \sqrt{-\tilde g} \, \bigg\{ - \frac{1}{2} \tilde g^{\mu\nu} K_5 \delta_{\hat{A}\hat{B}} \mathcal P_{5 \, \mu}^{(1)}{}^{\hat{A}} \mathcal P_{5 \, \nu}^{(1)}{}^{\hat{B}} \bigg\} \bigg|_{\rho = \rho_2} \,. 
\eeq

The expansion of $\mathcal V_4$ to quadratic order can be written formally as
\beq
	S_{\mathcal V_4,2}^{(2)} = - \int \dd^4 x \sqrt{-\tilde g} \, \bigg\{ \mathcal V_4^{(0)}(\phi,\chi,v) + \mathcal V_4^{(2)}(\phi,\chi,v,P_5^4,\Pi^4) \bigg\} \bigg|_{\rho = \rho_2} \,,
\eeq
where the zeroth-order contribution, $\mathcal V_4^{(0)}$, is evaluated on the background solutions, while at second order
\beq
	\mathcal V_4^{(2)} = \frac{1}{2} \partial_v^2 \mathcal V_4 \left( \Pi^4 - \frac{P_5^4}{v_5} \right)^2 \,,
	\qquad {\rm with} \qquad \partial_v^2 \mathcal V_4 = \sin^2(v) \phi^2 v_5^2 \frac{\partial^2 \mathcal V_4}{\partial \psi^2} \,.
\eeq

The final contributions to  the boundary actions, $\mathcal S_{4,i}^{(2)}$, take the form
\begin{align}
	\mathcal S_{4,i}^{(2)} = (-)^i & \int \dd^4 x \sqrt{-\tilde g} \, \bigg\{ \frac{K}{2} + \lambda_i - \frac{1}{2} \tilde g^{\mu\nu} K_{\mathcal X,i} \partial_\mu \phi \partial_\nu \phi \nonumber \\ & - \frac{1}{2} \tilde g^{\mu\nu} C_{i \, AB}^{(1)} {\cal H}^{(1)}_\mu{}^A {\cal H}^{(1)}_\nu{}^B -\frac{1}{4} \tilde g^{\mu\sigma} \tilde g^{\nu\gamma} D_{i \, AB}^{(1)} F_{\mu\nu}{}^{A}F_{\sigma\gamma}{}^B \bigg\} \bigg|_{\rho = \rho_i} \,,
\end{align}
with $C_1^{(1)} =0$,  $D_1^{(1)}=0$, while 
\beq
C_2^{(1)} = K_{\mathcal X,2} \, \phi^2 \, { \left(\begin{array}{c|c|c|c|c}
0&&&&\cr
\hline
&\mathbb 1_{3 \times 3}&&&\cr
\hline
&& 1 &&\cr
\hline
&&& \mathbb 0_{3 \times 3} & \cr
\hline
&&&& \mathbb 0_{3 \times 3} \cr
\end{array}\right) } \,,
\eeq
and
\beq
D_2^{(1)} = { \left(\begin{array}{c|c|c|c|c}
D_{\chi,2} &&&&\cr
\hline
& \frac{1}{2} \big[ \bar D_2 + \hat D_2+ \cos(2v) ( \hat D_2 - \bar D_2 ) \big] \, \mathbb{1}_{3 \times 3} && \frac{1}{2} \sin(2v) ( \bar D_2 - \hat D_2 ) \, \mathbb 1_{3 \times 3} &\cr
\hline
&& \hat D_2 &&\cr
\hline
& \frac{1}{2} \sin(2v) ( \bar D_2 - \hat D_2 ) \, \mathbb 1_{3 \times 3} && \frac{1}{2} \big[ \bar D_2 + \hat D_2+ \cos(2v) ( \bar D_2- \hat D_2) \big] \, \mathbb 1_{3 \times 3} & \cr
\hline
&&&& \bar D_2 \, \mathbb 1_{3 \times 3} \cr
\end{array}\right) } \,.
\eeq

\section{Fluctuation equations and the parameters of the model}
\label{Sec:vacuum}

\begin{table}
\caption{Summary table associating the fields in five-dimensional language (left) to their fluctuations
in the four-dimensional, ADM formalism (right). Notice the existence of mixing in the physical states denoted by the parenthesis:
the mass eigenstates for $(\mathfrak{a}^{\phi},\mathfrak{a}^{\chi})$ and 
 $\left(\mathfrak{v}_{\mu}{}^{\tilde{\mathcal A}},\mathfrak{v}_{\mu}{}^{\hat{\mathcal A}}\right)$ 
 are admixtures of the original fluctuations in the theory.
For some of the spin-0 fluctuations, mass degeneracies survive even after the spontaneous 
 breaking $SO(4)\rightarrow SO(3)$, hence the eigenstates can be grouped together  in the $SO(4)$ language.
}
\label{Fig:Fluctuations}
\begin{tabular}{||c||c|c|}
\hline\hline
{\rm Field}  &\multicolumn{2}{|c||}{  {\rm Fluctuation}} \cr
\hline\hline
$g_{MN}$ &\multicolumn{2}{|c||}{ $\mathfrak{e}_{\mu\nu}$} \cr
$\chi_M{}$ & \multicolumn{2}{|c||}{ $\mathfrak{v}_{\mu}$} \cr
$(\phi,\,\chi)$ & \multicolumn{2}{|c||}{ $(\mathfrak{a}^{\phi},\mathfrak{a}^{\chi})$}\cr
\hline
$\left.\begin{array}{c}
\mathcal B_6^{\hat{\mathcal A}} \cr
 \mathcal A_6^{4}\end{array}\right\}
$ & 
\multicolumn{2}{|c||}{ 
$
\mathfrak{a}^{\hat{A}}=\left\{\begin{array}{c}
\mathfrak{a}^{\hat{\cal A}}
\cr
\mathfrak{a}^{4}
\end{array}\right.$}
\cr
$\left.\begin{array}{c}
\mathcal B_6^{\tilde{\mathcal A}} \cr
 \mathcal A_6^{\bar{\mathcal A}}\end{array}\right\}
$ & 
\multicolumn{2}{|c||}{ 
$
\mathfrak{a}^{\bar{A}}=\left\{\begin{array}{c}
\mathfrak{a}^{\tilde{\cal A}}
\cr
\mathfrak{a}^{\bar{\cal A}}
\end{array}\right.$}
\cr
\hline
$\left({\cal B}_M{}^{\hat{\mathcal A}},{\cal B}_M{}^{\tilde{\mathcal A}}\right)$ &
\multicolumn{2}{|c||}{$\left(\mathfrak{v}_{\mu}{}^{\hat{\mathcal A}},\mathfrak{v}_{\mu}{}^{\tilde{\mathcal A}}\right)$} \cr
${\cal A}_M{}^4$ & 
\multicolumn{2}{|c||}{ $\mathfrak{v}_{\mu}{}^4$} \cr
 ${\cal A}_M{}^{\bar{\mathcal A}}$ &
 \multicolumn{2}{|c||}{  $\mathfrak{v}_{\mu}{}^{\bar{\mathcal A}}$} \cr
\hline
$\left.\begin{array}{c}
\pi^{\hat{\mathcal A}} \cr
 \Pi^{4}\end{array}\right\}$
& 
\multicolumn{2}{|c||}{ 
$
\mathfrak{p}^{\hat{A}}=\left\{\begin{array}{c}
\mathfrak{p}^{\hat{\cal A}}
\cr
\mathfrak{p}^{4}
\end{array}\right.$}
\cr

\hline\hline
\end{tabular}
\end{table}

In this section, we discuss the fluctuations of all the fields in the backgrounds of interest, and summarise the salient features of the 
gauge-invariant  formalism that we use to compute the mass spectrum reported  in Sect.~\ref{Sec:spectrum} (for further details, 
including our use of the ADM formalism~\cite{Arnowitt:1962hi} and the introduction of gauge-invariant combinations of the fluctuations,
see Appendix~\ref{Sec:formalism}).
We find it convenient to switch between the $SO(4)$ language (the $SO(4)$ indices are $\hat A = 1, \cdots, 4$ and $\bar A = 5, \cdots, 10$), and the $SO(3)$ language (with indices $\hat{\mathcal A}= 1,2,3$, $\tilde{\mathcal A}= 5,6,7$, and $\bar{\mathcal A} = 8,9,10$).
We denote the original fields in the action and the gauge-invariant combinations
of fluctuations corresponding to them with different symbols,  summarising the correspondences in Table~\ref{Fig:Fluctuations}.

We apply the gauge-invariant formalism of Refs.~\cite{Bianchi:2003ug,Berg:2005pd,Berg:2006xy,Elander:2009bm,Elander:2010wd,Elander:2014ola,Elander:2017cle,Elander:2017hyr}
to the treatment of  tensor, ${\mathfrak e}_{\mu\nu}$, and 
 scalar fluctuations of fields carrying no $SO(4)$ quantum numbers, $\mathfrak a^{\phi}$ and  $\mathfrak a^{\chi}$. 
  Scalar fluctuations associated with non-trivial $SO(3)$
irreducible representations, denoted as
$\{ \mathcal B_6^{\tilde{\mathcal A}}, \mathcal A_6^{\bar{\mathcal A}} \}$ and $\{ \mathcal B_6^{\hat{\mathcal A}}, \mathcal A_6^4 \}$,
  form $SO(4)$ multiplets transforming in the adjoint and fundamental representations, respectively---see Table~\ref{Fig:Fluctuations}.
We treat them in the same way as $\mathfrak a^{\phi}$ and $\mathfrak a^{\chi}$,  although they do not mix with components of the metric,
and they do not introduce significant elements of novelty in the paper---see
Appendix~\ref{Sec:scalars}  and  Refs.~\cite{Elander:2023aow,Elander:2022ebt}.

  The vector fluctuations, ${\mathfrak v}_{\mu}$ ($\mathfrak{v}$ in the following), associated with the $U(1)$ gauge field, 
  $\chi_M$, complete the set of $SO(4)$ singlets in the model. 
  Their treatment requires gauge fixing, 
  but ultimately the study of the mass spectrum is carried out by focusing
  on the (gauge-invariant) transverse part of the fluctuations, which obeys the differential equation~\cite{Elander:2022ebt}
\beqs
    0 &=& \bigg[ \partial_\rho^2 + (2 \partial_\rho A + 7 \partial_\rho \chi) \partial_\rho - e^{2\chi - 2 A} q^2 \bigg] \mathfrak{v} \,,
\eeqs
 subject to the UV boundary condition
\beq
	0 = \bigg[ e^{7\chi} \partial_\rho + D_{\chi,2} e^{-2 A} q^2 \bigg] \mathfrak v^{\bar{\mathcal A}} \Big|_{\rho = \rho_2}\,,
\eeq
together with the Neumann boundary condition, $\partial_\rho \mathfrak{v} |_{\rho = \rho_1} = 0$, in the IR. The mass spectrum is given by those $M^2 \equiv-q^2$ for which solutions exist that satisfy both the bulk equations of motion and boundary conditions above. In the following, we make the choice $D_{\chi,2} = 0$, as  this  $U(1)$ is a global symmetry in the field theory dual.

We devote central  attention to the fluctuations that are affected by  $SO(5)$ and $SO(4)$ symmetry breaking, given by the vectors $\mathfrak v^{\tilde{\mathcal A}}$, $\mathfrak v^{\bar{\mathcal A}}$, $\mathfrak v^{\hat{\mathcal A}}$, and $\mathfrak v^4$, associated with the fields defined in Eq.~\eqref{eq:flucbasis3} (except $\chi_M$), together with the pseudoscalars $\mathfrak p^{\hat{\mathcal A}}$ and $\mathfrak p^4$, associated with ${\cal H}^{(1)}_M{}^{\hat{\mathcal A}}$ and ${\cal H}^{(1)}_M{}^4$ defined in Eq.~\eqref{eq:flucbasis4}. 
The treatment of these vector and pseudoscalar states requires adding appropriate gauge-fixing terms. We  adopt the $R_{\xi}$-gauge,
and report details of the procedure in Appendix~\ref{sec:vectorspseudoscalars}.
The (gauge-invariant) bulk equations of motion associated with the transverse polarisations of the spin-1 fluctuations in the 
symmetry-breaking backgrounds, as well as the pseudoscalar ones,  are manifestly 
SO(4) invariant, and can be written  as follows~\cite{Elander:2023aow}:
\beqs \label{eq:barv}
    0 &=& \bigg[ \partial_\rho^2 + (2 \partial_\rho A + \partial_\rho \chi) \partial_\rho - e^{2\chi - 2 A} q^2 \bigg] \mathfrak v^{\bar A} \,,\\
    \label{eq:hatv}
    0 &=& \bigg[ \partial_\rho^2 + (2 \partial_\rho A + \partial_\rho \chi) \partial_\rho - \frac{g^2 \phi^2}{4} - e^{2\chi - 2 A} q^2 \bigg] \mathfrak v^{\hat A} \,,\\
    0 &=& \bigg[ \partial_\rho^2 - \left( 2 \partial_\rho A + \partial_\rho \chi + \frac{2\partial_\rho \phi}{\phi} \right) \partial_\rho - \frac{g^2 \phi^2}{4} - e^{2\chi - 2 A} q^2 \bigg] \mathfrak p^{\hat A} \label{eq:hatp}\,.
 \eeqs

The UV boundary conditions, at $\rho=\rho_2$, induce spontaneous symmetry breaking to $SO(3)\subset SO(4)$.  
The vectors that transform as triplets, $3$, of $SO(3)$, but live along the $SO(5)/SO(3)$ broken directions, are denoted as $\mathfrak v^{\tilde{\mathcal A}}$ and $\mathfrak v^{\hat{\mathcal A}}$. They mix and obey
the following boundary conditions:
{
\beqs
\label{eq:vectorBCs}
    0 &=& \bigg[ e^{\chi} \partial_\rho + \frac{g^2}{4} \sin(v)^2 K_5 v_5^2 + \frac{1}{2} \Big( \bar D_2 + \hat D_2 + \cos(2v) ( \bar D_2 - \hat D_2 ) \Big) e^{-2A} q^2 \bigg] \mathfrak v^{\tilde{\mathcal A}} \Big|_{\rho = \rho_2} \nonumber \\
    && + \frac{\sin(2v)}{2} \bigg[ - \frac{g^2}{4} K_5 v_5^2 + (\bar D_2 - \hat D_2) e^{-2 A} q^2 \bigg] \mathfrak v^{\hat {\mathcal A}} \Big|_{\rho = \rho_2} \,,\\
    0 &=& \bigg[ e^{\chi} \partial_\rho + \frac{g^2}{4} \Big( K_{\mathcal X,2} \phi^2 + \cos(v)^2 K_5 v_5^2 \Big) + \frac{1}{2} \Big( \bar D_2 + \hat D_2 + \cos(2v) ( \hat D_2 - \bar D_2 ) \Big) e^{-2A} q^2 \bigg] \mathfrak v^{\hat{\mathcal A}} \Big|_{\rho = \rho_2} \nonumber \\
    && + \frac{\sin(2v)}{2} \bigg[ - \frac{g^2}{4} K_5 v_5^2 + (\bar D_2 - \hat D_2) e^{-2 A} q^2  \bigg] \mathfrak v^{\tilde {\mathcal A}} \Big|_{\rho = \rho_2} \,.
\eeqs
The  boundary conditions for the other vectors, $\mathfrak v^{\bar{\mathcal A}}$ and $\mathfrak v^4$, are given by
\beqs \label{eq:barbv}
    0 &=& \bigg[ e^{\chi} \partial_\rho + \bar D_2 e^{-2 A} q^2 \bigg] \mathfrak v^{\bar{\mathcal A}} \Big|_{\rho = \rho_2} \,,\\
    0 &=& \bigg[ e^{\chi} \partial_\rho + \frac{g^2}{4} \Big( K_{\mathcal X,2} \phi^2 + K_5 v_5^2 \Big) + \hat D_2 e^{-2A} q^2 \bigg] \mathfrak v^4 \Big|_{\rho = \rho_2} \,.\label{eq:4bv}
\eeqs
The presence of the spurion, $P_5$, does not affect the boundary conditions for the pseudoscalar triplet, 
$\mathfrak p^{\hat{\mathcal A}}$: 
\beq\label{eq:hatbp}
    0 = \bigg[ K_{\mathcal X,2} e^{-\chi} \partial_\rho + 1 \bigg] \mathfrak p^{\hat{\mathcal A}} \Big|_{\rho = \rho_2}\,,
\eeq
yet, it modifies the boundary condition for the $SO(3)$ singlet, $\mathfrak p^4$, that  at $\rho = \rho_2$ obeys:
\beqs
    0 &=& \bigg[ \left( K_{\mathcal{X},2} + \frac{K_5 v_5^2}{\phi^2} + \frac{K_5 v_5^2}{\partial_v^2 \mathcal V_4} K_{\mathcal{X},2} e^{-2 A} q^2 \right) e^{-\chi} \partial_\rho + \left(1+\frac{K_5 v_5^2}{\partial_v^2 \mathcal V_4} e^{-2 A} q^2 \right) \bigg] \mathfrak p^4 \Big|_{\rho = \rho_2} \,.
    \label{eq:p4BCs}
 \eeqs
}

In the IR, the boundary conditions obeyed by the fluctuations at $\rho=\rho_1$ are significantly simpler. They reduce to Neumann boundary conditions for the vectors, with $\partial_\rho v^{\tilde{\mathcal A}}(\rho_1)=\partial_\rho v^{\hat{\mathcal A}}(\rho_1)
=\partial_\rho v^{\bar{\mathcal A}}(\rho_1)=\partial_\rho v^{4}(\rho_1)=0$, and to Dirichlet boundary conditions for the pseudoscalars, with
$ \mathfrak p^{\hat{\mathcal A}} (\rho_1)=\mathfrak p^4 (\rho_1) = 0$.
The IR boundary conditions do not introduce additional symmetry breaking, and could be recast,
 equivalently,  in terms of 
$SO(4)$ multiplets.

\subsection{Model parameters and $SO(4)$ gauging }
\label{Sec:4}

It is instructive to compute some of the two-point functions, in particular 
$\langle \mathcal A_\mu^{\bar{\mathcal A}}(q) \mathcal A_\nu^{\bar{\mathcal A}}(-q) \rangle$ 
and $\langle \mathcal A_\mu^4(q) \mathcal A_\nu^4(-q) \rangle$, and exhibit
their structure. We separate the transverse and longitudinal polarisations, and write the results in terms of the projectors,
$P_{\mu\nu}=\eta_{\mu\nu}-\frac{q_{\mu}q_{\nu}}{q^2}$ and $\frac{q_{\mu}q_{\nu}}{q^2}$, highlighting the dependence on gauge-fixing
 parameters, $\bar{M}_2$ and $M_2^4$ (defined in Appendix~\ref{sec:vectorspseudoscalars}), 
 in the longitudinal part of the propagators only:
\beqs
	\langle \mathcal A_\mu^{\bar{\mathcal A}}(q) \mathcal A_\nu^{\bar{\mathcal A}}(-q) \rangle &=& (-i) \lim_{\rho_2 \rightarrow \infty} \Bigg\{ \hspace{-0.08cm} \, e^{-2A} \bigg( \bar D_2 e^{-2A} q^2 + e^\chi \frac{\partial_\rho \mathfrak v^{\bar{\mathcal A}}}{\mathfrak v^{\bar{\mathcal A}}} \bigg)^{-1} \bigg|_{\rho = \rho_2} P_{\mu\nu} \nonumber \\ && \hspace{1.5cm} + e^{-2A} \bigg( \frac{1}{\bar M_2} e^{-2A} q^2 + e^\chi \frac{\partial_\rho \mathfrak v_L^{\bar{\mathcal A}}}{\mathfrak v_L^{\bar{\mathcal A}}} \bigg)^{-1} \bigg|_{\rho = \rho_2} \frac{q_\mu q_\nu}{q^2} \Bigg\} \,, \\
	\langle \mathcal A_\mu^4(q) \mathcal A_\nu^4(-q) \rangle &=& (-i) \lim_{\rho_2 \rightarrow \infty} \Bigg\{ \hspace{-0.08cm} \, e^{-2A} \bigg( \hat D_2 e^{-2A} q^2 + \frac{g^2}{4} K_5 v_5^2 + \frac{g^2}{4} K_{\mathcal X,2} \phi^2 + e^\chi \frac{\partial_\rho \mathfrak v^4}{\mathfrak v^4} \bigg)^{-1} \bigg|_{\rho = \rho_2} P_{\mu\nu} \nonumber \\ && \hspace{1.5cm} + e^{-2A} \bigg( \frac{1}{M_2^4} e^{-2A} q^2 + \frac{g^2}{4} K_5 v_5^2 + \frac{g^2}{4} K_{\mathcal X,2} \phi^2 + e^\chi \frac{\partial_\rho \mathfrak v_L^4}{\mathfrak v_L^4} \bigg)^{-1} \bigg|_{\rho = \rho_2} \frac{q_\mu q_\nu}{q^2} \Bigg\}  \,.
\eeqs
Here, $\mathfrak v^{4,\bar{\mathcal A}}_L$ stand the longitudinal ($L$) parts of the corresponding gauge fields. They appear in the unphysical, longitudinal parts of the two-point functions, that contain gauge-fixing parameters. We include them for completeness, and their bulk equations of motion and boundary conditions can be found in Appendix~\ref{sec:vectorspseudoscalars}.

A careful analysis of the (UV) expansions in powers of   small $z \equiv e^{-\rho}$,  
shows that
$A\simeq 4\chi  \simeq -\frac{4}{3} \log(z)$  and $\phi \simeq \phi_J z^{\Delta_J}$---see Eqs.~(\ref{eq:backgroundUVexp1})--(\ref{eq:backgroundUVexp3}). The second and third terms contribute
to  $\langle \mathcal A_\mu^4(q) \mathcal A_\nu^4(-q) \rangle$, if we impose the scalings 
\beq
\label{eq:K5KXscalings}
	K_5 = \frac{k_5}{v_5^2} \, z^{8/3} \,, \qquad
	K_{\mathcal X,2} = k_{\mathcal X} \, z^{8/3 - 2\Delta_J} \,,
\eeq
where we introduced new parameters, $k_5$ and $k_{\mathcal X}$, that do not depend on $z$.

In order to see how to fix $\bar D_2$ and $\hat D_2$, we work out the example of $\Delta=\Delta_J = 2$---the generalisation to any $\Delta$ requires a case-by-case analysis, but is straightforward. By building upon the small-$z$ expansions reported in Appendix~\ref{sec:IRUVexpansions}, one sees that
 in order to cancel divergences one must choose
\beq
\label{eq:D2s}
	\bar D_2 = - z^{-1} + \frac{1}{\bar \varepsilon^2} \,, \qquad
	\hat D_2 = - z^{-1} + \frac{1}{\hat \varepsilon^2} \,,
\eeq
with ${\bar \varepsilon^2}$ and ${\hat \varepsilon^2}$ two free parameters, independent of $z$. 
With these replacements, one can take the $z\rightarrow 0$ limit, to  find:
\beqs
\label{eq:twopointAbAb}
	P^{\mu\sigma} P^{\nu\gamma} \langle \mathcal A_\mu^{\bar{\mathcal A}}(q) \mathcal A_\nu^{\bar{\mathcal A}}(-q) \rangle &=& -i \, \Bigg( \frac{q^2}{\bar \varepsilon^2} - \frac{3 v^{\bar{\mathcal A}}_3}{v^{\bar{\mathcal A}}_0} \Bigg)^{-1} P^{\sigma\gamma} \,, \\
 \label{eq:twopointA4A4}
	P^{\mu\sigma} P^{\nu\gamma} \langle \mathcal A_\mu^4(q) \mathcal A_\nu^4(-q) \rangle &=& -i \, \Bigg( \frac{q^2}{\hat \varepsilon^2} - \frac{3 v^4_3}{v^4_0} + \frac{g^2 k_5}{4} + \frac{g^2 k_{\mathcal X} \phi_J^2}{4} \Bigg)^{-1} P^{\sigma\gamma} \,.
\eeqs
Choosing the values of  $\bar \varepsilon$ and $\hat \varepsilon$ corresponds to choosing the strength of the (weak) gauge couplings.\footnote{If one sets $\bar \varepsilon = \hat \varepsilon \equiv \varepsilon$, then the full $SO(5)$ is (weakly) gauged in the dual field theory, and, provided $\varepsilon$ is small, the gauge coupling in four dimensions, measured at $q^2 \simeq 0$, is approximately given by $g_4 \equiv \varepsilon \, g$. }
After   rescaling the normalisation of the fields, we set $\hat \varepsilon \rightarrow 0$.
Only the $SO(4)$ subgroup is gauged in the dual field theory, with coupling strength 
that, for small $\bar \varepsilon$, is approximately equal to
$g_4 \equiv \bar \varepsilon \, g$.\footnote{This choice could also be implemented by picking $\hat D_2 = 0$, and letting $\hat \varepsilon \rightarrow 0$ as a function of $z \rightarrow 0$. Either choice is analogous  to the limit $\tilde{\kappa}\rightarrow \infty$, discussed in Sect.~\ref{Sec:SO(5)oSO(4)}.} 
 These expressions are valid for small $q^2$ and $\bar{\varepsilon}$; we 
 discuss how to improve these results, and obtain the physical two-point function valid for all $q^2$,
 in Appendix~\ref{sec:RGimprovement}.

Finally, in order to obtain a non-trivial contribution to Eq.~\eqref{eq:p4BCs}, from $\partial_v^2 \mathcal V_4$,
 for any $\Delta$, we must impose the scaling:
\beq
\label{eq:m4sqscaling}
	\partial_v^2 \mathcal V_4 = m_4^2 \, z^{16/3} \,,
\eeq
which introduces the parameter $m_4^2$. 
This parameter is the analogue of the combination $\lambda f^2\sin^2\theta$, in Eq.~(\ref{Eq:VSO5}).

\begin{table}
\caption{Summary table of boundary terms and parameters
 appearing in our model. The first column contains the list of boundary terms added to the theory, and the second the parameters appearing in them before specifying the lmits needed for holographic renormalisation and gauging. The free parameters affecting the observables are reported in the third column. The last column references where the relevant terms and quantities appear for the first time in the text.}
\label{Fig:bTerms}
\begin{tabular}{||c||c|c|c|c|c|}
\hline\hline
{\rm Boundary term}  &\multicolumn{2}{|c||}{  {\rm Parameters}} & \multicolumn{2}{|c||}{ Free parameters }& Definition\cr
\hline\hline
$\mathcal S_{{\rm GHY},i}$&\multicolumn{2}{|c||}{ --- } & \multicolumn{2}{|c||}{ --- }& \eqref{Eq:GHY} \cr 
$\mathcal S_{\lambda,i}$&\multicolumn{2}{|c||}{ $m_{\mathcal X,i}^2$ } & \multicolumn{2}{|c||}{ --- }& \eqref{Eq:Slambdai}, text after \eqref{Eq:window}\cr 
  $\mathcal S_{\chi,2}$ &\multicolumn{2}{|c||}{ $D_{\chi,2}$ } & \multicolumn{2}{|c||}{ --- }& \eqref{Eq:Schi2} \cr
$\mathcal S_{P_5,2}$ &\multicolumn{2}{|c||}{ $K_5 (k_5), \lambda_5, v_5$} & \multicolumn{2}{|c||}{ --- }& \eqref{eq:SP5}, \eqref{eq:K5KXscalings} \cr
$\mathcal S_{\mathcal V_4,2}$ & \multicolumn{2}{|c||}{ $\partial_v^2 \mathcal V_4(v,m_4^2)$} & \multicolumn{2}{|c||}{ $v, m_4^2$} & \eqref{eq:SV}, text after \eqref{Eq:v}, \eqref{eq:m4sqscaling}\cr
$\mathcal S_{\mathcal A,2}$ & \multicolumn{2}{|c||}{ $\bar D_2 (\bar \varepsilon), \hat D_2 (\hat \varepsilon)$}& \multicolumn{2}{|c||}{ $\bar \varepsilon$ }& \eqref{Eq:SA2}, \eqref{eq:D2s}\cr
$\mathcal S_{\mathcal X,2}$ & \multicolumn{2}{|c||}{ $K_{\mathcal X,2}(k_{\mathcal X})$}&\multicolumn{2}{|c||}{ $k_{\mathcal X}$} & \eqref{Eq:SX2}, \eqref{eq:K5KXscalings} \cr
\hline\hline
\end{tabular}
\end{table}

For any choices of the free parameters discussed above, one is describing the spontaneous breaking of an exact, gauge symmetry in the dual field theory (as in Sect.~\ref{Sec:SO(5)oSO(4)}), without violating the gauge principle (and unitarity). But let us now 
summarise the order  of limits that allows to make contact with phenomenological requirements, 
along the lines of what we did in Sect.~\ref{Sec:SO(5)oSO(4)}.
 Table~\ref{Fig:bTerms} may help to follow this process.
By first taking the limit $\lambda_5 \rightarrow \infty$, which  introduces an  infinite mass term in the action $\mathcal S_{P_5}$ of Eq.~\eqref{eq:SP5}, the absolute value of 
$P_5$ is frozen. Next, we remove the IR and UV regulators by taking the limits $\rho_1 \rightarrow \rho_o$ and $\rho_2 \rightarrow \infty$. The parameter $k_5$ survives this limiting procedure, and by further taking the limit $k_5 \rightarrow \infty$, the couplings of the remaining degrees of freedom in $P_5$ vanish as well. At this point, $P_5$ is a genuine spurion: although we introduced it as a field, obeying its characteristic transformation rules under a symmetry transformation, it has been reduced to a vector of real numbers.
Finally, the limit $\hat \varepsilon \rightarrow 0$ incorporates the gauging of the $SO(4)$ subgroup, by effectively freezing the gauge bosons along the coset $SO(5)/SO(4)$.\footnote{One may verify that these two limits, $k_5 \rightarrow \infty$ and $\hat \varepsilon \rightarrow 0$, commute, and that, moreover, the results are consistent with putting $\hat D_2 = 0$.}

We conclude by providing 
an explicit example of how to set up  the UV boundary conditions for
the vector and pseudoscalar modes.  
We specify to the particular case of $\Delta = 2$; other cases requiring a case-by-case analysis.
We make use of the UV expansions for the fluctuations, as reported 
 in Appendix~\ref{sec:IRUVexpansions}, which we replace into the boundary conditions given in Eqs.~(\ref{eq:vectorBCs}--\ref{eq:p4BCs}), and obtain relations between the leading and subleading coefficients
  appearing  in the general solution of the bulk second-order linearised equations.
This results in the following relations:\footnote{The substitution $q^2 \rightarrow e^{2\chi_U - 2 A_U} q^2$ reinstates the dependence on $A_U$ and $\chi_U$ in these expressions. In order for this to be the case, one needs to also have properly reinstated $A_U$ and $\chi_U$ in Eqs.~(\ref{eq:K5KXscalings}--\ref{eq:D2s}).}
\beqs
	0 &=& \frac{q^2}{\bar \varepsilon^2}  v^{\bar{\mathcal A}}_0 - 3 \mathfrak v^{\bar{\mathcal A}}_3 \,, \\
	0 &=& \cos(v) \mathfrak v^{\hat{\mathcal A}}_0 - \sin(v) \mathfrak v^{\tilde{\mathcal A}}_0 \,, \\
	0 &=& -3 \cos(v) \mathfrak v^{\tilde{\mathcal A}}_3 + \sin(v) \left( \frac{g^2}{4} k_{\mathcal X} \phi_J^2 \mathfrak v^{\hat{\mathcal A}}_0 - 3 \mathfrak v^{\hat{\mathcal A}}_3 \right) + \frac{q^2}{\bar \varepsilon^2} \left( \cos(v) \mathfrak v^{\tilde{\mathcal A}}_0 + \sin(v) \mathfrak v^{\hat{\mathcal A}}_0 \right) \,, \\
	0 &=& \mathfrak v^4_0 \,, \\
	0 &=& \mathfrak p^{\hat{\mathcal A}}_0 - k_{\mathcal X} \mathfrak p^{\hat{\mathcal A}}_1 \,, \\
	0 &=& \mathfrak p^4_0 - \left( k_{\mathcal X} + \frac{m_4^2}{q^2 \phi_J^2} \right)  \mathfrak p^4_1 \,.
\eeqs

The body of our numerical study, exemplified by the plots in Sect.~\ref{Sec:spectrum}, consists of an extensive study of the parameter space of the model, in which we studied in detail the dependence of the whole spectrum of fluctuations on the remaining free parameters of the model, which we summarise as follows.
\begin{itemize}
\item The background functions are determined by the parameters $\Delta$ (associated with the dimension of the dual operator inducing $SO(5)$ breaking) and $\phi_I$ (the parameter in the IR expansion that controls  the profile for $\phi$, including how much it departs from $0$). We restrict the former to lie in the range $\frac{3}{2}\leq \Delta  \leq \frac{7}{2}$, and the latter to choices that lie
along the stable 
branch of background solutions, as identified in Ref.~\cite{Elander:2022ebt}, by requiring that   $\phi_I \leq \phi_I(c)$, with $\phi_I(c)$ the critical value at which a first-order phase transition is taking place.

\item The $SO(4)$ gauge coupling 
in the dual field theory interpretation is approximately given by $g_4 \equiv \bar \varepsilon g$, where
 $g$ is the bulk coupling. We require the renormalisation constant $\bar \varepsilon$  to
be small enough to apply perturbation theory.

\item The vacuum misalignment angle, $v$, and the parameter, $m_4^2$, that encodes the explicit breaking of $SO(5)$ after the appropriate limits have been taken, are both dialled to produce an appreciable separation of mass scales between parametrically light states and towers of heavy resonances.

\item The constant $k_{\mathcal X}$ deserves further discussion, and we devote the short subsection~\ref{Sec:comments} to it.

\item In presenting numerical results, we specify the values of $\rho_1-\rho_o$ and $\rho_2-\rho_o$ used in generating them. We verified that the results do not depend on these choices, within the set numerical accuracy of our study.

\end{itemize}

We highlight again that we adopt boundary conditions for the gauge fields along the unbroken $SO(3)$ that contain additional terms, in respect to earlier literature, and allow to remove the cutoff dependence. The boundary conditions for other modes are more complicated, though, as a result of the symmetry-breaking effects and Higgs phenomena. One of the important differences between this study and the literature is indeed the implementation of the gauging and symmetry-breaking related phenomena directly in the gravity calculation.

 We conclude with two clarifications.
First of all, the bulk equations and boundary conditions we discussed here are used to identify the eigenstates with $M^2= - q^2\neq 0$.
For the massless modes, we already have the spectrum: the multiplicity is determined by the symmetries of the system.
For example, the unbroken $SO(3)$ leads to three massless vector states.
Secondly, the gauging process we outlined, in the presence of a non-trivial bulk profile for $\phi$ in a $D$-dimensional gravity
background that is asymptotically AdS,
hence reproducing the lines of thought of Sect.~\ref{Sec:SO(5)oSO(4)}, relies on the value of $\Delta$ lying in the range
 $\frac{D-3}{2} \leq \Delta \leq \frac{D+1}{2}$. 
 In this range of $\Delta$, the scalar field in the gravity description admits 
more than one interpretation in terms of operators in the dual field theory \cite{Klebanov:1999tb}. Indeed, in the numerical examples we will display,  we choose $\Delta=2$, with $D=6$.
We elaborate further on this point in the next, short subsection.


\subsection{More on the gauging of $SO(4)$ and the role of $k_{\mathcal X}$}
\label{Sec:comments}

 We find it useful to pause and digress, in order  to clarify  the role of the parameter $k_{\mathcal X}$.
In the gravity theory,  the action is  $SO(5)$ gauge-invariant, but the background solutions have  $\mathcal X_\alpha\neq 0$, breaking (spontaneously) the $SO(5)$ symmetry to $SO(4)$. 
The standard field-theory  interpretation invokes an admixture of explicit and spontaneous breaking 
of a global $SO(5)$ symmetry, encoded in the two non-vanishing parameters, $\phi_J$ and $\phi_V$, appearing in the 
UV expansion of the background solutions in Eqs.~(\ref{eq:backgroundUVexp1})--(\ref{eq:backgroundUVexp3}). 
The (asymptotic) boundary values of the bulk fields become (up to normalisations) the sources for composite operators on the field theory side, appearing as a  deformation of the form
\beq \label{eq:operator}
	\int \dd^5x \, \mathcal X^{(0)}_\alpha \mathcal O^\alpha \,, \qquad  
	{\rm where} \qquad \mathcal X^{(0)}_\alpha \equiv \lim_{z \rightarrow 0} \left( z^{-\Delta_J} \mathcal X_\alpha \right) \,,
\eeq
 where $\mathcal O^\alpha$ is the (composite) operator dual to $\mathcal X_\alpha$.

The gauged  $SO(4)$ subgroup of $SO(5)$ is not the same $SO(4)$ preserved by the aforementioned deformation.
But explicitly broken symmetries cannot be gauged. This obstruction requires turning the boundary value, $\mathcal X^{(0)}_\alpha$, into a  dynamical field, to restore the full $SO(5)$ invariance of the field theory. One can then take the appropriate limits to turn the field is a non-dynamical spurion. But, to do so,  the field/spurion (before freezing) must  admit a (unitary) field theory description, which restricts its
scaling dimension to lie above the unitarity bound. This is possible only if 
\beqs
\label{Eq:window}
\frac{D-3}{2} \leq \Delta \leq \frac{D+1}{2}\,,
\eeqs
with $D = 6$ in the present case.

The boundary-localised potentials of the sigma-model fields, $\lambda_i$, include mass terms, and we take the limit of infinite boundary masses, $m_{\mathcal X,i}^2 \rightarrow \infty$, when setting up the boundary conditions for the corresponding fluctuations, to
freeze the modulus of $\mathcal X_\alpha$ at the boundaries.
This is analogous to the limit $\lambda_5 \rightarrow +\infty$ for the spurion $P_5$,
but also related to the limit $\lambda_{\Sigma}\rightarrow +\infty$ discussed in Sect.~\ref{Sec:SO(5)oSO(4)}.
Similarly, the parameter $k_{\mathcal X}$ is analogous to $k_5$. In Ref.~\cite{Elander:2023aow}, the boundary conditions were chosen such that the limit $k_{\mathcal X} \rightarrow +\infty$ was implemented, leading the four extra massless degrees of freedom (the PNGBs associated with the spontaneous breaking of $SO(5)$ to $SO(4)$ due to the VEV of $\mathcal X^{(0)}_\alpha$) to decouple.  Here, instead, we keep $k_{\mathcal X}$ as a (finite) free parameter.  Three of 
the massless components of the scalar are eaten by the Higgs mechanism, and become the longitudinal components for the massive gauge fields along the $SO(4)/SO(3)$  
coset. The fourth component (morally corresponding to the Higgs boson in a CHM implementation of these ideas) acquires a mass, due to the explicit breaking encoded in $m_4^2\neq 0$.

The essential elements of novelty discussed in this subsection can be summarised as follows. As a result of the fact that  we made $\mathcal X^{(0)}_\alpha$ dynamical,
the standard quantisation interpretation dictates that the breaking of SO(5) associated with the bulk scalar field is, in field theory terms, spontaneous, and not explicit---unlike in our previous 
work~\cite{Elander:2022ebt,Elander:2023aow}.
Hence, we can gauge a (spontaneously) broken subgroup.
Nevertheless, we find that this is admissible only in the window of $\Delta$ in which both standard and alternative quantisations are allowed, as this is a necessary requirement to allow $\mathcal X^{(0)}_\alpha$ to be dynamical.

\section{Spectrum}
\label{Sec:spectrum}

\begin{figure}[th]
\begin{center}
\includegraphics[width=16cm]{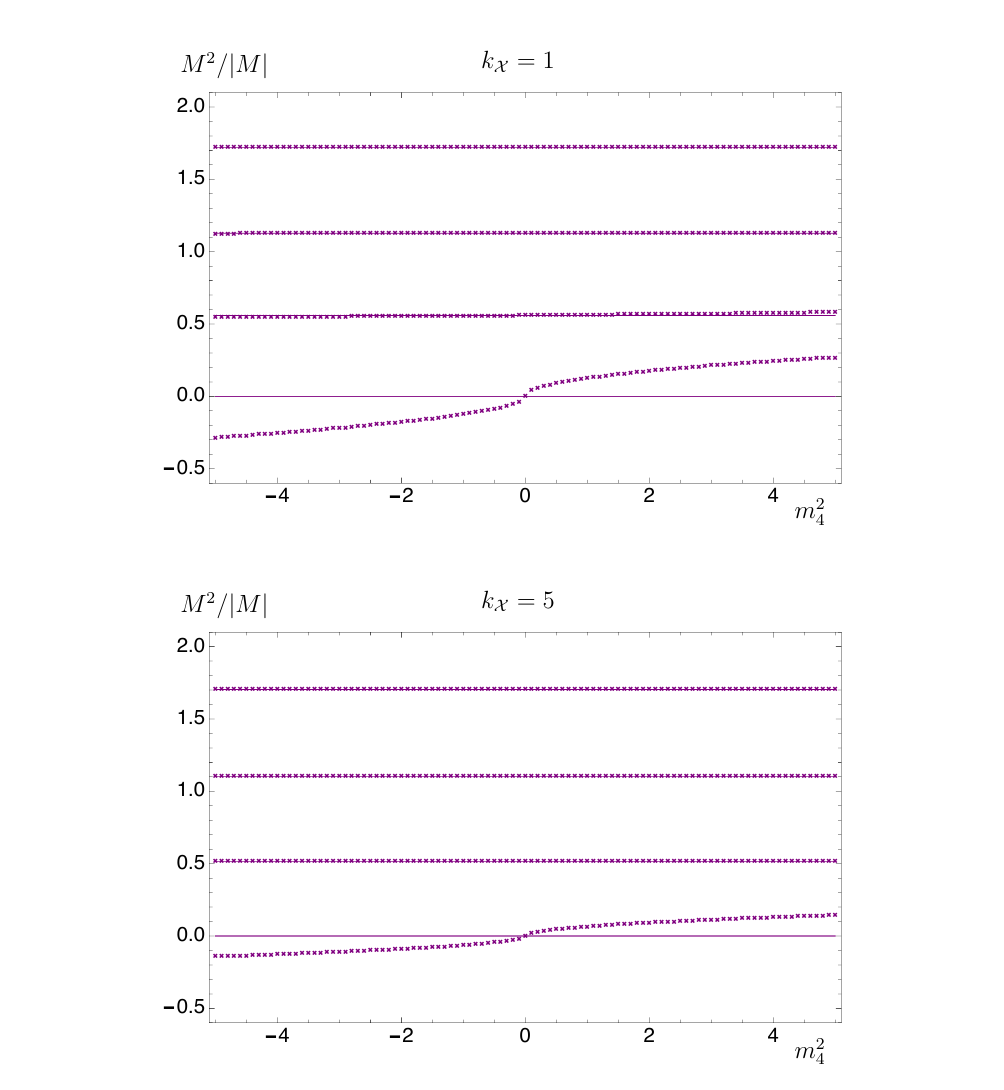}
\caption{Mass spectrum, $\frac{M^2}{|M|}$, of pseudoscalar fluctuations, $\mathfrak p^{\hat {\mathcal A}}$ (lines) and $\mathfrak p^4$ (crosses), as a function of the parameter $m_4^2$, for two representative choices, $k_{\mathcal X}=1$ (top panel) and $k_{\mathcal X} =5$ (bottom panel). The spectrum shown is that of the theory prior to gauging the $SO(4)$ subgroup, hence the $\mathfrak p^{\hat {\mathcal A}}$ sector contains massless modes, higgsed away in the gauged case. All plots have $\Delta = 2$, $g = 5$, $\phi_I = \phi_I(c) \approx 0.3882$, and the values of the IR and UV cutoffs are, respectively,  $\rho_1 - \rho_o = 10^{-9}$ and  $\rho_2 - \rho_o = 5$.}
\label{Fig:Spectrum1}
\end{center}
\end{figure}

\begin{figure}[th]
\begin{center}
\includegraphics[width=16cm]{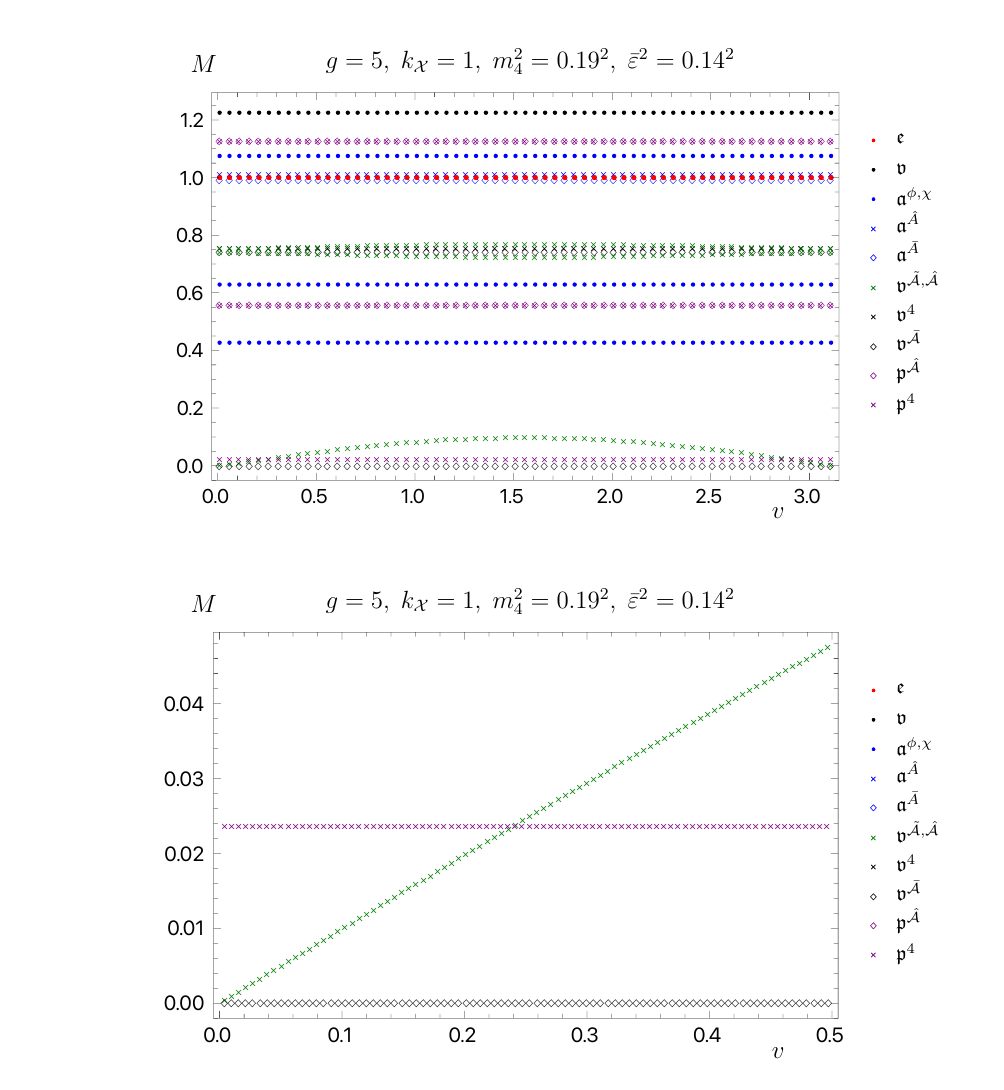}
\caption{Mass spectrum, $M$, of fluctuations, including pseudoscalar, $\mathfrak{p}^4$ (purple crosses), $\mathfrak{p}^{\hat {\mathcal A}}$ (purple diamonds), vectors $\mathfrak{v}^{\bar {\mathcal A}}$ (black diamonds), $\mathfrak{v}^{4}$ (black crosses),  $\mathfrak{v}^{\hat {\mathcal A}}$ and $\mathfrak{v}^{\tilde {\mathcal A}}$ (green crosses), scalars $\mathfrak{a}^{\bar A}$ (blue diamond), and  $\mathfrak{a}^{\hat A}$ (blue crosses),  active scalars, $\mathfrak{a}^{\phi},\mathfrak{a}^{\chi}$ (blue dots), graviphoton $\mathfrak{v}$ (black dots), and spin-2 tensors, $\mathfrak{e}$ (red dots), as a function of the misalignement angle, $v$, for fixed values of $k_{\mathcal X}, m_4^2$, and $\bar \varepsilon$. The bottom panel is a detail of the top one. All plots have $\Delta = 2$, $g = 5$, and $\phi_I = \phi_I(c) \approx 0.3882$, and the values of the IR and UV cutoffs are, respectively, given by $\rho_1 - \rho_o = 10^{-9}$ and  $\rho_2 - \rho_o = 5$.}
\label{Fig:Spectrum2a}
\end{center}

\end{figure}
\begin{figure}[th]
\begin{center}
\includegraphics[width=16cm]{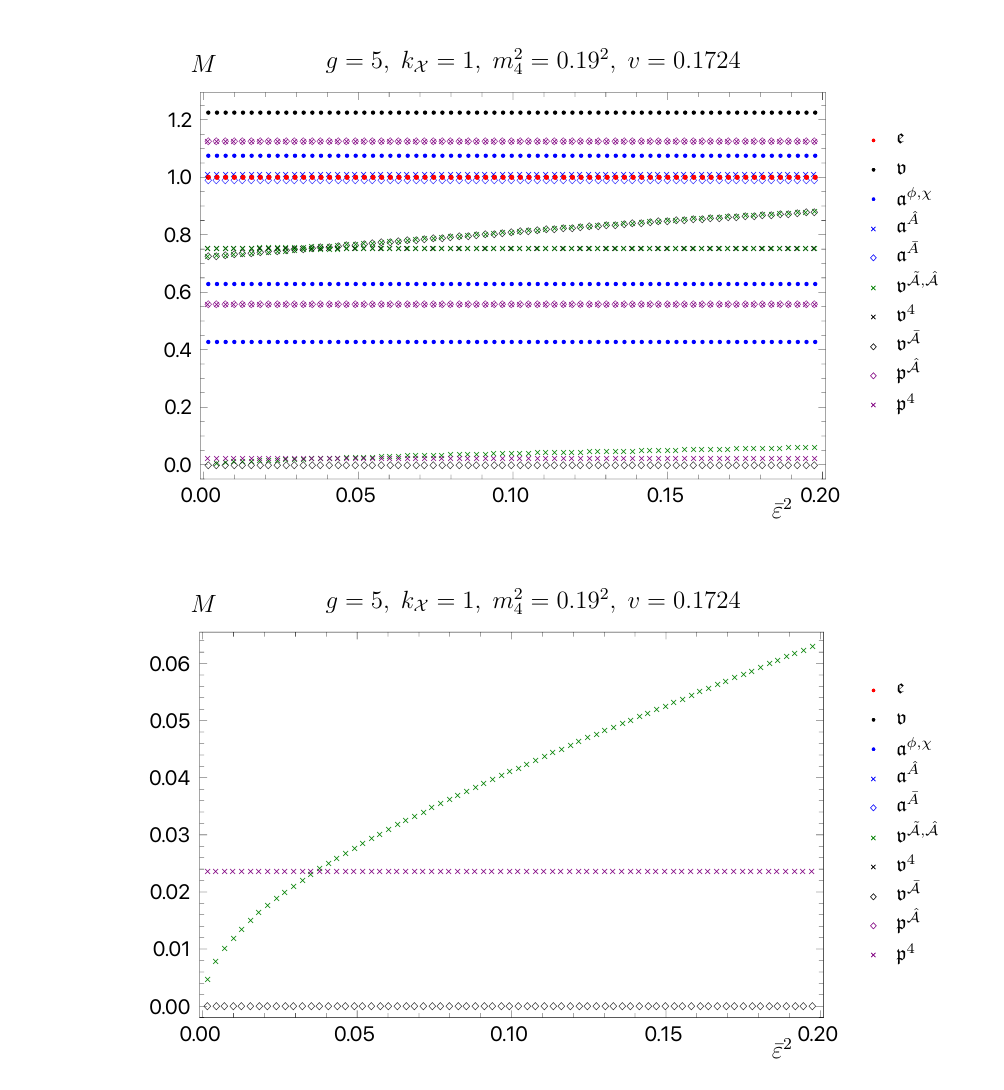}
\caption{Mass spectrum, $M$, of fluctuations, including pseudoscalar, $\mathfrak{p}^4$ (purple crosses), $\mathfrak{p}^{\hat {\mathcal A}}$ (purple diamonds), vectors $\mathfrak{v}^{\bar {\mathcal A}}$ (black diamonds), $\mathfrak{v}^{4}$ (black crosses),  $\mathfrak{v}^{\hat {\mathcal A}}$ and $\mathfrak{v}^{\tilde {\mathcal A}}$ (green crosses), scalars $\mathfrak{a}^{\bar A}$ (blue diamond), and  $\mathfrak{a}^{\hat A}$ (blue crosses),  active scalars, $\mathfrak{a}^{\phi},\mathfrak{a}^{\chi}$ (blue dots), graviphoton $\mathfrak{v}$ (black dots), and spin-2 tensors, $\mathfrak{e}$ (red dots),
as a function of the parameter $\bar \varepsilon^2$, for fixed values of $k_{\mathcal X}, m_4^2$, and $v$. The bottom panel is a detail of the top one. All plots have $\Delta = 2$, $g = 5$, $\phi_I = \phi_I(c) \approx 0.3882$, and the values of the IR and UV cutoffs are, respectively, given by $\rho_1 - \rho_o = 10^{-9}$ and  $\rho_2 - \rho_o = 5$.}
\label{Fig:Spectrum2b}
\end{center}
\end{figure}

\begin{figure}[th]
\begin{center}
\includegraphics[width=17.5cm]{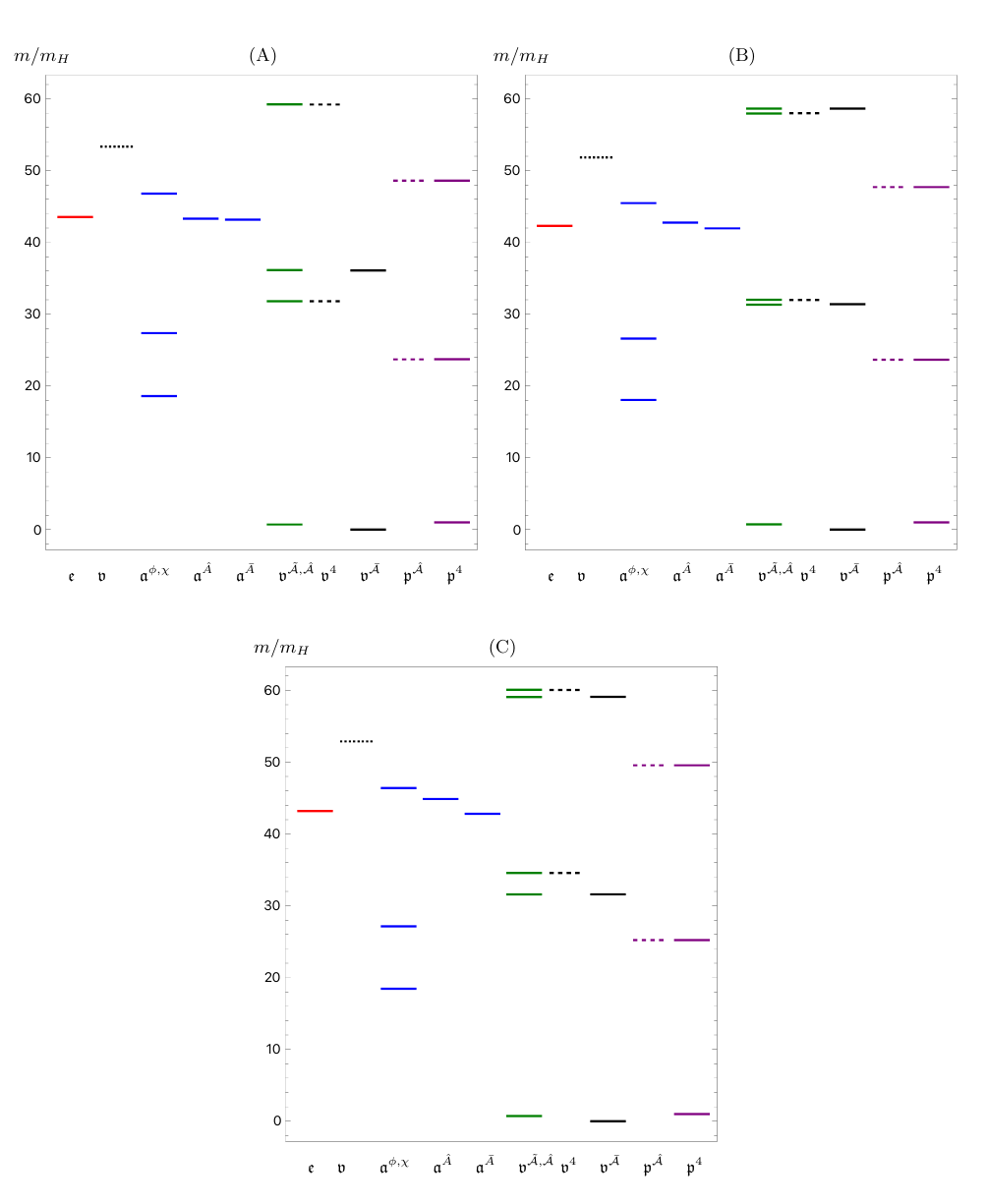}
\caption{Mass spectrum, $M$, of fluctuations, normalized to the mass, $m_H$,  of the lightest $\mathfrak{p}^4$ state, including  pseudoscalars, $\mathfrak{p}^{\hat {\mathcal A}}$ (purple), $\mathfrak{p}^4$ (purple dashed), vectors, $\mathfrak{v}^{\hat {\mathcal A}}$ and $\mathfrak{v}^{\tilde {\mathcal A}}$ (green), $\mathfrak{v}^{\bar {\mathcal A}}$ (black), $\mathfrak{v}^{4}$ (black dahsed), $v$ (black dotted), scalars, $\mathfrak{a}^{\phi}, \mathfrak{a}^{\chi}, \mathfrak{a}^{\bar A}, \mathfrak{a}^{\hat A}$ (blue) and spin-2 tensors, $\mathfrak{e}$ (red), for three values of the coupling $g$: (A) $g=2, \;v=0.15, \;g_4=0.7, \;m_4^2=0.185^2$, (B) $g=5,\; v=0.1724, \;g_4=0.7,\;m_4^2=0.19^2$,  (C)  $g=8,\; v=0.17, \;g_4=0.7,\; m_4^2=0.185^2$.  All plots have $\Delta = 2$, $\phi_I = \phi_I(c) \approx 0.3882$, and $k_{\mathcal X}=1$. The IR and UV cutoffs are given by $\rho_1 - \rho_o = 10^{-9}$ and  $\rho_2 - \rho_o = 5$, respectively.}
\label{Fig:Spectrum3}
\end{center}
\end{figure}

In this section, we present examples of the numerical results we obtained for the complete mass spectrum of fluctuations, and its dependence on the model parameters. For concreteness, we set $\Delta=2$, $\phi_I = \phi_I(c) \approx 0.3882$, 
$\rho_2-\rho_o=5$, and $\rho_1-\rho_o=10^{-9}$, throughout the section.
Figures~\ref{Fig:Spectrum1}--\ref{Fig:Spectrum3} illustrate the spectrum dependence on the residual freedom encoded 
 in the choices of $\bar{\varepsilon}$, $g$, $v$, $m_4^2$, and $k_{\mathcal X}$.  All spectra, except for Fig.~\ref{Fig:Spectrum3}, are normalised so that the lightest of the $SO(3)$-singlet spin-2 fluctuations, $\mathfrak{e}_{\mu\nu}$, has unit mass.

Figure~\ref{Fig:Spectrum1} displays the dependence of the mass spectrum of the pseudoscalar fluctuations,
$\mathfrak p^{\hat {\mathcal A}}$ and $\mathfrak p^4$, as a function of the parameter $m_4^2$, for 
two representative choices of $k_{\mathcal X}$. The spectra in these two sectors do not depend on $\bar{\varepsilon}$ or $v$. Several interesting features are worth commenting about.
We notice that the spectrum of $\mathfrak p^{\hat {\mathcal A}}$ contains three exactly massless states, that disappear
when the $SO(4)$ subgroup is gauged, and we work in unitary gauge, as they provide the longitudinal components for three of the vector bosons.
A tachyonic mode appears in $\mathfrak p^4$ if one chooses $m_4^2<0$, which is hence forbidden.
Positive, but small values of $m_4^2$, close to zero, lead to a small mass for the physical state corresponding to the PNGB responsible for $SO(4)$ spontaneous symmetry breaking.
(If this were a composite Higgs model, such a state would be identified with the Higgs boson.)

Figures~\ref{Fig:Spectrum2a} and~\ref{Fig:Spectrum2b} show the complete spectra, for representative choices of $g$, $k_{\mathcal X}$, $\Delta$, $\phi_I$, and $m_4^2$, as a function of $v$ and $\bar\varepsilon^2$, respectively.
Superficially, the spectra appear to be quite complicated, due to the large number of states, 
and details depend on the choices of parameters.
Yet, the figures display a few important general features.
First, only a small number of states are light: the massless vectors corresponding to zero modes in the unbroken, gauged $SO(3)$ sector, the lightest of the $\mathfrak{p}^4$ pseudoscalars, and the lightest combination of
vectors in the $SO(4)/SO(3)$ coset. All other states have masses that are of the order of the typical scale in the theory,
that we identify with the mass of the lightest spin-2 state.
A small hierarchy emerges between these two groups of states.
Furthermore, as expected on general grounds, the mass of the lightest vector states grows when either of $v$ or $\bar \varepsilon^2$ is small and growing, and vanishes when either of these two parameters vanishes,
 which are the natural expectations for vector bosons associated with the Higgs mechanism
 for spontaneous breaking of a weakly coupled gauge theory.

In Figure~\ref{Fig:Spectrum3}, we show three numerical examples of the spectrum, aimed at illustrating how a semi-realistic implementation of this model as a CHM would look like. 
We set aside the differences with the standard model, purely for illustration purposes, and interpret the lightest state in the $\mathfrak{p}^4$ fluctuations as the Higgs boson. For the sole purposes of this exercise, we call its mass $m_H$, and measure the other masses in units of $m_H$. We fix $\Delta = 2$ and $\phi_I = \phi_I(c) \approx 0.3882$, impose the requirement that $g_4=\bar{\varepsilon} g=0.7$---of the order of the $SU(2)_L$ coupling in the standard model---and adjust the other parameters so that  the ratio of mass between the lightest fluctuation in the spin-1 ($v^{\hat {\mathcal A}}$ and $v^{\tilde {\mathcal A}}$) sector, and the spin-0
($\mathfrak{p}^4$) sector, be approximately given by $M_Z/m_H\simeq 0.73$---the ratio between the experimental values of the  mass of the  $Z$ and Higgs bosons.

This model is not realistic, and this final exercise should be taken with a grain of salt.
Yet, it meets its purpose, and demonstrates that it is possible, within this model, to produce a small hierarchy between the Higgs and $Z$ mass on one side, and, on the other side,  the towers of new bound states predicted by the theory. It is also worth noting that the next-to-lightest states appear to be the spin-0 states, with spin-1 states and spin-2 states significantly heavier.

\section{Outlook}
\label{Sec:outlook}

In summary, we demonstrated how to build a bottom-up, holographic model that, at low energies, can be interpreted in dual field theory terms as a sigma-model with $SO(5)$ global symmetry broken to $SO(4)$. An $SO(4)$ subgroup is gauged. The presence of explicit $SO(5)$ symmetry-breaking interactions leads to vacuum misalignment and to the spontaneous breaking of the gauged $SO(4)$ to its $SO(3)$ gauge group.
Much of this paper is devoted to the non-trivial development of the formalism, showing that symmetry breaking can be consistently triggered by a bulk scalar field in the gravity theory, in which the $SO(5)$ is gauged. It is worth noticing that this can be done without violating the gauge principle, despite the presence of explicit symmetry breaking in the dual field theory. But this can be done consistently only if the bulk field triggering symmetry breaking corresponds in the dual field theory to an operator with scaling dimension restricted to the range $\frac{D-3}{2}\leq \Delta \leq \frac{D+1}{2}$, for reasons described in the body of the paper.

A significant distinctive feature of this proposal is that the gravity background is completely smooth, in a way that mimics confinement in the dual field theory and leads to the introduction of a mass gap.
 Although more sophisticated holographic descriptions of confinement may require 
further extensions, and although we did not yet implement  a realistic realisation of the electroweak model,
 the study of the spectrum we performed and reported here indicates that the  phenomenology is quite simple, as expected in CHMs based on the $SO(5)/SO(4)$ coset.  All the new particles are parametrically heavy in respect to the bosons that play the role of the $Z$, $W$, and Higgs boson. Because a (custodial)  $SO(4)$ symmetry is built into the model,  one expects  a realistic realisation of a CHM based on this model to have escaped indirect detection although a detailed calculation of all precision on electroweak parameters is needed.

To build a fully realistic model, that might be detectable in direct collider searches, requires additional non-trivial steps. First, the model has a gauged $SO(4)$ symmetry, while  the SM gauge symmetry is $SU(2)_L\times U(1)_Y$, and, furthermore, 
 the quantum assignments of the standard-model fermions require identifying an additional $U(1)$ global symmetry, related to baryon and lepton number, so that hypercharge assignments  are realistic.
We leave this task for future work.

Second, this theory does not contain fermions. We anticipated in the body of the paper that 
we could proceed in two ways towards their introduction:
  either by assuming that all fermions are  localised on the UV boundary,
or that there are additional bulk fermions, transforming in the spinorial representation, $4$, of $SO(5)$.
These ingredients would then determine the mechanism for mass generation for the SM fermions, and in turn the contribution of the fermions to the effective potential triggering spontaneous symmetry breaking of the gauge symmetry via vacuum misalignment. Also this task is left for future investigations.

Finally, the techniques illustrated in this paper can be applied to a large class of holographic models in which bulk scalar fields implement symmetry breaking. In particular, there are clear similarities between the gravity set up we discussed here and the one in Ref.~\cite{Elander:2021kxk}, which is based on maximal supergravity in $D=7$ dimensions, and has bulk $SO(5)$ gauge symmetry. As discussed in the Introduction, it is
 still an open challenge to identify
 a UV-complete CHM model based on the $SO(5)/SO(4)$ coset, in which the gravity theory is
 embedded into a known supergravity.  The results of this paper provide a major step in this direction. 

\begin{acknowledgments}

 We would like to thank Carlos Hoyos, Ronnie Rodgers, and Javier Subils, for useful comments on an earlier version of 
the manuscript.

The work of AF has been supported by the STFC Consolidated Grant No. ST/V507143/1
and by the
EPSRC Standard Research Studentship (DTP)  EP/T517987/1.

The work of  MP has been supported in part by the STFC 
Consolidated Grants No. ST/T000813/1 and ST/X000648/1.
  MP received funding from
the European Research Council (ERC) under the European
Union’s Horizon 2020 research and innovation program
under Grant Agreement No.~813942.

\vspace{1.0cm}

{\bf Open Access Statement ---} For the purpose of open access, the authors have applied a Creative Commons 
Attribution (CC BY) licence  to any Author Accepted Manuscript version arising.

\vspace{1.0cm}

{\bf Research Data Access Statement ---} The data generated for this manuscript can be downloaded from Ref.~\cite{EFP}.

\end{acknowledgments}

\appendix
\allowdisplaybreaks

\section{Basis of $SO(5)$ generators}
\label{Sec:so5}

We present here an example of a basis of $SO(5)$ generators,
used in the body of the paper, taken from Ref.~\cite{Elander:2023aow}.
\beqs
t^1 &=& \frac{i}{2} \left(
\begin{array}{ccccc}
0 & 0 & 0 & 0 &-1 \\
0 & 0 & 0 & 0 & 0 \\
0 & 0 & 0 & 0 & 0 \\
0 & 0 & 0 & 0 & 0 \\
1 & 0 & 0 & 0 & 0
\end{array}\right) \,, \quad
t^2 = \frac{i}{2} \left(\begin{array}{ccccc}
0 & 0 & 0 & 0 & 0 \\
0 & 0 & 0 & 0 & -1 \\
0 & 0 & 0 & 0 & 0 \\
0 & 0 & 0 & 0 & 0 \\
0 & 1 & 0 & 0 & 0
\end{array}\right) \,, \quad
t^3 = \frac{i}{2} \left(\begin{array}{ccccc}
0 & 0 & 0 & 0 & 0 \\
0 & 0 & 0 & 0 & 0 \\
0 & 0 & 0 & 0 & -1 \\
0 & 0 & 0 & 0 & 0 \\
0 & 0 & 1 & 0 & 0
\end{array}\right) \,, \quad
t^4 = \frac{i}{2} \left(\begin{array}{ccccc}
0 & 0 & 0 & 0 & 0 \\
0 & 0 & 0 & 0 & 0 \\
0 & 0 & 0 & 0 & 0 \\
0 & 0 & 0 & 0 & -1 \\
0 & 0 & 0 & 1 & 0
\end{array}\right) \,, \nonumber \\
t^5 &=& \frac{i}{2} \left(\begin{array}{ccccc}
0 & 0 & 0 & -1 & 0 \\
0 & 0 & 0 & 0 & 0 \\
0 & 0 & 0 & 0 & 0 \\
1 & 0 & 0 & 0 & 0 \\
0 & 0 & 0 & 0 & 0
\end{array}\right) \,, \quad
t^6 = \frac{i}{2} \left(\begin{array}{ccccc}
0 & 0 & 0 & 0 & 0 \\
0 & 0 & 0 & -1 & 0 \\
0 & 0 & 0 & 0 & 0 \\
0 & 1 & 0 & 0 & 0 \\
0 & 0 & 0 & 0 & 0
\end{array}\right) \,, \quad
t^7 = \frac{i}{2} \left(\begin{array}{ccccc}
0 & 0 & 0 & 0 & 0 \\
0 & 0 & 0 & 0 & 0 \\
0 & 0 & 0 & -1 & 0 \\
0 & 0 & 1 & 0 & 0 \\
0 & 0 & 0 & 0 & 0
\end{array}\right) \,, \nonumber \\
t^8 &=& \frac{i}{2} \left(\begin{array}{ccccc}
0 & 0 & -1 & 0 & 0 \\
0 & 0 & 0 & 0 & 0 \\
1 & 0 & 0 & 0 & 0 \\
0 & 0 & 0 & 0 & 0 \\
0 & 0 & 0 & 0 & 0
\end{array}\right) \,, \quad
t^9 = \frac{i}{2} \left(\begin{array}{ccccc}
0 & 0 & 0 & 0 & 0 \\
0 & 0 & -1 & 0 & 0 \\
0 & 1 & 0 & 0 & 0 \\
0 & 0 & 0 & 0 & 0 \\
0 & 0 & 0 & 0 & 0
\end{array}\right) \,, \quad
t^{10} = \frac{i}{2} \left(\begin{array}{ccccc}
0 & -1 & 0 & 0 & 0 \\
1 & 0 & 0 & 0 & 0 \\
0 & 0 & 0 & 0 & 0 \\
0 & 0 & 0 & 0 & 0 \\
0 & 0 & 0 & 0 & 0
\end{array}\right) \,, \quad
\eeqs

We find it convenient to present also a basis of $SU(4)$, adapted from Ref.~\cite{Bennett:2017kga},
written in terms of $4\times 4$ Hermitian matrices.
The adjoint representation
of $SU(4)$  decomposes as $15=5 \oplus 10$ in $Sp(4)\sim SO(5)$, both being used in the body of the paper.
We order the basis so that
$\Gamma^A$, for $A=1,\,\cdots,\,5$, spans the coset $SU(4)/Sp(4)$, and hence can be used to describe the $5$ of $Sp(4)$, while the generators of $Sp(4)$ are denoted $T^A$, with $A=1,\,\cdots,\,10$.
We conventionally normalise  $\Tr \, (T^AT^B)=\frac{1}{4}\delta^{AB}=\Tr \, (\Gamma^A\Gamma^B)$, for all the $SU(4)$ matrices.
The $\Gamma^A$ matrices are
\beqs
\nonumber
\Gamma^1 &=&\frac{1}{4}
\left(
\begin{array}{cccc}
 0 & 1 & 0 & 0 \cr 1 & 0 & 0 & 0 \cr 0 & 0 & 0 & 1 \cr 0 & 0 & 1 &  0
 \end{array}\right)\,,\qquad
 \Gamma^2\,=\,\frac{1}{4}\left(
\begin{array}{cccc}
  0 & -i & 0 & 0 \cr i & 0 &  0 & 0 \cr  0 &  0 &  0 & i \cr  0 &  0 & -i & 0
  \end{array}\right)\,,\qquad
\Gamma^3\,=\,\frac{1}{4} \left(
\begin{array}{cccc}
  0 &  0 &  0 & -i \cr  0 &  0 & i&0 \cr  0 & -i& 0 & 0 \cr i&0 &  0 & 0
  \end{array}\right)\,,\\
\Gamma^4&=&\frac{1}{4} \left(
\begin{array}{cccc}
  0 &  0 &  0 & 1 \cr  0 &  0 & -1&0 \cr  0 & -1& 0 & 0 \cr 1& 0 &  0 &  0
  \end{array}\right)\,,\qquad
\Gamma^5\,=\,\frac{1}{4} \left(
\begin{array}{cccc}
1& 0 &  0 & 0 \cr  0 & -1& 0 & 0 \cr  0 &  0 & 1&0 \cr  0 &  0 &  0 & -1
  \end{array}\right)\,.
  \eeqs
  
  The generators of $Sp(4)$, the $10$ of $Sp(4)\sim SO(5)$ are:
    \beqs
    \nonumber
T^1 &=&
\frac{1}{4} \left(
\begin{array}{cccc}
  0 & i& 0 & 0 \cr -i & 0 &  0 & 0 \cr  0 &  0 &  0 & i \cr  0 &  0 & -i& 0
  \end{array}\right)\,,\qquad
  T^{2}\,=\,\frac{1}{4} \left(
\begin{array}{cccc}
  0 & 1& 0 & 0 \cr 1& 0 &  0 & 0 \cr  0 &  0 &  0 & -1 \cr  0 &  0 & -1& 0
  \end{array}\right)\,,\qquad
  T^{3}\,=\,\frac{1}{4} \left(
\begin{array}{cccc}
  0 &  0 &  0 & 1 \cr  0 &  0 & 1&0 \cr  0 & 1& 0 & 0 \cr 1& 0 &  0 &  0
  \end{array}\right)\,,\\
    \nonumber
T^4&=&\frac{1}{4} \left(
\begin{array}{cccc}
  0 &  0 &  0 &i \cr  0 &  0 & i&0 \cr  0 & -i& 0 & 0 \cr -i& 0 &  0 &  0
  \end{array}\right)\,,\qquad
  T^5\,=\,\frac{1}{4} \left(
\begin{array}{cccc}
  0 &  0 & i&0 \cr  0 &  0 &  0 & -i \cr -i& 0 &  0 & 0 \cr  0 & i& 0 &  0
  \end{array}\right)\,,\qquad
  T^{6}\,=\,\frac{1}{4} \left(
\begin{array}{cccc}
 0 &  0 & 1 &0 \cr  
 0 &  0 &  0 &  1 \cr
1& 0 &  0 & 0 \cr
    0 &  1 &  0 &  0
  \end{array}\right)\,,\\
  \nonumber
  T^{7}&=&\frac{1}{4} \left(
\begin{array}{cccc}
 -1& 0 &  0 & 0 \cr  0 & -1& 0 & 0 \cr  0 &  0 & 1&0 \cr  0 &  0 &  0 & 1
  \end{array}\right)\,,\qquad
   T^{8}\,=\,\frac{1}{4} \left(
\begin{array}{cccc}
 0 &  0 & 1 &0 \cr  
 0 &  0 &  0 &  -1 \cr
1& 0 &  0 & 0 \cr
    0 &  -1 &  0 &  0
  \end{array}\right)\,\qquad
     T^{9}\,=\,\frac{1}{4} \left(
\begin{array}{cccc}
 0 &  0 & -i &0 \cr  
 0 &  0 &  0 &  -i \cr
i& 0 &  0 & 0 \cr
    0 &  i &  0 &  0
  \end{array}\right)\,\\
  T^{10}&=&\frac{1}{4} \left(
\begin{array}{cccc}
 1& 0 &  0 & 0 \cr  0 & -1& 0 & 0 \cr  0 &  0 & -1& 0 \cr  0 &  0 &  0 & 1
  \end{array}\right)\,.
\eeqs
They can  be written as commutators of two $\Gamma^A$ matrices. For example, $T^1=-2 i \left[\Gamma^1,\,\Gamma^5\right]$. Similar relations hold for the other generators.

\section{Five-dimensional formalism}
\label{Sec:formalism}

In this Appendix, we report  technical details on the treatment of gauge-invariant fluctuations.
Most of this material is borrowed from the literature, but we find it useful to summarise it here,
 to make the exposition self-contained, and the notation coherent and self-consistent.

\subsection{Scalars coupled to gravity}
\label{Sec:scalars}

We report here the salient features of the  
gauge-invariant formalism developed in 
Refs.~\cite{Bianchi:2003ug,Berg:2005pd,Berg:2006xy,Elander:2009bm,Elander:2010wd,Elander:2014ola,Elander:2017cle,Elander:2017hyr}.
Borrowing from Refs.~\cite{Berg:2005pd,Elander:2010wd}, consider $n$ 
real scalars, $ \Phi^a$, with $a=1\,,\,\cdots\,,\,n$;  the action, ${\cal S}_D$, is written in general as follows:
\beqs
{\cal S}_D &=&\int\dd^Dx\sqrt{-g}\left[\frac{{ R} }{4}
-\frac{1}{2} G_{ab}g^{MN}\partial_M  \Phi^a\partial_N  \Phi^b - {\mathcal V}( \Phi^a)\right]\,.
\label{Eq:ActionD}
\eeqs
 (In this paper, the relevant scalars are denoted as $\{\chi,\phi,{\cal B}_6^{\tilde{\cal A}}, 
 {\cal A}_6^{\bar{\cal A}},
{\cal B}_6^{\hat{\cal A}}, {\cal A}_{6}^{4}\}$, hence $n=12$, and the dimensionality of the system is $D=5$.)
The backgrounds are described by the ansatz 
\beqs
\dd s^2_D&=&\dd r^2 +e^{2{ A}(r)}\, \eta_{\mu\nu}\dd x^{\mu}\dd x^{\nu}\,,\\
 \Phi^a&=& \Phi^a(r)\,,
\eeqs
in which the background functions depend only on the radial direction, $r$.
The connection symbols are
\beqs
\Gamma^P_{\,\,\,\,MN}&\equiv&\frac{1}{2}g^{PQ}\left(\frac{}{}\partial_Mg_{NQ}+\partial_Ng_{QM}-\partial_Qg_{MN}\right)\,,
\eeqs
while the Riemann tensor is
\beqs
R_{MNP}^{\,\,\,\,\,\,\,\,\,\,\,\,\,\,\,\,\,Q}&\equiv&\partial_N\Gamma^Q_{\,\,\,\,MP}
-\partial_M\Gamma^Q_{\,\,\,\,NP}+\Gamma^S_{\,\,\,\,MP}\Gamma^Q_{\,\,\,\,SN}
-\Gamma^S_{\,\,\,\,NP}\Gamma^Q_{\,\,\,\,SM}\,,
\eeqs
the Ricci tensor is
\beqs
R_{MN}&\equiv&R_{MPN}^{\,\,\,\,\,\,\,\,\,\,\,\,\,\,\,\,\,P}\,,
\eeqs
and  the Ricci scalar is
\beqs
{R}&\equiv&R_{MN}g^{MN}\,.
\eeqs
The conventions are such that the (gravity) covariant derivative for a $(1,1)$-tensor takes the form
\beqs
\nabla_M T^{P}_{\,\,\,\,N}&\equiv&\partial_MT^{P}_{\,\,\,\,N}+\Gamma^P_{\,\,\,\,MQ}T^{Q}_{\,\,\,\,N}-\Gamma^Q_{\,\,\,\,MN}T^{P}_{\,\,\,\,Q}\,,
\eeqs
and generalises to any $(m,n)$-tensors. 

Because of the presence of boundaries, at which the five-dimensional  manifold ends on
two four-dimensional manifolds, one must introduce the  induced metric, $\tilde{g}_{MN}$, 
for which we adopt the following conventions:
\beqs
\tilde{g}_{MN}&\equiv& g_{MN}-N_MN_M\,,
\eeqs
where $N_M$ is the  ortho-normalised vector to the boundary,
which satisfies the defining properties:
\beqs
g_{MN}N^MN^N&=&1\,,\qquad {\rm and} \qquad \tilde{g}_{MN} N^N\,=\,0\,.
\eeqs
It is conventional to orient the ortho-normalised vector so that it points outside of the space. Yet,
in the body of the paper, 
we use a definition of $N_M$ which aligns it along the holographic direction, so that it points outwards from the boundary at the UV, but inside the space  at the IR boundary. For this reason,  different signs appear in front of the terms localised at the two boundaries in Eq.~(\ref{Eq:GHY}).
The extrinsic curvature, $K$, is given by $K\equiv \tilde{g}^{MN}K_{MN}$,
in terms of the symmetric tensor
\beqs
K_{MN}&\equiv&\nabla_MN_N\,=\,\partial_MN_N-\Gamma^Q_{\,\,\,\,MN}N_Q\,.
\eeqs

In  parallel to the space-time, in the internal space, the sigma-model connection 
 descends from the sigma-model metric, $ G_{ab}$, and the
 sigma-model derivative,  $\partial_b=\frac{\partial}{\partial \Phi^b}$, to read 
\beqs
 {\cal G}^d_{\,\,\,\,ab}&\equiv& \frac{1}{2} G^{dc}\left(\frac{}{}\partial_a G_{cb}
+\partial_b G_{ca}-\partial_c G_{ab}\right)\,.
\eeqs
The sigma-model Riemann tensor is 
\beqs
 {\cal R}^a_{\,\,\,\,bcd}
&\equiv& \partial_c {\cal G}^a_{\,\,\,\,bd}-\partial_d {\cal G}^a_{\,\,\,\,bc}
+ {\cal G}^e_{\,\,\,\,bd} {\cal G}^a_{\,\,\,\,ce}- {\cal G}^e_{\,\,\,\,bc} {\cal G}^a_{\,\,\,\,de}\,,
\eeqs
while the sigma-model covariant derivative is
\beqs
D_b X^d_{\,\,\,\,a}&\equiv& \partial_b X^d_{\,\,\,\,a}+{\cal G}^d_{\,\,\,\,cb}X^c_{\,\,\,\,a}
- {\cal G}^c_{\,\,\,\,ab}X^d_{\,\,\,\,c}\,.
\eeqs

The equations of motion, satisfied by the background scalars,
 are the following:
\beqs
\label{eq:backgroundEOM1}
\partial_r^2\Phi^a\,+\,(D-1)\partial_r {A}\partial_r\Phi^a\,+\, {\cal G}^a_{\,\,\,\,bc}\partial_r\Phi^b\partial_r\Phi^c\,-\,\mathcal V^a
&=&0\,,
\eeqs
where the sigma-model derivatives are given by $\mathcal V^a\equiv  G^{ab}\partial_b \mathcal V$, and  
$\partial_b \mathcal V\equiv \frac{\partial \mathcal V}{\partial \Phi^b}$. The Einstein equations reduce to
\beqs
\label{eq:backgroundEOM2}
(D-1)(\partial_r { A})^2\,+\,\partial_r^2 { A}\,+\,\frac{4}{D-2} \mathcal V &=&
0\,,\\
(D-1)(D-2)(\partial_r { A})^2\,-\,2 G_{ab}\partial_r\Phi^a\partial_r\Phi^b\,+\,4 \mathcal V&=&0\,.
\eeqs

\subsubsection{Tensor and  scalar fluctuations}
\label{Sec:fluctuations}

Following Refs.~\cite{Bianchi:2003ug,Berg:2006xy,Berg:2005pd,Elander:2009bm,Elander:2010wd},
the scalar fields can be written  as
\beq
	\Phi^a(x^\mu,r) =  \Phi^a(r) + \varphi^a(x^\mu,r) \,,
\eeq
where $\varphi^a(x^\mu,r)$ are small fluctuations around the background 
solutions, $\Phi^a(r)$.
The metric fluctuations are decomposed with the ADM formalism~\cite{Arnowitt:1962hi}:
\beqs
	\dd s_D^2 &=& \left( (1 + \nu)^2 + \nu_\sigma \nu^\sigma \right) \dd r^2 + 2 \nu_\mu \dd x^\mu \dd r 
	+ e^{2 {A}(r)} \left( \eta_{\mu\nu} + h_{\mu\nu} \right) \dd x^\mu \dd x^\nu \,,
	\eeqs
	where 
	\beqs
	h^\mu{}_\nu &=& \mathfrak e^\mu{}_\nu + i q^\mu \epsilon_\nu + i q_\nu \epsilon^\mu 
	+ \frac{q^\mu q_\nu}{q^2} H + \frac{1}{D-2} \delta^\mu{}_\nu h.
\eeqs
Here, $\nu(x^\mu,r)$, $\nu^\mu(x^\mu,r)$,
 $\mathfrak e^\mu{}_\nu(x^\mu,r)$, $\epsilon^\mu(x^\mu,r)$, $H(x^\mu,r)$, and $h(x^\mu,r)$ are small
  fluctuations around the background metric, of which $\mathfrak e^\mu{}_\nu$ is transverse and traceless (while
  $\epsilon^\mu$ is transverse), and gauge invariant.

The other (diffeomorphism)  gauge-invariant combinations  are
\beqs
	\mathfrak a^a &=& \varphi^a - \frac{\partial_r  \Phi^a}{2(D-2)\partial_r { A}} h \,, \\
	\mathfrak b &=& \nu - \partial_r \left( \frac{h}{2(D-2)\partial_r {A}} \right) \,, \\
	\mathfrak c &=& e^{-2{ A}} \partial_\mu \nu^\mu - \frac{e^{-2{ A}} q^2 h}{2(D-2) \partial_r A} 
	- \frac{1}{2} \partial_r H \,, \\
	\mathfrak d^\mu &=& e^{-2{A}} P^\mu{}_\nu \nu^\nu - \partial_r \epsilon^\mu \,.
\eeqs
The algebraic  nature of the equations 
 for $\mathfrak{b}$ and $\mathfrak{c}$ allows us  to decouple the equations and solve them.

The tensor fluctuations $\mathfrak e^\mu{}_\nu$   obey the equation of motion
\beq
\label{eq:tensoreom}
	\left[ \partial_r^2 + (D-1) \partial_r {A} \partial_r - e^{-2{ A}(r)} q^2 \right] \mathfrak e^\mu_{\,\,\,\nu} = 0 \,,
\eeq
with boundary conditions given by 
\beq
\label{eq:tensorbc}
\left.\frac{}{}	\partial_r \mathfrak e ^\mu_{\,\,\,\nu} \right|_{r=r_i}= 0 \, .
\eeq
 The equation of motion for $\mathfrak d^\mu$ is algebraic and does not lead to a spectrum of states. 
 The equations of motion for $\mathfrak a^a$ obey the following equations of motion:
\beqs
\label{eq:scalareom}
	0 &=& \Big[ {\cal D}_r^2 + (D-1) \partial_{r}{A} {\cal D}_r - e^{-2{A}} q^2 \Big] \mathfrak{a}^a \,\,\\ \nonumber
	&& - \Big[  {\mathcal V}^{\,\,a}{}_{\,|c} - \mathcal{R}^a{}_{bcd} \partial_{r}\Phi^b \partial_{r}\Phi^d + 
	\frac{4 (\partial_{r}\Phi^a  {\mathcal V}^{\,b} +  {\mathcal V}^{\,a} 
	\partial_{r}\Phi^b) G_{bc}}{(D-2) \partial_{r} {A}} + 
	\frac{16  {\mathcal V} \partial_{r}\Phi^a \partial_{r}\Phi^b G_{bc}}{(D-2)^2 (\partial_{r}{A})^2} \Big] \mathfrak{a}^c\,,
\eeqs
while the boundary conditions are given by
\beqs
\label{eq:scalarbc}
 \frac{2  e^{2A}\partial_{r} \Phi^a }{(D-2)q^2 \partial_{r}{ A}}
	\left[ \partial_{r} \Phi^b{\cal D}_r -\frac{4  {\cal V} \partial_{r} \Phi^b}{(D-2) 
	\partial_r { A}} - {\cal V}^b \right] \mathfrak a_b - \mathfrak a^a\Big|_{r_i} = 0 \, .
\eeqs
The background covariant derivative is  
$\mathcal D_r \mathfrak a^a \equiv \partial_r \mathfrak a^a +
 \mathcal G^a_{\ bc} \partial_r  \Phi^b \mathfrak a^c$,
 and ${\mathcal V}^a{}_{|b} \equiv \frac{\partial {\mathcal V}^a}{\partial \Phi^b} + \mathcal G^a_{\ bc} {\mathcal V}^c$.

\subsection{Vectors, pseudoscalars, and spurions}
\label{sec:vectorspseudoscalars}

In this appendix, we  provide the gauge fixing terms and equations of motion for the spin-1 and spin-0 fluctuations introduced in Sect.~\ref{Sec:addboundaryterms}. The resulting equations for the fields defined in Eq.~\eqref{eq:flucbasis3} (except for $\chi_M$) and other fields interacting with them are distributed in the following subsections. Section ~\ref{subsec:hattilde} reports the
equations of motion  and boundary conditions for $ \mathcal B_M{}^{\tilde{\mathcal A}}$ and $\mathcal B_M{}^{\hat{\mathcal A}}$,
 and for the associated  spin-0 states. Section~\ref{subsec:4} discusses $\mathcal A_M{}^{4}$ and Sect.~\ref{Sec:bar} focuses on
 $\mathcal A_M{}^{\bar{\mathcal A}}$, respectively.

\subsubsection{The $\mathcal B_M{}^{\hat{\mathcal A}}$ and $\mathcal B_M{}^{\tilde{\mathcal A}}$ sectors}
\label{subsec:hattilde}

Following the procedure in Ref.~\cite{Elander:2018aub}, we choose the gauge fixing terms for $\mathcal B_M{}^{\hat{\mathcal A}}$ and $\mathcal B_M{}^{\tilde{\mathcal A}}$ to be
\beqs
\label{eq:hatgauge}
\mathcal{S}_{\hat \xi}^{(1)}&=& \int \mathrm{d}^{4} q \mathrm{~d} r\left\{-\frac{H^{(1)}_{\hat{\mathcal A}\hat{\mathcal A}}}{2 \hat\xi}\left[i q^{\mu} \mathcal B_{\mu}{}^{\hat{\mathcal A}}(-q)-\frac{g \sin (v)}{2 v} \frac{\hat\xi}{H^{(1)}_{\hat{\mathcal A}\hat{\mathcal A}}} G^{(1)}_{\hat{\mathcal A}\hat{\mathcal A}} e^{2 A} \pi^{\hat{\mathcal A}}(-q)- \frac{\hat\xi}{H^{(1)}_{\hat{\mathcal A}\hat{\mathcal A}}} \partial_{r}\left(H^{(1)}_{\hat{\mathcal A}\hat{\mathcal A}} e^{2 A} \mathcal B_{5}{}^{\hat{\mathcal A}}(-q)\right)\right] \times\right.\nonumber\\
& &\times  {\left.\left[-i q^{\nu}  \mathcal B_{\nu}{}^{\hat{\mathcal A}}(q)-\frac{g \sin (v)}{2 v} \frac{\hat\xi}{H^{(1)}_{\hat{\mathcal A}\hat{\mathcal A}}} G^{(1)}_{\hat{\mathcal A}\hat{\mathcal A}} e^{2 A} \pi^{\hat{\mathcal A}}(q)- \frac{\hat\xi}{H^{(1)}_{\hat{\mathcal A}\hat{\mathcal A}}} \partial_{r}\left(H^{(1)}_{\hat{\mathcal A}\hat{\mathcal A}} e^{2 A} \mathcal B_{5}{}^{\hat{\mathcal A}}(q)\right)\right]\right\}\,, }
\eeqs
and
\beqs
\label{eq:tildegauge}
\mathcal{S}_{\tilde \xi}^{(1)}&=& \int \mathrm{d}^{4} q \mathrm{~d} r\left\{-\frac{H^{(1)}_{\tilde{\mathcal A}\tilde{\mathcal A}}}{2 \tilde\xi}\left[i q^{\mu} \mathcal B_{\mu}{}^{\tilde{\mathcal A}}(-q)- \frac{\tilde \xi}{H^{(1)}_{\tilde{\mathcal A}\tilde{\mathcal A}}} \partial_{r}\left(H^{(1)}_{\tilde{\mathcal A}\tilde{\mathcal A}} e^{2 A}\mathcal B_{5}{}^{\tilde{\mathcal A}}(-q)\right)\right] \times\right.\nonumber\\
& &\times {\left.\left[-i q^{\nu} \mathcal B_{\nu}{}^{\tilde{\mathcal A}}(q)- \frac{\tilde \xi}{H^{(1)}_{\tilde{\mathcal A}\tilde{\mathcal A}}} \partial_{r}\left(H^{(1)}_{\tilde{\mathcal A}\tilde{\mathcal A}} e^{2 A} \mathcal B_{5}{}^{\tilde{\mathcal A}}(q)\right)\right]\right\}\,,  }
\eeqs
where $\hat \xi$ and $\tilde \xi$ are gauge-fixing parameters. The boundary-localised gauge fixing terms at $r=r_2$ are
\begin{align}\label{eq:hatbgauge}
\mathcal{S}_{\hat M}^{(1)}=& \int \mathrm{d}^{4} q \mathrm{~d} r \delta\left(r-r_2\right)\left\{-\frac{1}{2 \hat M_2}\left[i q^{\mu} \mathcal B_{\mu}{}^{\hat{\mathcal A}}(-q)+ \hat M_2 H^{(1)}_{\hat{\mathcal A}\hat{\mathcal A}} e^{2 A} \mathcal B_{5}{}^{\hat{\mathcal A}}(-q)- \hat M_2 K_5\frac{gv_5}{2 } e^{2 A}\cos(v)P^{\hat{\mathcal A}}_5(-q)\right.\right.\\\nonumber
&\left.\left. - \frac{g \sin (v)}{2 v} \hat M_2 C_{2 \, \hat{\mathcal A}\hat{\mathcal A}}^{(1)} e^{2 A}\pi^{\hat{\mathcal A}}(-q)\right]\times\left[(q \rightarrow -q)\right]\right\}\,,
\end{align}
and 
\begin{align}\label{eq:tildebgauge}
\mathcal{S}_{\tilde M}^{(1)}=& \int \mathrm{d}^{4} q \mathrm{~d} r \delta\left(r-r_2\right)\left\{-\frac{1}{2 \tilde M_2}\left[i q^{\mu} \mathcal B_{\mu}{}^{\tilde{\mathcal A}}(-q)+ \tilde M_2 H^{(1)}_{\tilde{\mathcal A}\tilde{\mathcal A}} e^{2 A} \mathcal B_5{}^{\tilde{\mathcal A}}(-q)+ \tilde M_2 K_5\frac{gv_5}{2} e^{2 A}\sin(v) P^{\hat{\mathcal A}}_5(-q)\right]\right.\times \\\nonumber
&\left.\times\left[(q \rightarrow -q)\right]\right\}\,.
\end{align}

The boundary-localised gauge fixing parameters, $\hat M_2$ and $\tilde M_2$, are independent of the bulk dynamics. 
Gauge fixing at $r=r_1$ can be done in a similar manner.
Gathering terms from the action Eq.~\eqref{eq:5d} and gauge fixing terms in Eqs.~\eqref{eq:hatgauge}, \eqref{eq:tildegauge}, \eqref{eq:hatbgauge}, and~\eqref{eq:tildebgauge}, the equations of motion and boundary conditions for $\mathcal B_{\mu}{}^{\hat{\mathcal A}}$ and $\mathcal B_{\mu}{}^{\tilde{\mathcal A}}$ read
\begin{align}
0=&{\left[q^{2} H^{(1)}_{\hat{\mathcal A}\hat{\mathcal A}}-\partial_{r}\left(H^{(1)}_{\hat{\mathcal A}\hat{\mathcal A}} e^{2 A} \partial_{r}\right)+\left(\frac{g}{2}\right)^2 G^{(1)}_{\hat{\mathcal A}\hat{\mathcal A}} e^{2 A}\right] P^{\mu \nu} \mathcal B_{\mu}{}^{\hat{\mathcal A}}(q, r) } , \label{eq:hatB} \\
0=&{\left.\left[
q^{2} D_{2 \, \hat{\mathcal A}\hat{\mathcal A}}^{(1)}+H^{(1)}_{\hat{\mathcal A}\hat{\mathcal A}} e^{2 A} \partial_{r}+ e^{2 A}\left(\frac{g}{2}\right)^2 C_{2 \, \hat{\mathcal A}\hat{\mathcal A}}^{(1)} +e^{2 A}(\frac{gv_5}{2})^{2} K_{5}\cos(v)^2\right] P^{\mu \nu}\mathcal B_{\nu}{}^{\hat{\mathcal A}}(q, r)\right|_{r=r_{2}} } \nonumber\\
&+\left.\left[
q^{2} D_{2 \, \hat{\mathcal A}\tilde{\mathcal A}}^{(1)}- e^{2 A}(\frac{gv_5}{2})^{2} K_{5}\cos(v)\sin(v)\right]P^{\mu \nu}  \mathcal B_{\nu}{}^{\tilde{\mathcal A}}(q, r)\right|_{r=r_{2}} \,,\\
\label{eq:hatbB}
 0=&{\left[\frac{q^{2}}{\hat \xi} H^{(1)}_{\hat{\mathcal A}\hat{\mathcal A}}-\partial_{r}\left(H^{(1)}_{\hat{\mathcal A}\hat{\mathcal A}} e^{2 A} \partial_{r}\right)+\left(\frac{g}{2}\right)^2 G^{(1)}_{\hat{\mathcal A}\hat{\mathcal A}} e^{2 A}\right] \frac{q^{\mu} q^{\nu}}{q^{2}} \mathcal B_{\mu}{}^{\hat{\mathcal A}}(q, r) }, \\
0=&{\left.\left[\frac{q^2}{\hat M_2}+H^{(1)}_{\hat{\mathcal A}\hat{\mathcal A}} e^{2 A} \partial_{r}+ e^{2 A}\left(\frac{g}{2}\right)^2 C_{2 \, \hat{\mathcal A}\hat{\mathcal A}}^{(1)} +e^{2 A}(\frac{gv_5}{2})^{2} K_{5}\cos(v)^2\right] \frac{q^{\mu} q^{\nu}}{q^{2}}\mathcal B_{\nu}{}^{\hat{\mathcal A}}(q, r)\right|_{r=r_{2}} } \nonumber\\
&{\left.- \left[e^{2 A}(\frac{gv_5}{2})^{2} K_{5}\cos(v)\sin(v)\right]\frac{q^{\mu} q^{\nu}}{q^{2}} \mathcal B_{\nu}{}^{\tilde{\mathcal A}}(q, r)\right|_{r=r_{2}} }  \,,\\
\label{eq:tildeB}
0=&{\left[q^{2} H^{(1)}_{\tilde{\mathcal A}\tilde{\mathcal A}}-\partial_{r}\left(H^{(1)}_{\tilde{\mathcal A}\tilde{\mathcal A}} e^{2 A} \partial_{r}\right)\right] P^{\mu \nu}\mathcal B_{\mu}{}^{\tilde{\mathcal A}}(q, r) } \,, \\
0=&{\left.\left[
q^{2} D_{2 \, \tilde{\mathcal A}\tilde{\mathcal A}}^{(1)}+H^{(1)}_{\tilde{\mathcal A}\tilde{\mathcal A}} e^{2 A} \partial_{r}+e^{2 A}(\frac{gv_5}{2})^{2} K_{5}\sin(v)^2\right] P^{\mu \nu}\mathcal B_{\nu}{}^{\tilde{\mathcal A}}(q, r)\right|_{r=r_{2}} } \nonumber\\
&{\left.+\left[
q^{2} D_{2 \, \tilde{\mathcal A}\hat{\mathcal A}}^{(1)}-e^{2 A}(\frac{gv_5}{2})^{2} K_{5}\cos(v)\sin(v)\right]P^{\mu \nu} 
\mathcal B_{\nu}{}^{\hat{\mathcal A}}(q, r)\right|_{r=r_{2}} }\,, \label{eq:tildebB} \\
0=&{\left[\frac{q^{2}}{\tilde \xi} H^{(1)}_{\tilde{\mathcal A}\tilde{\mathcal A}}-\partial_{r}\left(H^{(1)}_{\tilde{\mathcal A}\tilde{\mathcal A} }e^{2 A} \partial_{r}\right)\right] \frac{q^{\mu} q^{\nu}}{q^{2}}\mathcal B_{\mu}{}^{\tilde{\mathcal A}}(q, r) } \,, \\
0=&{\left.\left[\frac{q^2}{\tilde M_2}+H^{(1)}_{\tilde{\mathcal A}\tilde{\mathcal A}} e^{2 A} \partial_{r}+e^{2 A}(\frac{gv_5}{2})^{2} K_{5}\sin(v)^2\right] \frac{q^{\mu} q^{\nu}}{q^{2}}\mathcal B_{\nu}{}^{\tilde{\mathcal A}}(q, r)\right|_{r=r_{2}} } \nonumber\\
&-{\left.\left[e^{2 A}(\frac{gv_5}{2})^{2} K_{5}\cos(v)\sin(v)\right]\frac{q^{\mu} q^{\nu}}{q^{2}} \mathcal B_{\nu}{}^{\hat{\mathcal A}}(q, r)\right|_{r=r_{2}} } \,.
\end{align}
Equations~\eqref{eq:hatB}, \eqref{eq:hatbB}, \eqref{eq:tildeB}, and \eqref{eq:tildebB} can be compared to Eqs.~\eqref{eq:barv}, \eqref{eq:hatv}, and \eqref{eq:vectorBCs} in the body of the paper.

For pseudoscalar and spurion fields, the equations are obtained from the variation with respect to
 $\mathcal B_{5}{}^{\tilde{\mathcal A}}$, $\mathcal B_{5}{}^{\hat{\mathcal A}}$, $\pi^{\hat{\mathcal A}}$ 
 in the bulk and boundary and $P_5^{\hat {\mathcal A}}$ in the boundary, respectively:
\begin{align}
0=&{\left[q^{2}-\partial_{r}\left(\frac{\tilde \xi}{H^{(1)}_{\tilde{\mathcal A}\tilde{\mathcal A}}}  \partial_{r}\left(H^{(1)}_{\tilde{\mathcal A}\tilde{\mathcal A}}e^{2A}\right)\right)\right]\mathcal B_{5}{}^{\tilde{\mathcal A}} (q, r) } \,, \\
0=&{\left.\left[\tilde \xi \frac{e^{-2A}}{H^{(1)}_{\tilde{\mathcal A}\tilde{\mathcal A}}}\partial_{r} \left(H^{(1)}_{\tilde{\mathcal A}\tilde{\mathcal A}}e^{2A}\right)+\tilde M_2 H^{(1)}_{\tilde{\mathcal A}\tilde{\mathcal A}}\right] \mathcal B_{5}{}^{\tilde{\mathcal A}} (q, r)\right|_{r=r_{2}} +\left.\left[\frac{gv_5}{2}\tilde M_2 K_5 \sin (v) \right]P_5^{\hat{\mathcal A}}(q)\right|_{r=r_{2}} } \,, \\
0=&{\left[q^{2}H^{(1)}_{\hat{\mathcal A}\hat{\mathcal A}}-H^{(1)}_{\hat{\mathcal A}\hat{\mathcal A}}\partial_{r}\left(\frac{\hat \xi}{H^{(1)}_{\hat{\mathcal A}\hat{\mathcal A}}}  \partial_{r}\left(H^{(1)}_{\hat{\mathcal A}\hat{\mathcal A}}e^{2A}\right)\right)+\frac{g}{2}^{2} G^{(1)}_{\hat{\mathcal A}\hat{\mathcal A}} e^{2 A}\right]\mathcal B_{5}{}^{\hat{\mathcal A}}(q, r) } \nonumber\\
&+{\frac{ g\sin (v)}{ 2v}\left[G^{(1)}_{\hat{\mathcal A}\hat{\mathcal A}}e^{2 A}\partial_{r}-H^{(1)}_{\hat{\mathcal A}\hat{\mathcal A}}\partial_{r}\left(\hat \xi\frac{G^{(1)}_{\hat{\mathcal A}\hat{\mathcal A}}e^{2 A}}{H^{(1)}_{\hat{\mathcal A}\hat{\mathcal A}}}\right)\right] \pi^{\hat{\mathcal A}}(q, r)} \,, \\
0=&{\left.\left[ \frac{e^{-2A}\hat \xi}{H^{(1)}_{\tilde{\mathcal A}\tilde{\mathcal A}}}\partial_{r} \left( H^{(1)}_{\hat{\mathcal A}\hat{\mathcal A}}e^{2A}\right)+\hat M_2 H^{(1)}_{\hat{\mathcal A}\hat{\mathcal A}} \right] \mathcal B_{5}{}^{\hat{\mathcal A}}(q, r)-\left[\frac{g v_5}{2}\hat M_2   \cos (v) K_5 \right]P_5^{\hat{\mathcal A}}(q)\right|_{r=r_{2}} }  \nonumber \\
&+{\left.\frac{g \sin (v)}{ 2 v}\left[\hat \xi \frac{G^{(1)}_{\hat{\mathcal A}\hat{\mathcal A}}}{H^{(1)}_{\tilde{\mathcal A}\tilde{\mathcal A}}} -\hat M_2   C_{2 \, \hat{\mathcal A}\hat{\mathcal A}}^{(1)}\right] \pi^{\hat{\mathcal A}}(q, r)\right|_{r=r_{2}} } \,, \\
0=&{\frac{ \sin (v)}{ v}\left[\partial_{r}\left(G^{(1)}_{\hat{\mathcal A}\hat{\mathcal A}}e^{4 A}\partial_{r}\right)-G^{(1)}_{\hat{\mathcal A}\hat{\mathcal A}}e^{2 A}q^2-\left(\frac{g}{2}\right)^2\frac{\hat \xi}{H^{(1)}_{\hat{\mathcal A}\hat{\mathcal A}}}e^{4A}(G^{(1)}_{\hat{\mathcal A}\hat{\mathcal A}})^2\right] \pi^{\hat{\mathcal A}}(q, r)}  \nonumber \\
&+{\frac{g}{2}\left[\partial_{r}\left(G^{(1)}_{\hat{\mathcal A}\hat{\mathcal A}}e^{4 A}\right)-\frac{\hat \xi}{H^{(1)}_{\hat{\mathcal A}\hat{\mathcal A}}}e^{2A}G^{(1)}_{\hat{\mathcal A}\hat{\mathcal A}}\partial_{r}\left(H^{(1)}_{\hat{\mathcal A}\hat{\mathcal A}}e^{2A}\right)\right]\mathcal B_{5}{}^{\hat{\mathcal A}}(q, r) } \,, \\
0=&{\left.\frac{ \sin (v)}{v}\left[C_{2 \, \hat{\mathcal A}\hat{\mathcal A}}^{(1)}e^{-2 A} q^2+\hat M_2\left(\frac{g}{2}\right)^{2} (C_{2 \, \hat{\mathcal A}\hat{\mathcal A}}^{(1)})^2 + G^{(1)}_{\hat{\mathcal A}\hat{\mathcal A}}\partial_{r}\right] \pi^{\hat{\mathcal A}}(q, r)\right|_{r=r_{2}} }  \nonumber\\
&+{\left.\left[\hat M_2 v_5 \left(\frac{g}{2}\right)^2 \cos(v) K_5 C_{2 \, \hat{\mathcal A}\hat{\mathcal A}}^{(1)}\right]P_5^{\hat{\mathcal A}}(q)\right|_{r=r_{2}} -\left.\left[\frac{g}{2}\hat M_2(C_{2 \, \hat{\mathcal A}\hat{\mathcal A}}^{(1)})H^{(1)}_{\hat{\mathcal A}\hat{\mathcal A}}-\frac{g}{2}G^{(1)}_{\hat{\mathcal A}\hat{\mathcal A}}\right] \mathcal B_{5}{}^{\hat{\mathcal A}}(q, r)\right|_{r=r_{2}} } \,,\\
0=&{\left.\left[K_5 e^{-2 A} q^2+\left(\frac{g v_5}{2}\right)^2 K_5^2 \left( \cos(v)^2 \hat M_2 + \sin(v)^ 2\tilde M_2 \right) \right]P_5^{\hat{\mathcal A}}(q)\right|_{r=r_{2}} }  \nonumber \\
&+\left.\left[\hat M_2 K_5 v_5 \left(\frac{g}{2}\right)^2 \frac{ \cos(v) \sin (v)}{ v} C_{2 \, \hat{\mathcal A}\hat{\mathcal A}}^{(1)}\right] \pi^{\hat{\mathcal A}}(q, r)\right|_{r=r_{2}}   \nonumber \\
&-{\left.\left[K_5 v_5 \cos(v)\frac{g}{2}\hat M_2H^{(1)}_{\hat{\mathcal A}\hat{\mathcal A}}\right] \mathcal B_{5}{}^{\hat{\mathcal A}}(q, r)\right|_{r=r_{2}} + \left.\left[K_5\sin(v)\frac{g v_5}{2}\tilde M_2H^{(1)}_{\tilde{\mathcal A}\tilde{\mathcal A}}\right] \mathcal B_{5}{}^{\tilde{\mathcal A}} (q, r) \right|_{r=r_{2}} } \,.
\end{align}

We introduce convenient redefinitions
\beqs
\mathcal B_{5}{}^{\hat{\mathcal A}}&\equiv&\frac{X^{\hat{\mathcal A}}}{e^{4A} G^{(1)}_{\hat{\mathcal A}\hat{\mathcal A}}} - \frac{2}{g}\frac{ \sin (v)}{ v} \partial_r \pi^{\hat{\mathcal A}}\,,\\
\pi^{\hat{\mathcal A}}&\equiv& \frac{v}{ \sin (v)} \left(Y^{\hat{\mathcal A}} + \frac{(g/2) \partial_r X^{\hat{\mathcal A}}}{q^2 e^{2A} G^{(1)}_{\hat{\mathcal A}\hat{\mathcal A}}}\right)\,,
\eeqs
to separate the physical states from the gauge-dependent ones. The equations for the gauge-independent scalar fields, $X^{\hat{\mathcal A}}$, and gauge dependent, non physical $Y^{\hat{\mathcal A}}$ are
\begin{align}
    0=&{\label{eq:hatX}\left[\partial^2_{r}+\left(-2 \partial_{r}A(r)-\frac{\partial_{r}G^{(1)}_{\hat{\mathcal A}\hat{\mathcal A}}}{G^{(1)}_{\hat{\mathcal A}\hat{\mathcal A}}}\right)\partial_{r}+ \left(-q^2 e^{-2 A(r)}-\frac{g^2 G^{(1)}_{\hat{\mathcal A}\hat{\mathcal A}}}{4H^{(1)}_{\hat{\mathcal A}\hat{\mathcal A}}}\right)\right]X^{\hat{\mathcal A}}(q, r) } \,, \\
0=&{\left[\partial^2_{r}+\left(2 \partial_{r}A(r)+\frac{\partial_{r}H^{(1)}_{\hat{\mathcal A}\hat{\mathcal A}}}{H^{(1)}_{\hat{\mathcal A}\hat{\mathcal A}}}\right)\partial_{r}+ \left(-\frac{q^2 e^{-2 A(r)}}{\hat \xi }-\frac{g^2 G^{(1)}_{\hat{\mathcal A}\hat{\mathcal A}}}{4 H^{(1)}_{\hat{\mathcal A}\hat{\mathcal A}}}\right)\right] Y^{\hat{\mathcal A}}(q, r)} \,.
\end{align}
For the boundary terms we only mention the physical $X^{\hat{\mathcal A}}$ boundary condition:
\begin{align}\label{eq:hatbX}
    0=&{\left.\left[\partial_r+ \frac{G^{(1)}_{\hat{\mathcal A}\hat{\mathcal A}}}{C_{2 \, \hat{\mathcal A}\hat{\mathcal A}}^{(1)}} \right]X^{\hat{\mathcal A}}(q, r)\right|_{r=r_{2}} } .
\end{align}
Equations~\eqref{eq:hatX} and~\eqref{eq:hatbX} can be compared to Eqs.~\eqref{eq:hatp} and~\eqref{eq:hatbp}.

\subsubsection{The $\mathcal A_M{}^{4}$ sector}
\label{subsec:4}

Appropriate gauge fixing terms for $\mathcal A_M{}^{4}$ in the bulk are the following:
\begin{align}
\mathcal{S}_{\xi}^{(1)}=& \int \mathrm{d}^{4} q \mathrm{~d} r\left\{-\frac{H^{(1)}_{4 4}}{2 \xi}\left[i q^{\mu} \mathcal A_{\mu}{}^{4}(-q)-\frac{g}{2} \frac{\xi}{H^{(1)}_{4 4}} G^{(1)}_{44} e^{2 A} \Pi^{4}(-q)- \frac{\xi}{H^{(1)}_{4 4}} \partial_{r}\left(H^{(1)}_{4 4} e^{2 A} \mathcal A_{5}{}^{4}(-q)\right)\right] \times\right.\nonumber\\
 &\times  {\left.\left[-i q^{\nu}  \mathcal A_{\nu}{}^{4}(q)-\frac{g}{2}  \frac{\xi}{H^{(1)}_{4 4}} G^{(1)}_{44} e^{2 A} \Pi^{4}(q)- \frac{\xi}{H^{(1)}_{4 4}} \partial_{r}\left(H^{(1)}_{4 4} e^{2 A} \mathcal A_{5}{}^{4}(q)\right)\right]\right\}, }
\end{align}
where $\xi$ is the gauge-fixing parameter. The boundary-localised, gauge-fixing terms at $r=r_2$ are
\begin{align}
\mathcal{S}_{M}^{(1)}=& \int \mathrm{d}^{4} q \mathrm{~d} r \delta\left(r-r_2\right)\left\{-\frac{1}{2  M_2}\left[i q^{\mu} \mathcal A_{\mu}{}^{4}(-q)+  M_2 H^{(1)}_{4 4} e^{2 A} \mathcal A_{5}{}^{4}(-q)-  M_2 K_5\frac{g v_5}{2 } e^{2 A}P^{4}_5(-q)\right.\right.\nonumber \\
&\left.\left. - \frac{g}{2}   M_2 C_{2 \, 44}^{(1)} e^{2 A}\Pi^{4}(-q)\right]\times\left[(q \rightarrow -q)\right]\right\}\,,
\end{align}
with $M_2$ a free parameter. Thus, the equations of motion and boundary conditions for $\mathcal A_{\mu}{}^{4}$ read as
\begin{align}
\label{eq:4A}
0=&{\left[q^{2} H^{(1)}_{4 4}-\partial_{r}\left(H^{(1)}_{4 4} e^{2 A} \partial_{r}\right)+\left(\frac{g}{2}\right)^2 G^{(1)}_{44} e^{2 A}\right] P^{\mu \nu}\mathcal A_{\mu}{}^{4}(q, r) }, \\
\label{eq:4bA}
0=&{\left.\left[H^{(1)}_{4 4} e^{2 A} \partial_{r}+q^{2} D_{2 \, 44}^{(1)}+e^{2 A}\left(\frac{g}{2}\right)^2 C_{2 \, 44}^{(1)}+(\frac{gv_5}{2})^{2} K_5 e^{2 A}\right] P^{\mu \nu} \mathcal A_{\nu}{}^{4}(q, r) \right|_{r=r_{2}} }, \\
0=&{\left[\frac{q^{2}}{\xi} H^{(1)}_{4 4}-\partial_{r}\left(H^{(1)}_{4 4} e^{2 A} \partial_{r}\right)+\left(\frac{g}{2}\right)^2 G^{(1)}_{44} e^{2 A}\right] \frac{q^{\mu} q^{\nu}}{q^{2}}\mathcal A_{\mu}{}^{4}(q, r) }, \\
0=&{\left.\left[H^{(1)}_{4 4} e^{2 A} \partial_{r}+\frac{q^{2}}{M_2}+ e^{2 A}\left(\frac{g}{2}\right)^2 C_{2 \, 44}^{(1)}+(\frac{gv_5}{2})^{2} K_5 e^{2 A}\right] \frac{q^{\mu} q^{\nu}}{q^{2}} \mathcal A_{\nu}{}^{4}(q, r) \right|_{r=r_{2}} } .
\end{align}
Equations~\eqref{eq:4A} and \eqref{eq:4bA} are restated in Eqs.~\eqref{eq:hatv} and \eqref{eq:4bv}.

For the pseudoscalars and the spurion, the equations are obtained from the variation with respect to $\mathcal A_{5}{}^{4}$ and $\Pi^{4}$
 in bulk and boundary and $P_5^{4}$ in the boundary, respectively. They are the following:
\begin{align}
0=&{\left[q^{2}H^{(1)}_{4 4}-H^{(1)}_{4 4}\partial_{r}\left(\frac{\xi}{H^{(1)}_{4 4}} \partial_{r}\left(H^{(1)}_{4 4}e^{2A}\right)\right)+\left(\frac{g}{2}\right)^2 G^{(1)}_{44} e^{2 A}\right]\mathcal A_{5}{}^{4}(q, r) } \nonumber\\
&+\frac{g}{2}{\left[G^{(1)}_{44}e^{2 A}\partial_{r}-H^{(1)}_{4 4}\partial_{r}\left( \frac{\xi G^{(1)}_{44}e^{2 A}}{H^{(1)}_{4 4}}\right)\right] \Pi^{4}(q, r)} \,, \\
0=&{\left.\left[\xi \frac{e^{-2A}}{H^{(1)}_{4 4}}\partial_{r} \left(H^{(1)}_{4 4}e^{2A}\right)+M_2 H^{(1)}_{4 4} \right] \mathcal A_{5}{}^{4}(q, r)\right|_{r=r_{2}} } \nonumber\\
&+{\left.\frac{g}{2}\left[\frac{\xi }{H^{(1)}_{4 4}}G^{(1)}_{44}- M_2  C_{2 \, 44}^{(1)}\right] \Pi^{4}(q, r)-\left[M_2 \frac{g}{2}  K_{5}v_5\right]P^{4}_{5}(q) \right|_{r=r_{2}} }\,, \\
0=&{\left[\partial_{r}\left(G^{(1)}_{44}e^{4 A}\partial_{r}\right)-G^{(1)}_{44}e^{2 A}q^2-\left(\frac{g}{2}\right)^2\frac{\xi}{H^{(1)}_{4 4}}e^{4A}(G^{(1)}_{44})^2\right] \Pi^{4}(q, r)} \nonumber \\
&+{\frac{g}{2}\left[\partial_{r}\left(G^{(1)}_{44}e^{4 A}\right)-\frac{\xi}{H^{(1)}_{4 4}}e^{2A}G^{(1)}_{44}\partial_{r}\left(H^{(1)}_{4 4}e^{2A}\right)\right]\mathcal A_{5}{}^{4}(q, r) } \,, \\
0=&{\left.\left[C_{2 \, 44}^{(1)}e^{-2 A} q^2+ M_2\left(\frac{g}{2}\right)^2 (C_{2 \, 44}^{(1)})^2+ \partial_v^2 \mathcal V_4 +G^{(1)}_{44}\partial_{r}\right] \Pi^{4}(q, r)\right|_{r=r_{2}} } \nonumber\\
&+{\left.\frac{g}{2}\left[G^{(1)}_{44}- M_2   C_{2 \, 44}^{(1)} H^{(1)}_{44}\right] \mathcal A_{5}{}^{4}(q, r)+\left[\frac{g^2 v_5}{4 }M_2 K_5 C_{2 \, 44}^{(1)}-\frac{1}{v_5} \partial_v^2 \mathcal V_4\right]P^{4}_{5}(q)\right|_{r=r_{2}}} \,, \\
0=&{\left.\left[K_5e^{-2 A} q^2+M_2 \left(\frac{gv_5}{2 }\right)^{2} K_{5}^2 +\frac{1}{ v_5^2} \partial_v^2 \mathcal V_4 \right] P^{4}_{5}(q)\right|_{r=r_{2}}}\nonumber\\
&{\left.-\left[M_2 \frac{g v_5}{2} K_{5} H^{(1)}_{44}\right]\mathcal A_{5}{}^{4}(q, r)+\left[\frac{g^2v_5}{4} M_2 K_5 C_{2 \, 44}^{(1)}- \frac{1}{ v_5}\partial_v^2 \mathcal V_4 \right]\Pi^{4}(q, r)\right|_{r=r_{2}} } \,.
\end{align}

By applying the convenient definition
\beqs
\mathcal A_{5}{}^{4}&\equiv&\frac{ X^4}{e^{4A} G^{(1)}_{44}} - \frac{2}{g}\partial_r \Pi^4\,,
\eeqs
one can derive the equations for gauge invariant field $X^4$:
\begin{align} \label{eq:4X}
0=&{\left[\partial^2_{r}+\left(-2 \partial_{r}A-\frac{\partial_{r}G^{(1)}_{44}}{G^{(1)}_{44}}\right)\partial_{r}+ \left(-q^2 e^{-2 A}-\frac{g^2 G^{(1)}_{44}}{4H^{(1)}_{44}}\right)\right]X^{4}(q, r) },\\
0=&{\left.\left[\partial_{r}+G^{(1)}_{44} \left(\frac{\partial_v^2 \mathcal V_4 e^{2 A(r)}+K_5 v_5^2 q^2}{\partial_v^2 \mathcal V_4 K_5 v_5^2  e^{2 A(r)}+C_{2 \, 44}^{(1)}  \partial_v^2 \mathcal V_4 e^{2 A(r)}+K_5 C_{2 \, 44}^{(1)} v_5^2 q^2}\right)\right]X^{4}(q, r) \right|_{r=r_{2}} }. \label{eq:4bX}
\end{align}
Equations~\eqref{eq:4X} and~\eqref{eq:4bX} are restated in Eqs.~\eqref{eq:hatp} and~\eqref{eq:p4BCs}.

\subsubsection{The $\mathcal A_M{}^{\bar {\mathcal{A}}}$ sector}
\label{Sec:bar}
The gauge fixing terms for $\mathcal A_M{}^{\bar {\mathcal{A}}}$ in the bulk are chosen to be
\begin{align}
\mathcal{S}_{\bar \xi}^{(1)}=& \int \mathrm{d}^{4} q \mathrm{~d} r\left\{-\frac{H^{(1)}_{\bar{\mathcal A}\bar{\mathcal A}}}{2 \bar \xi}\left[i q^{\mu} \mathcal A_{\mu}{}^{\bar{\mathcal A}}(-q)- \frac{\bar \xi}{H^{(1)}_{\bar{\mathcal A}\bar{\mathcal A}}} \partial_{r}\left(H^{(1)}_{\bar{\mathcal A}\bar{\mathcal A}} e^{2 A} \mathcal A_{5}{}^{\bar {\mathcal{A}}}(-q)\right)\right] \times\right.\nonumber\\
 & {\left.\left[-i q^{\nu}  \mathcal A_{\nu}{}^{\bar{\mathcal A}}(q)- \frac{\bar \xi}{H^{(1)}_{\bar{\mathcal A}\bar{\mathcal A}}} \partial_{r}\left(H^{(1)}_{\bar{\mathcal A}\bar{\mathcal A}} e^{2 A} \mathcal A_{5}{}^{\bar {\mathcal{A}}}(q)\right)\right]\right\}, }
\end{align}
where $\bar \xi$ is the gauge fixing parameter. The boundary-localised gauge fixing terms at $r=r_2$ reads
\begin{align}
\mathcal{S}_{\bar M}^{(1)}=& \int \mathrm{d}^{4} q \mathrm{~d} r \delta\left(r-r_2\right)\left\{-\frac{1}{2  \bar M_2}\left[i q^{\mu} \mathcal A_{\mu}{}^{\bar{\mathcal A}}(-q)+  \bar M_2 H^{(1)}_{\bar{\mathcal A}\bar{\mathcal A}} e^{2 A} \mathcal A_{5}{}^{\bar{\mathcal A}}(-q)\right]\times\left[-iq^{\nu} \mathcal A_{\nu}{}^{\bar{\mathcal A}}(q)+  \bar M_2 H^{(1)}_{\bar{\mathcal A}\bar{\mathcal A}} e^{2 A} \mathcal A_{5}{}^{\bar{\mathcal A}}(q))\right]\right\}\,,
\end{align}
with $\bar M_2$, the boundary gauge fixing parameter.

Similar to the previous sections, the equations of motion and boundary conditions for $\mathcal A_{\mu}{}^{\bar{\mathcal A}}$ are
\begin{align} \label{eq:barA}
0=&{\left[q^{2} H^{(1)}_{\bar{\mathcal A}\bar{\mathcal A}}-\partial_{r}\left(H^{(1)}_{\bar{\mathcal A}\bar{\mathcal A}} e^{2 A} \partial_{r}\right)\right] P^{\mu \nu}\mathcal A_{\mu}{}^{\bar{\mathcal A}}(q, r) }, \\
0=&{\left.\left[H^{(1)}_{\bar{\mathcal A}\bar{\mathcal A}} e^{2 A} \partial_{r}+q^{2}  D_{2 \, \bar{\mathcal A}\bar{\mathcal A}}^{(1)}\right] P^{\mu \nu} \mathcal A_{\nu}{}^{\bar{\mathcal A}}(q, r) \right|_{r=r_{2}} }, \label{eq:barbA} \\
0=&{\left[\frac{q^{2}}{\bar \xi} H^{(1)}_{\bar{\mathcal A}\bar{\mathcal A}}-\partial_{r}\left(H^{(1)}_{\bar{\mathcal A}\bar{\mathcal A}} e^{2 A} \partial_{r}\right)\right] \frac{q^{\mu} q^{\nu}}{q^{2}}\mathcal A_{\mu}{}^{\bar{\mathcal A}}(q, r) }, \\
0=&{\left.\left[\frac{q^{2}}{\bar M_2}+H^{(1)}_{\bar{\mathcal A}\bar{\mathcal A}} e^{2 A} \partial_{r}\right] \frac{q^{\mu} q^{\nu}}{q^{2}} \mathcal A_{\nu}{}^{\bar{\mathcal A}}(q, r) \right|_{r=r_{2}} } .
\end{align}
Equations ~\eqref{eq:barA} and \eqref{eq:barbA} are restated in Eqs.~\eqref{eq:barv}, and \eqref{eq:barbv}. The fifth component of the gauge field is pure gauge in this case.

\subsection{Asymptotic expansions of the fluctuations}
\label{sec:IRUVexpansions}

We present here some of the asymptotic expansions for the fluctuations, see also Ref.~\cite{Elander:2023aow}.

\subsubsection{IR expansions}

For convenience, we put $\rho_o = 0$ and $A_I = 0$ in this subsection,\footnote{The dependence on $\rho_o$ and $A_I$ can be reinstated by making the substitutions $\rho \rightarrow \rho - \rho_o$ and $q^2 \rightarrow e^{-2A_I} q^2$ in the expressions for the IR expansions.} while $\chi_I = 0$ in order to avoid a conical singularity.

For the fluctuations of the scalars, we have
\begin{align}
\mathfrak a^{\phi} =&\ \mathfrak a^{\phi}_{I,0}+\mathfrak a^{\phi}_{I,l} \log (\rho )+\frac{1}{4} \rho ^2 \bigg[ -\frac{1}{4} \Delta  \left(\mathfrak a^{\phi}_{I,0} \left(\Delta  \left(15 \phi_I^2-4\right)+20\right)+6 \phi_I (\mathfrak a^{\chi}_{I,0}-\mathfrak a^{\chi}_{I,l}) \left(\Delta  \left(5 \phi_I^2-4\right)+20\right)\right)
\nonumber \\ &
+q^2 (\mathfrak a^{\phi}_{I,0}-\mathfrak a^{\phi}_{I,l}) -\frac{1}{48} \mathfrak a^{\phi}_{I,l} \left(\Delta  \left(25 \Delta  \phi_I^4+20 (10-11 \Delta ) \phi_I^2+48 (\Delta -5)\right)+400\right)
\nonumber \\ &
+\log (\rho ) \left(\mathfrak a^{\phi}_{I,l} \left(-\frac{15 \Delta ^2 \phi_I^2}{4}+(\Delta -5) \Delta +q^2\right)-\frac{3}{2} \mathfrak a^{\chi}_{I,l} \Delta  \phi_I \left(\Delta  \left(5 \phi_I^2-4\right)+20\right)\right) \bigg]+\mathcal O \left(\rho ^4\right) \,, \\ 
\mathfrak a^{\chi} =&\ \mathfrak a^{\chi}_{I,0}+\mathfrak a^{\chi}_{I,l} \log (\rho )+\frac{1}{4} \rho ^2 \bigg[-\frac{1}{4} \Delta  \phi_I (\mathfrak a^{\phi}_{I,0}-\mathfrak a^{\phi}_{I,l}) \left(\Delta  \left(5 \phi_I^2-4\right)+20\right) +q^2 (\mathfrak a^{\chi}_{I,0}-\mathfrak a^{\chi}_{I,l})
\nonumber \\ &
-\frac{3}{8} \mathfrak a^{\chi}_{I,0} \left(\Delta  \phi_I^2 \left(\Delta  \left(5 \phi_I^2-8\right)+40\right)+80\right)+\frac{13}{48} \mathfrak a^{\chi}_{I,l} \left(\Delta  \phi_I^2 \left(\Delta  \left(5 \phi_I^2-8\right)+40\right)+80\right)
\nonumber \\ &
+\log (\rho ) \left(-\frac{5}{4} \mathfrak a^{\phi}_{I,l} \Delta ^2 \phi_I^3+\mathfrak a^{\phi}_{I,l} (\Delta -5) \Delta  \phi_I+\mathfrak a^{\chi}_{I,l} \left(-\frac{15}{8} \Delta ^2 \phi_I^4+3 (\Delta -5) \Delta  \phi_I^2+q^2-30\right)\right)\bigg]+\mathcal O \left(\rho ^4\right) \,, \\
\mathfrak a^{\bar A} =&\ \mathfrak a^{\bar A}_{I,0}+\rho ^2 \left(\frac{1}{2} \mathfrak a^{\bar A}_{I,0} q^2 \log (\rho )+\mathfrak a^{\bar A}_{I,2}\right)+\mathcal O \left(\rho ^4\right) \,, \\ 
\mathfrak a^{\hat A} =&\ \mathfrak a^{\hat A}_{I,0}+\rho ^2 \left(\frac{1}{2} \mathfrak a^{\hat A}_{I,0} \left( q^2+ \frac{g^2\phi_I^2}{4} \right) \log (\rho )+\mathfrak a^{\hat A}_{I,2}\right)+\mathcal O \left(\rho ^4\right) \,.
\end{align}

For the pseudoscalar fluctuations, we have
\beq
\mathfrak p^{\hat A} = \mathfrak p^{\hat A}_{I,0}+\rho ^2 \left[\mathfrak p^{\hat A}_{I,2} + \frac{1}{2} \mathfrak p^{\hat A}_{I,0} \left(q^2+\frac{g^2}{4}\phi_I^2\right) \log (\rho )\right]+\mathcal O \left(\rho ^4\right) \,.
\eeq

For the vector fluctuations, we have
\begin{align}
\mathfrak v =&\ \mathfrak v_{I,-2} \rho ^{-2}+\frac{1}{2} q^2 \mathfrak v_{I,-2} \log (\rho )+\mathfrak v_{I,0} +\frac{1}{12288} \rho ^2 \Big[1536 q^2 \mathfrak v_{I,0}+80 \Delta ^2 \mathfrak v_{I,-2} \phi_I^4 \left(2 \left(8 \Delta ^2-50 \Delta +75\right)-3 q^2\right)
\nonumber \\ &
+128 (\Delta -5) \Delta  \mathfrak v_{I,-2} \phi_I^2 \left(-3 (\Delta -5) \Delta +3 q^2-50\right)-64 \left(9 q^4+60 q^2-500\right) \mathfrak v_{I,-2}+125 \Delta ^4 \mathfrak v_{I,-2} \phi_I^8
\nonumber \\ &
-1000 (\Delta -2) \Delta ^3 \mathfrak v_{I,-2} \phi_I^6+ 768 q^4 \mathfrak v_{I,-2} \log (\rho )\Big]+\mathcal O \left(\rho ^4\right) \,, \\ 
\mathfrak v^{\bar  A} =&\ \mathfrak v^{\bar  A}_{I,0}+\mathfrak v^{\bar  A}_{I,l} \log (\rho )+\frac{1}{96} \rho ^2 \Big[24 q^2 (\mathfrak v^{\bar  A}_{I,0}-\mathfrak v^{\bar  A}_{I,l})+\mathfrak v^{\bar  A}_{I,l} \left(-5 \Delta ^2 \phi_I^4+8 (\Delta -5) \Delta  \phi_I^2-80\right)
\nonumber \\ &
+24 q^2 \mathfrak v^{\bar  A}_{I,l} \log (\rho )\Big]+\mathcal O \left(\rho ^4\right) \,, \\
\mathfrak v^{\hat A} =&\ \mathfrak v^{\hat A}_{I,0}+\mathfrak v^{\hat A}_{I,l} \log (\rho )+\frac{1}{96} \rho ^2 \Big[\left(24 q^2+6 g^2 \phi_I^2\right) \mathfrak v^{\hat A}_{I,0}+\left(-80-24 q^2-6 g^2 \phi_I^2-40 \Delta  \phi_I^2+\Delta ^2 \left(8 \phi_I^2-5 \phi_I^4\right)\right) \mathfrak v^{\hat A}_{I,l}\nonumber \\ &+\left(24 q^2+6 g^2 \phi_I^2\right) \log (\rho ) \mathfrak v^{\hat A}_{I,l}
   \Big]+\mathcal O \left(\rho ^4\right) \,.
\end{align}

For the tensor fluctuations, we have
\beq
\mathfrak e = \mathfrak e_{I,0}+\mathfrak e_{I,l} \log (\rho )+\frac{1}{192} \rho ^2 \Big[48 q^2 (\mathfrak e_{I,0}-\mathfrak e_{I,l})-25 \Delta ^2 \mathfrak e_{I,l} \phi_I^4+40 (\Delta -5) \Delta  \mathfrak e_{I,l} \phi_I^2-400 \mathfrak e_{I,l}+48 \mathfrak e_{I,l} q^2 \log (\rho )\Big]+\mathcal O \left(\rho ^4\right) \,.
\eeq


\subsubsection{UV expansions}

In this subsection, we put $\Delta = 2$, and $A_U = 0 = \chi_U$.\footnote{The dependence on $\chi_U$ and $A_U$ can be reinstated by making the substitution $q^2 \rightarrow e^{2\chi_U-2A_U} q^2$ in the expressions for the UV expansions.} We write the expansions in terms of $z \equiv e^{-\rho}$.

For the fluctuations of the scalars, we have
\begin{align}
\mathfrak a^{\phi} =&\ \mathfrak a^{\phi}_2 z^2+\mathfrak a^{\phi}_3 z^3+\frac{1}{2} \mathfrak a^{\phi}_2 q^2 z^4+\frac{1}{6} \mathfrak a^{\phi}_3 q^2 z^5+\frac{1}{24} \mathfrak a^{\phi}_2 \left(q^4-12 \phi_V^2\right) z^6 +\mathcal O \left(z^7\right) \,, \\ 
\mathfrak a^{\chi} =&\ \mathfrak a^{\chi}_0-\frac{1}{6} \mathfrak a^{\chi}_0 q^2 z^2 +\frac{1}{24} \mathfrak a^{\chi}_0 q^4 z^4+\mathfrak a^{\chi}_5 z^5+\frac{1}{144} \mathfrak a^{\chi}_0 q^2 \left(q^4-14 \phi_V^2\right) z^6 +\mathcal O \left(z^7\right) \,, \\
\mathfrak a^{\bar A} =&\ \mathfrak a^{\bar A}_0-\frac{1}{2} \mathfrak a^{\bar A}_0 q^2 z^2 +\mathfrak a^{\bar A}_3 z^3-\frac{1}{8} \mathfrak a^{\bar A}_0 q^4 z^4 +\frac{1}{10} \mathfrak a^{\bar A}_3 q^2 z^5-\frac{1}{144} \mathfrak a^{\bar A}_0 q^2 \left(q^4+10 \phi_V^2\right) z^6 +\mathcal O \left(z^7\right) \,, \\ 
\mathfrak a^{\hat A} =&\ \mathfrak a^{\hat A}_0-\frac{1}{2} \mathfrak a^{\hat A}_0 q^2 z^2 +\mathfrak a^{\hat A}_3 z^3+\frac{1}{16} \mathfrak a^{\hat A}_0 \left(g^2\phi_J^2-2 q^4\right) z^4 +\frac{1}{20} \left(\mathfrak a^{\hat A}_0 g^2\phi_J \phi_V+2 \mathfrak a^{\hat A}_3 q^2\right) z^5
\nonumber \\ &
-\frac{1}{288} \mathfrak a^{\hat A}_0 \left(2 q^6+(20+g^2) q^2 \phi_J^2-4g^2\phi_V^2\right) z^6 +\mathcal O \left(z^7\right) \,.
\end{align}

For the pseudo-scalar fluctuations, we have
\beq
\mathfrak p^{\hat A} = \mathfrak p^{\hat A}_0+\mathfrak p^{\hat A}_1 z+ \left(\frac{\mathfrak p^{\hat A}_0 q^2}{2}+\frac{\mathfrak p^{\hat A}_1 \phi_V}{\phi_J}\right) z^2 +\frac{2 \mathfrak p^{\hat A}_0 q^2 \phi_J \phi_V+\mathfrak p^{\hat A}_1 q^2 \phi_J^2+2 \mathfrak p^{\hat A}_1 \phi_V^2}{6 \phi_J^2} z^3 +\mathcal O \left(z^4\right) \,.
\eeq

For the vector fluctuations, we have
\begin{align}
\mathfrak v =&\ \mathfrak v_0-\frac{1}{6} q^2 \mathfrak v_0 z^2 +\frac{1}{24} q^4 \mathfrak v_0 z^4+\mathfrak v_5 z^5+\frac{1}{144} q^2 \mathfrak v_0 \left(q^4-14 \phi_V^2\right) z^6
\nonumber \\ &
+\frac{1}{350} q^2 (350 \mathfrak v_0 \chi_5-38 \mathfrak v_0 \phi_V \phi_J+25 \mathfrak v_5) z^7 +\mathcal O \left(z^8\right) \,, \\ 
\mathfrak v^{\bar A}  =&\ \mathfrak v^{\bar A} _0-\frac{1}{2} q^2 \mathfrak v^{\bar A} _0 z^2 +\mathfrak v^{\bar A} _3 z^3-\frac{1}{8} q^4 \mathfrak v^{\bar A} _0 z^4 +\frac{1}{10} q^2 \mathfrak v^{\bar A} _3 z^5-\frac{1}{144} q^2 \mathfrak v^{\bar A} _0 \left(q^4+10 \phi_V^2\right) z^6
\nonumber \\ &
+\frac{1}{280} \left(q^4 \mathfrak v^{\bar A} _3+6 q^2 \mathfrak v^{\bar A} _0 (15 \chi_5-\frac{26}{5} \phi_V \phi_J)+45 \mathfrak v^{\bar A} _3 \phi_V^2\right) z^7 +\mathcal O \left(z^8\right) \,, \\
\mathfrak v^{\hat A} =&\ \mathfrak v^{\hat A}_0-\frac{1}{2} q^2 \mathfrak v^{\hat A}_0 z^2 +\mathfrak v^{\hat A}_3 z^3-\frac{1}{8} \mathfrak v^{\hat A}_0 \left(q^4- \frac{g^2}{2}\phi_J^2\right) z^4 +\frac{1}{10} \left(q^2 \mathfrak v^{\hat A}_3+\frac{g^2}{2} \mathfrak v^{\hat A}_0 \phi_J \phi_V\right) z^5 +\mathcal O \left(z^6\right) \,.
\end{align}

For the tensor fluctuations, we have
\beq
\mathfrak e = \mathfrak e_0- \frac{1}{6} \mathfrak e_0 q^2 z^2 +\frac{1}{24} \mathfrak e_0 q^4 z^4+\mathfrak e_5 z^5+\mathcal O \left(z^6\right) \,.
\eeq

\section{Of two-point functions }
\label{sec:RGimprovement}

 In this short appendix,  we perform an exercise to demonstrate the relation between
mass spectra of vector fluctuations and holographically renormalised  two-point functions.
To this purpose, we digress and analyse a simplified system, 
in which analytical calculations  
can be carried out, and symmetries are manifest.
We consider a new $U(1)$ gauge theory living in a new six-dimensional spacetime,
yet adopt the same ansatz as in Eqs.~\eqref{eq:metricansatz} and~\eqref{eq:5dmetricansatz}.
At variance with the body of the paper, we consider solutions for which $A = 4\chi =4 \rho / 3$, resulting in an AdS$_6$ geometry. To introduce a scale, we hold the IR cutoff, $\rho_1=0$, fixed, and interpret it as a confinement scale, as in hard-wall models in the AdS/QCD literature~\cite{Erlich:2005qh,DaRold:2005mxj}.
The (reduced) five-dimensional action for the new $U(1)$ gauge field, $\mathcal A_M$, is\footnote{This action closely resembles the one for $\mathcal A_M^{\bar{\mathcal A}}$, in the body of the paper.}
\beq
	\mathcal S = \int \dd \rho \int \dd^4 x \sqrt{-g} \, \bigg\{ - \frac{1}{4} \, e^{2\chi} \, g^{MP} g^{NQ} \mathcal F_{MN} \mathcal F_{PQ} \bigg\} + \int \dd^4 x \sqrt{-\tilde g} \, \bigg\{ -\frac{1}{4} D_2 \,
\tilde{g}^{\mu\rho}\tilde{g}^{\nu\sigma}\mathcal F_{\mu\nu}\mathcal F_{\rho\sigma} \bigg\} \bigg|_{\rho = \rho_2} \,,
\eeq
where $\mathcal F_{MN} =\partial_M {\cal A}_N - \partial_N {\cal A}_M $, and  ${\cal A}_M$ has trivial background profile. After Fourier transforming in the Minkowski directions, the transverse part of the gauge field satisfies an equation of motion that can be written as follows:
\beq
\label{eq:EOMv}
    \bigg[ \partial_\rho^2 + 3\partial_\rho - e^{-2\rho} q^2 \bigg] P_{\mu\nu} \mathcal A^\nu(q,\rho) = 0 \,.
\eeq
In the following, we call $v(q,\rho)$ the solutions to this linear equation, 
which depend on the four-momentum, $q_{\mu}$, and on the radial direction, $\rho$.
In the IR, we impose  Neumann boundary conditions, $\partial_\rho v(q,\rho) \big|_{\rho = \rho_1}=0$.

The presence of the boundary-localised term, 
proportional to $D_2$, introduces (weak) gauging of the global $U(1)$ symmetry of the dual field theory.
The transverse part of the two-point function (propagator) is
\beqs
P^{\mu\sigma} P^{\nu\gamma} \langle {\cal A}_{\mu}(q){\cal A}_{\nu}(-q) \rangle &=&\lim_{\rho_2\rightarrow +\infty}
\frac{-i\,P^{\sigma\gamma}}{q^2}\,\frac{1}{\Pi(q^2,\rho_2)}\,,
\eeqs
where we anticipated  that we will take the limit $  \rho_2\rightarrow +\infty$, and where
\beqs
\label{Eq:ppp}
\Pi(q^2,\rho)&\equiv& D_2 +\frac{1}{q^2} e^{3\rho} \frac{\partial_{\rho} v(q,\rho)}{v(q,\rho)}\,.
\eeqs
To remove a divergence, that appears when $\rho_2\rightarrow +\infty$, we require that
\beqs
  \label{Eq:Dd2}
D_2&\equiv&-e^{\rho_2} +\frac{1}{\varepsilon^2}\,,
\eeqs
where $\varepsilon$ is a renormalisation constant. 
The transverse part of the propagator  in the background has the closed form:
\beqs
P^{\mu\sigma} P^{\nu\gamma} \langle {\cal A}_{\mu}(q){\cal A}_{\nu}(-q) \rangle 
&=&
\label{Eq:AdSprop1}
\frac{-i\,P^{\sigma\gamma}}{q^2}\,\frac{\varepsilon^2}{1+\varepsilon^2 \Pi_o(q^2)}\,,
\eeqs
and the physical $q^2$-dependence is encoded in
\beqs
\Pi_o(q^2)&=&-\frac{\sqrt{-q^2}}{\tan (\sqrt{-q^2})}\,.
\eeqs

The poles in Eq.~(\ref{Eq:AdSprop1}) identify the mass spectrum. 
Equivalently, one can solve Eq.~(\ref{eq:EOMv}), 
subject to Neumann boundary condition in the IR, $\partial_\rho v(q,\rho) \big|_{\rho = \rho_1}=0$, 
impose in the UV the constraint
$\partial_\rho v(q,\rho) +q^2 D_2 e^{-3\rho}v(q,\rho)\big|_{\rho = \rho_2}=0$,
hence identifying the discrete spectrum in $q^2$, 
and afterwards take the limit $\rho_2\rightarrow +\infty$ to recover the physical spectrum.

As long as $\varepsilon^2$ and  $|q^2|$ are small, the process discussed after 
Eq.~(\ref{Eq:AdSprop1}) yields sensible physical results.
However, if one considers large values of $|q^2|$,
 an unphysical tachyonic mode appears, with $q^2 \simeq (\varepsilon^2)^{-2}$,
which is due to having taken the limit $\rho_2\rightarrow 0$. 

If one, conversely, chooses $D_2$ to be non-negative, and retains a large, but finite, value for $\rho_2$, there are no tachyons while the rest of the spectrum is shifted only by small effects.
The appearance of the tachyon is, hence, truly unphysical, and it should be ignored.
Is there a way to write a two-point function that does not depend explicitly on the (unphysical) cutoff, $\rho_2$,
nor leads to the appearance of an (unphysical) tachyon?
We devote these final paragraphs to propose a simple and elegant solution to this puzzle.

As in ordinary perturbation theory, the presence of large hierarchies between the renormalisation scale and physical scales of interest may lead to difficulties, which can be overcome by applying the renormalisation group to improve
perturbation theory.
Inspired by the work on the holographic Wilsonian renormalisation group~\cite{Heemskerk:2010hk,Faulkner:2010jy} 
(see also Refs.~\cite{Laia:2011wf,Elander:2011vh,Elander:2021bmt} and Ref.~\cite{Papadimitriou:2010as}), we proceed as follows.

Suppose  one  computes the spectrum with a finite cutoff, $\rho_2 = \rho_{\Lambda}$, for a given
choice of $D_{2}$. One can ask how to change  $D_2$ into a function that  
depends on $q^2$ and $\rho_2$ so that the same spectrum is reproduced for any choice of the finite cutoff, $\rho_2$. This can be achieved by requiring that $D_2(q^2,\rho_2)$ satisfies the first-order differential equation
\beq
\label{eq:RGflowD2}
	\partial_{\rho_2} D_2 - e^{-3 \rho_2} q^2 D_2^2 + e^{\rho_2} = 0 \,,
\eeq
with boundary condition $D_2(q^2,\rho_{\Lambda}) = D_{2,\,\Lambda}$.

For concreteness, we set $D_{2,\Lambda} = 0$. From Eq.~(\ref{Eq:Dd2}), we see that if one identifies $\varepsilon^2 = e^{-\rho_{\Lambda}}$, then the new $D_2$ matches the previous expression at $\rho_2 = \rho_\Lambda$. The solution is given by
\beq
\label{eq:D2sol}
	D_2(q^2,\rho_2) = - \frac{e^{\rho_2}}{1 + e^{-\rho_2} \sqrt{-q^2} 
	\cot\left(\sqrt{-q^2}\left(\varepsilon^2 - e^{-\rho_2}\right)\right)} \,,
\eeq
and we can replace it into Eq.~(\ref{Eq:ppp}).
One may verify that, for any $q^2$ such that the resulting  $\Pi(q^2,\rho_2) = 0$, one also has $\partial_{\rho_2} \Pi(q^2,\rho_2) = 0$. The zeroes of the new $\Pi(q^2,\rho_2)$ do not change as a function of $\rho_2$, leaving the spectrum invariant.

We may now take the limit $\rho_2 \rightarrow \infty$, evolving $D_2(q^2,\rho_2)$ towards the UV by using Eq.~\eqref{eq:D2sol}, after which the renormalised two-point function is given by
\beq
\label{eq:AAren2}
	P^{\mu\sigma} P^{\nu\gamma} \langle \mathcal A_\mu(q) \mathcal A_\nu(-q) \rangle = (-i) \,\left( \frac{1}{- \frac{1}{\varepsilon^2} \Pi_o(q^2 (\varepsilon^2)^2) + \Pi_o(q^2)} \right)\, \frac{1}{q^2} \, P^{\sigma\gamma} \,.
\eeq
 Equation~\eqref{eq:AAren2} is approximately equal to 
Eq.~\eqref{Eq:AdSprop1} for small $\varepsilon^2$ and $q^2$, which validates the approximation used in the body of the paper. Furthermore, the (unphysical) tachyon is no longer present, while the physical states are retained.
Finally, by expanding $\langle \mathcal A_\mu(q) \mathcal A_\nu(-q) \rangle$
 in powers of small $q^2$, one finds a normalisation factor for the gauge fields, 
 so that the four-dimensional gauge coupling is 
\beq
	{g_4^2}= \frac{\varepsilon^2}{1 - \varepsilon^2} \,{g^2} \approx \varepsilon^2{g^2}  \,.
\eeq


\end{document}